\documentclass[authoryear,12pt]{elsarticle}

\usepackage[margin=1in]{geometry}

\usepackage{setspace}
\usepackage{amsmath}
\usepackage{amsthm}
\usepackage{graphicx}
\usepackage{caption}
\usepackage{subcaption}
\usepackage{natbib}

\makeatletter
\def\ps@pprintTitle{
\let\@oddhead\@empty
\let\@evenhead\@empty
\let\@oddfoot\@empty
\let\@evenfoot\@oddfoot
}
\makeatother

\begin{document}

\title{Evolution of worker policing}
\author{Jason~W.~Olejarz$^\mathrm{a}$, Benjamin~Allen$^\mathrm{b,a,c}$, Carl~Veller$^\mathrm{a,d}$, Raghavendra~Gadagkar$^\mathrm{e,f}$, and Martin~A.~Nowak$^\mathrm{a,d,g}$}
\address{$^\mathrm{a}$Program for Evolutionary Dynamics, Harvard University, Cambridge, MA~02138 USA\\$^\mathrm{b}$Department of Mathematics, Emmanuel College, Boston, MA~02115 USA\\$^\mathrm{c}$Center for Mathematical Sciences and Applications, Harvard University, Cambridge, MA 02138 USA\\$^\mathrm{d}$Department of Organismic and Evolutionary Biology, Harvard University, Cambridge, MA~02138 USA\\$^\mathrm{e}$Centre for Ecological Sciences and Centre for Contemporary Studies, Indian Institute of Science, Bangalore~560~012, India\\$^\mathrm{f}$Indian National Science Academy, New Delhi~110~002, India\\$^\mathrm{g}$Department of Mathematics, Harvard University, Cambridge, MA~02138 USA}

\begin{abstract}
Workers in insect societies are sometimes observed to kill male eggs of other workers, a phenomenon known as worker policing. We perform a mathematical analysis of the evolutionary dynamics of policing. We investigate the selective forces behind policing for both dominant and recessive mutations for different numbers of matings of the queen. The traditional, relatedness-based argument suggests that policing evolves if the queen mates with more than two males, but does not evolve if the queen mates with a single male.  We derive precise conditions for the invasion and stability of policing alleles. We find that the relatedness-based argument is not robust with respect to small changes in colony efficiency caused by policing. We also calculate evolutionarily singular strategies and determine when they are evolutionarily stable. We use a population genetics approach that applies to dominant or recessive mutations of any effect size.
\end{abstract}
\maketitle

\vspace*{10mm}

\section*{Keywords}

\noindent
Sociobiology, Natural Selection, Evolutionary Dynamics, Models / Simulations

\vspace*{10mm}

\section{Introduction}
\label{sec:intro}

In populations with haplodiploid genetics, unfertilized female workers are capable of laying male eggs.  Thus, in a haplodiploid colony, male eggs can in principle originate from the queen or from the workers.  Worker policing is a phenomenon where female workers kill the male eggs of unmated female workers \citep{Ratnieks_1988,Ratnieks_1989,Ratnieks_2006,Gadagkar_2001,Wenseleers_2006_a}.  Worker policing is observed in many social insects, including ants, bees, and wasps.  Yet the precise conditions for the evolution of worker policing are still unclear.

Worker policing \citep{Ratnieks_1988,Ratnieks_2006,Gadagkar_2001,Wenseleers_2006_a} and worker sterility \citep{Wilson_1971,Hamilton_1972,Olejarz_2015} are two distinct phenomena that are widespread in the eusocial Hymenoptera.  In addition to worker policing, a subset of workers in a colony may forego their own reproductive potential to aid in raising their siblings.  Prior relatedness-based arguments have suggested that queen monogamy is important for the evolution of a non-reproductive worker caste \citep{Hughes_2008,Cornwallis_2010,Queller_1998,Foster_2006,Boomsma_2007,Boomsma_2009}.  In contrast, it is believed that polygamy---not monogamy---is important for the evolution of police workers.

Several papers have studied the evolution of policing. \cite{Starr_1984} explores various topics in the reproductive biology and sociobiology of eusocial Hymenoptera. He defines promiscuity as $1/(\sum_{i=1}^n f_i^2)$, where $n$ is the number of matings of each queen, and $f_i$ is the fractional contribution to daughters by the $i$-th male mate. He writes, regarding workers, that ``They are on average less related to nephews than brothers whenever [promiscuity is greater than two] and should prefer that the queen lay all the male eggs. Workers would therefore be expected to interfere with each other's reproduction.'' Thus, \cite{Starr_1984} was the first to suggest that workers should raise their nephews (sons of other workers) if the queen mates once, but should only raise their brothers (sons of the queen) if the queen mates more than twice. \cite{Starr_1984} uses a relatedness-based argument, but he does not provide any calculation of evolutionary dynamics in support of his argument; he uses neither population genetics nor inclusive fitness theory. In a book on honeybee ecology, \cite{Seeley_1985} also proposed, using a relatedness-based argument, that worker policing should occur in colonies with multiply mated queens, but that worker policing should be absent if queens are singly mated.

\cite{Woyciechowski_1987} perform a calculation based on population genetics and conclude that workers should raise their nephews (sons of other workers) if the queen mates once, but should only raise their brothers (sons of the queen) if the queen mates more than twice---the case of double mating is neutral with respect to preference. From this result, they claim that, under multiple mating of the queen, natural selection should favor non-reproductive workers. \cite{Woyciechowski_1987} consider both dominant and recessive alleles affecting worker behavior, but they do not consider colony efficiency effects.

\cite{Ratnieks_1988} considers the invasion of a dominant allele for policing. Using population genetics, he arrives at essentially the same conclusion as \cite{Woyciechowski_1987}: In the absence of efficiency effects, policing evolves with triple mating but not with single mating. But Ratnieks also considers colony efficiency effects, focusing mainly on the case where policing improves colony efficiency. Since policing occurs alongside other maintenance tasks (such as cleaning of cells, removal of pathogens, incubation of brood), and since eating worker-laid eggs might allow workers to recycle some of the energy lost from laying eggs, Ratnieks supposes that policing improves colony efficiency. He finds that worker policing with singly mated queens may evolve if policing improves colony reproductive efficiency. He also finds that worker policing with triply mated queens may not evolve if policing reduces colony reproductive efficiency, but he considered this case to be unlikely on empirical grounds. Ratnieks does not study recessive policing alleles. He also does not calculate evolutionary stability conditions.

Both papers \citep{Woyciechowski_1987,Ratnieks_1988} offer calculations based on population genetics without mentioning or calculating inclusive fitness. These early studies \citep{Starr_1984,Seeley_1985,Woyciechowski_1987,Ratnieks_1988} were instrumental in establishing the field of worker policing.

Testing theoretical predictions on the evolution of worker policing in the field or in the lab is difficult.  Due to the complexities inherent in insect sociality, published empirical results are not always easy to interpret.  While, so far, worker policing has been found in all species with multiple mating that have been studied, it has also been found in about $20\%$ of species with singly mated queens \citep{Hammond_2004,Wenseleers_2006_b,Bonckaert_2008}.  Herein lies the difficulty:  When worker policing is found in multiply mated species and found to be absent in singly mated species, this is taken as evidence supporting the relatedness argument, and when worker policing is found in singly mated species, it is explained away as not being evidence against the theory, but as having evolved for other reasons (such as colony efficiency).  See, for example, the following quotation by \cite{Bonckaert_2008}:  ``Nevertheless, our results are important in that they show that {\it V. germanica} forms no exception to the rule that worker reproduction should be effectively policed in a species where queens mate multiple times \citep{Ratnieks_1988}.  Indeed, any exception to this pattern would be a much bigger challenge to the theory than the occurrence of worker policing in species with single mating, which can be readily explained \citep{Ratnieks_1988,Foster_2001_b}.''  This is precisely why a careful simultaneous consideration of relatedness, male parentage, and colony efficiency is important for understanding worker policing.

We do not aim to provide an exhaustive catalog of all species in which worker policing has been studied.  We merely cite some specific examples to add context.  Policing is rampant in colonies of the honeybee \citep{Ratnieks_1989}, the wasp {\it Vespula vulgaris} \citep{Foster_2001_c}, and the wasp {\it Vespula germanica} \citep{Bonckaert_2008}, which are all multiply mated.  (As mentioned above, worker policing has been found in all of the studied species to date that are multiply mated.)  Worker removal of worker-laid eggs is much less prevalent in colonies of the bumblebee \citep{Velthuis_2002}, the stingless bee, \citep{Peters_1999}, and the wasp {\it Vespula rufa} \citep{Wenseleers_2005}, which are predominantly singly mated.  (As mentioned above, worker policing has been found only in about $20\%$ of the studied species to date that are singly mated.)  There are some studies based on observational evidence that find policing in singly mated species; examples of species with single mating and worker policing are {\it Vespa crabro} \citep{Foster_2002}, {\it Camponotus floridanus} \citep{Endler_2004}, {\it Aphaenogaster smythiesi} \citep{Wenseleers_2006_b}, and {\it Diacamma} \citep{Wenseleers_2006_b}.

Interspecies comparisons are somewhat problematic, because even though phylogeny can be controlled for, there are many (known and unknown) ways in which species differ in addition to mating frequency that may also affect the absence or presence of worker policing.  Furthermore, many empirical studies are based on genetic analyses of male parentage.  (Though studies of some species are based on actual observational evidence; see, e.g., \cite{Wenseleers_2006_b}.)  Regarding species for which the study of policing is based on genetic analyses, policing is often inferred if males are found to originate predominantly from the queen.  But such an inference, in cases where it is made, presupposes that workers actively try to lay male eggs in the first place.  It is therefore not clear how reliably genetic investigations can measure policing.

The small number of attempts at measuring the prevalence of worker policing in intraspecific experiments have also returned conflicting results.  \cite{Foster_2000} report that facultative worker policing in the saxon wasp, {\it Dolichovespula saxonica}, is more common in colonies headed by multiply mated queens.  But their sample size is only nine colonies.  The phenomenon was reinvestigated by \cite{Bonckaert_2011} who report no evidence of facultative worker policing depending on queen mating frequencies, and argue that the previous result may have been flawed or that there were interpopulational variations. 

Many empirical studies have emphasized that factors besides intracolony relatedness---including the effects of policing on a colony's rate of production of offspring---may play a role in explaining evolution of worker policing \citep{Foster_2001_a,Foster_2001_c,Hartmann_2003,Hammond_2004,Wenseleers_2006_b,Helantera_2007,Khila_2008,Zanette_2012}.  Yet reliable published data on the effect that policing has on colony reproductive efficiency are often hard to find.  (For some exceptions, see \cite{Wenseleers_2013} and references therein.)

In this paper, we derive precise conditions for the evolutionary invasion and evolutionary stability of police alleles. We consider any number of matings, changes in the proportion of queen-derived males, changes in colony efficiency, and both dominant and recessive mutations that affect the intensity of policing.

Our paper is based on an analysis of evolutionary dynamics and population genetics of haplodiploid species \citep{Nowak_2010,Olejarz_2015}.  It does not use inclusive fitness theory. Specifically, we adapt the mathematical approach that was developed by \cite{Olejarz_2015} for the evolution of non-reproductive workers.  We derive evolutionary invasion and stability conditions for police alleles. Mathematical details are given in \ref{appendix:stability}.

In Section~\ref{sec:model}, we present the basic model and state the general result for any number of matings for dominant policing alleles.  In Sections \ref{sec:single}, \ref{sec:double}, and \ref{sec:triple}, we specifically discuss single, double, and triple mating for dominant policing alleles. We take dominance of the policing allele to be the more realistic possibility because the policing phenotype is a gained function. Nonetheless, for completeness, we give the general result for recessive policing alleles in Section~\ref{sec:recessive}. In Section~\ref{sec:shape}, we discuss how the shape of the efficiency function determines whether or not policing is more likely to evolve for single or multiple matings. In Section~\ref{sec:gradual}, we analyze our results for the case where the phenotypic mutation induced by the mutant allele is weak (or, equivalently in our formalism, the case of weak penetrance). In this setting, the quantity of interest is the intensity of policing. We locate the evolutionarily singular strategies. These are the values of intensity of policing for which mutant workers with slightly different policing behavior are, to first order in the mutant phenotype, neither advantageous nor disadvantageous. We then determine if a singular strategy is an evolutionarily stable strategy (ESS). In Section~\ref{sec:IF}, we discuss the relationship between policing and inclusive fitness theory, together with the limitations of the relatedness-based argument. Section~\ref{sec:discussion} concludes.

\vspace*{10mm}

\section{The model}
\label{sec:model}

We investigate worker policing in insect colonies with haplodiploid genetics.  Each queen mates $n$ times.  We derive conditions under which a mutation that effects worker policing can spread in a population.  We make the simplifying assumption, as do \cite{Woyciechowski_1987} and \cite{Ratnieks_1988}, that the colony's sex ratio is not affected by the intensity of worker policing.

\begin{figure}
\begin{center}
\includegraphics[width=0.95\textwidth]{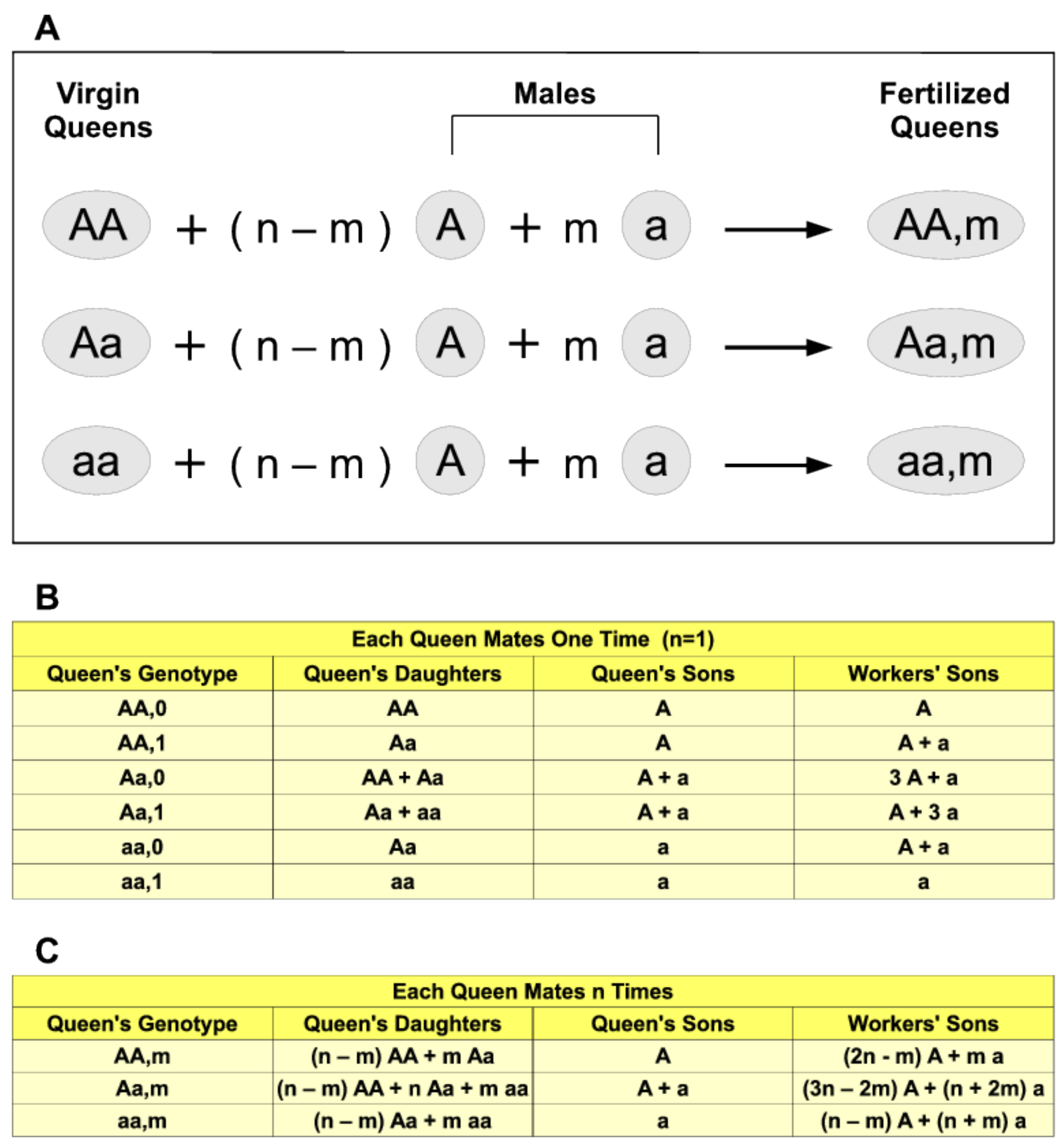}
\caption{(a)  The possible mating events with haplodiploid genetics are shown.  Each queen mates with $n$ males.  $m$ denotes the number of times that a queen mates with mutant type $a$ males and can take values between $0$ and $n$.  Thus, there are $3(n+1)$ types of colonies.  (b)  If each queen mates with only a single male, then there are six types of colonies.  The female and male offspring (right three columns) of each colony (leftmost column) are shown.  For example, $AA,1$ colonies arise from a type $AA$ female mating with a single mutant type $a$ male.  $AA,1$ queens produce female offspring of type $Aa$ and male offspring of type $A$.  $50\%$ of the offspring of workers in $AA,1$ colonies are of type $A$, while the remaining $50\%$ of the offspring of workers in $AA,1$ colonies are of type $a$.  (c)  The female and male offspring (right three columns) of each colony (leftmost column) when each queen mates $n$ times are shown.}
\label{fig:colonies}
\end{center}
\end{figure}

First we consider the case of a dominant mutant allele. Because the policing allele confers a gain of function on its bearer, the assumption that it is dominant is reasonable. There are two types of males, $A$ and $a$. There are three types of females, $AA$, $Aa$, and $aa$. If the mutant allele is dominant, then $Aa$ and $aa$ workers kill the male eggs of other workers, while $AA$ workers do not. (Alternatively, $AA$ workers police with intensity $Z_{AA}$, while $Aa$ and $aa$ workers police with intensity $Z_{Aa}=Z_{aa}=Z_{AA}+w$. We consider this case in Section~\ref{sec:gradual}.) For $n$ matings, there are $3(n+1)$ types of mated queens. We use the notation $AA,m$; $Aa,m$; and $aa,m$ to denote the genome of the queen and the number, $m$, of her matings that were with mutant males, $a$. The parameter $m$ can assume values $0,1,...,n$. A schematic of the possible mating events is shown in Figure~\ref{fig:colonies}(a).

There are three types of females, $AA$, $Aa$, and $aa$, and there are $n+1$ possible combinations of males that each queen can mate with. (For example, a queen that mates three times ($n=3$) can mate with three type $A$ males, two type $A$ males and one type $a$ male, one type $A$ male and two type $a$ males, or three type $a$ males.) Figure~\ref{fig:colonies}(b) shows the different colony types and the offspring of each type of colony when each queen is singly mated. Figure~\ref{fig:colonies}(c) shows the different colony types and the offspring of each type of colony when each queen mates $n$ times. The invasion of the mutant allele only depends on a subset of colony types. The calculations of invasion conditions are presented in detail in \ref{appendix:stability}.

\vspace*{10mm}

\subsection{Fraction of male offspring produced by the queen}

\begin{figure}
\begin{center}
\includegraphics[width=0.5\textwidth]{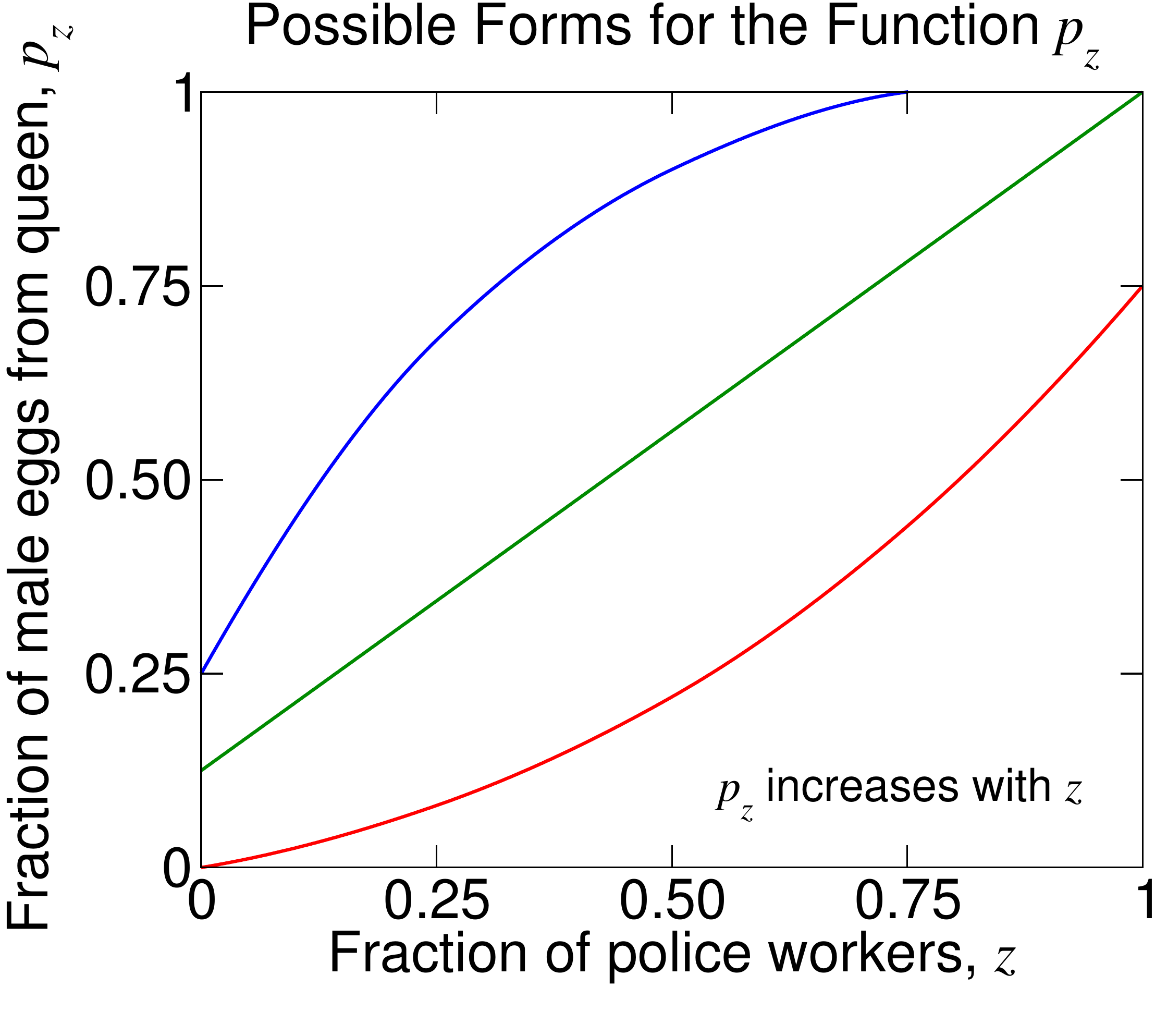}
\caption{The queen's production of male eggs, $p_z$, increases with the fraction of workers that are policing, $z$.  This is intuitive, since having a larger worker police force means that a greater amount of worker-laid eggs can be eaten or removed.  Three possibilities for a monotonically increasing function $p_z$ are shown.}
\label{fig:p}
\end{center}
\end{figure}

$p_z$ represents the fraction of males that are queen-derived if the fraction of police workers is $z$. (This quantity was already employed by \cite{Ratnieks_1988}.) The parameter $z$ can vary between 0 and 1. For $z=0$, there are no police workers in the colony, and for $z=1$, all workers in the colony are policing. We expect that $p_z$ is an increasing function of $z$. Increasing the fraction of police workers increases the fraction of surviving male eggs that come from the queen (Figure~\ref{fig:p}).

\vspace*{10mm}

\subsection{Colony efficiency as a function of policing}

\begin{figure}
\begin{center}
\includegraphics[width=0.5\textwidth]{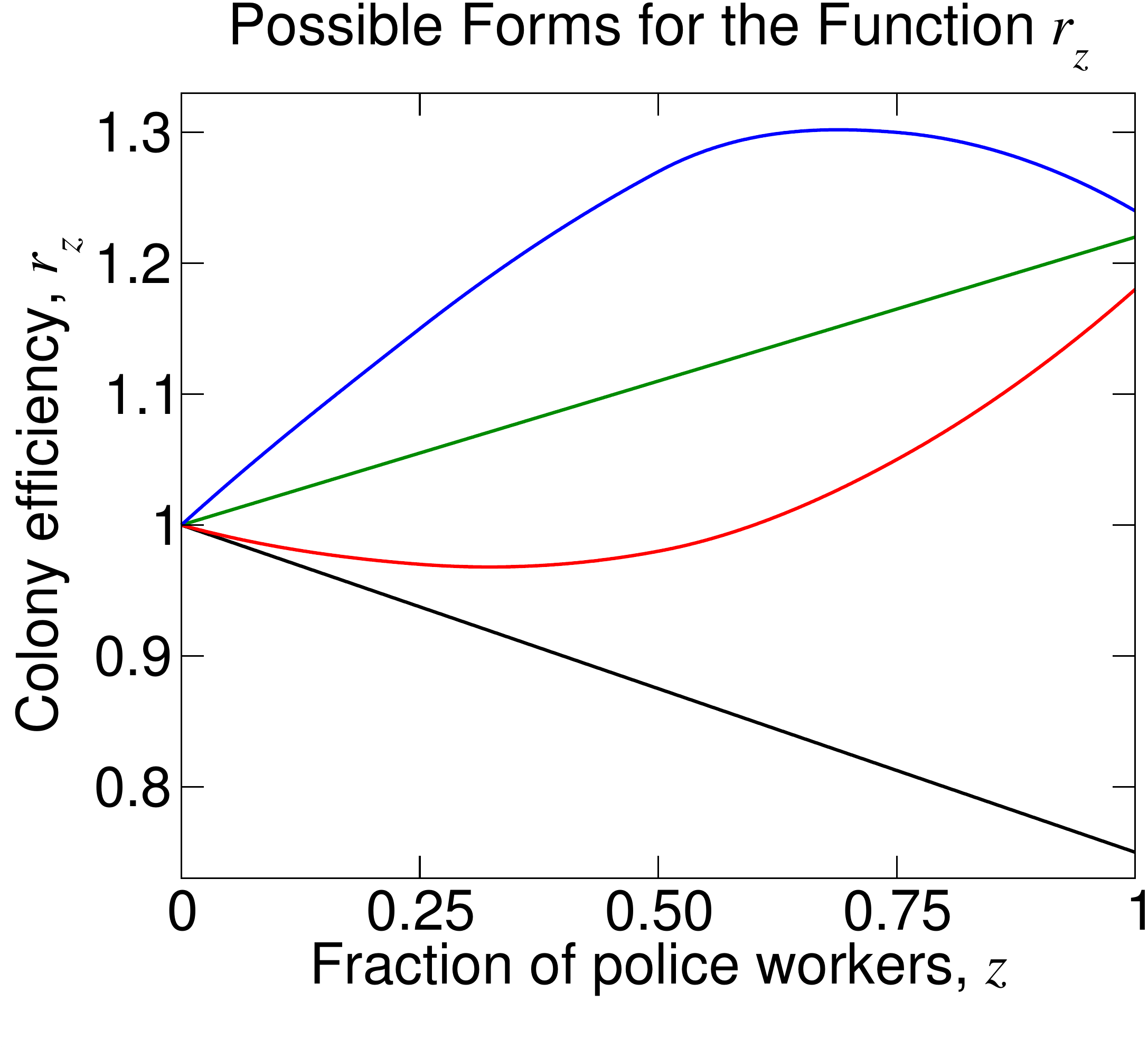}
\caption{The functional dependence of colony efficiency, $r_z$, on the fraction of workers that are policing, $z$, may take any one of many possibilities.}
\label{fig:r}
\end{center}
\end{figure}

$r_z$ represents the rate at which a colony produces offspring (virgin queens and males) if the fraction of police workers is $z$. (This quantity was also employed by \cite{Ratnieks_1988}.) Without loss of generality we can set $r_0=1$.  For a given mutation that affects the intensity of policing, and for a given biological setting, the efficiency function $r_z$ may take any one of a variety of forms (Figure~\ref{fig:r}).

Colony efficiency depends on interactions among police workers and other colony members. It also depends on the interactions of colonies and their environment. There are some obvious negative effects that policing can have on colony efficiency.  By the act of killing eggs, police workers are directly diminishing the number of potential offspring.  In the process of identifying and killing nephews, police workers may also be expending energy that could otherwise be spent on important colony maintenance tasks \citep{Cole_1986,Naeger_2013}.  Policing can also be costly if there are recognitional mistakes, i.e., queen-laid eggs may accidentally be removed by workers.  Recognitional errors could result in modifications to the sex ratio, which is an important extension of our model but is beyond the scope of this paper.

We can also identify positive effects that policing may have on colony efficiency.  It has been hypothesized that the eggs which are killed by police workers may be less viable than other male eggs \citep{Velthuis_2002,Pirk_2004,Gadagkar_2004,Nonacs_2006}, although this possibility has been disputed \citep{Beekman_2005,Helantera_2006,Zanette_2012}.  If less-viable worker-laid eggs are competing with more-viable queen-laid male eggs, then policing may contribute positively to overall colony efficiency.  Moreover, policing decreases the incentive for workers to expend their energy laying eggs in the first place \citep{Foster_2001_a,Wenseleers_2004_a,Wenseleers_2004_b,Wenseleers_2006_a}, which could be another positive influence on colony efficiency. (However, the decrease in incentive for workers to reproduce due to policing would only arise on a short time scale if there is a facultative response to policing, which is unlikely.)

As another speculative possibility:  Could it be that worker egg-laying and subsequent policing acts as a form of redistribution within the colony? That is, suppose that it is better for colony efficiency to have many average-condition workers than to have some in poor condition and some in good condition. Suppose further, as seems realistic, that good-condition workers are more likely to lay eggs (which are high in nutritional content, of course). If the average police worker is of condition below the average egg-laying worker, then worker egg-laying and policing serves to redistribute condition among the workers, improving overall colony efficiency.

The special case, where policing has no effect on colony efficiency and which has informed the conventional wisdom, is ungeneric, because policing certainly has energetic consequences for the colony that cannot be expected to balance out completely.  An early theoretical investigation of colony efficiency effects regarding invasion of dominant mutations that effect worker policing was performed by \cite{Ratnieks_1988}.

Although monotonically increasing or monotonically decreasing functions $r_z$ are the simplest possibilities, these cases are not exhaustive.  For example, a small or moderate amount of policing may be expected to improve colony efficiency.  However, the precise number of police workers that are needed to effectively police the entire worker population is unclear.  It is possible that a fraction $z<1$ of police workers can effectively police the entire population, and adding additional police workers beyond a certain point could result in wasted energy, inefficient use of colony resources, additional recognitional errors, etc. These effects may correspond to colony efficiency $r_z$ reaching a maximum value for some $0<z<1$.

As another possibility, suppose that police workers, when their number is rare, directly decrease colony efficiency by the act of killing male eggs.  It is possible that for some $z<1$, police workers are sufficiently abundant that their presence can be detected by other workers.  Assuming the possibility of some type of facultative response, the potentially reproductive workers may behaviorally adapt by reducing their propensity to lay male eggs, instead directing their energy at raising the queen's offspring.  In this scenario, colony efficiency $r_z$ may reach a minimum value for some $0<z<1$.

\vspace*{10mm}

\subsection{Main results for dominant police alleles}

We derive the following main results for dominant police alleles. If the queen mates with $n$ males, then the $a$ allele for policing can invade an $A$ resident population provided the following ``evolutionary invasion condition'' holds:
\begin{equation}
\frac{p_{1/n}+p_{1/2}}{2}\left(\frac{r_{1/n}}{r_0}\right)\left(\frac{r_{1/2}}{r_0}\right) > 2-\left(\frac{r_{1/2}}{r_0}\right)-\left(1-p_{1/n}\right)\left(\frac{r_{1/n}}{r_0}\right)
\label{eqn:dom_r}
\end{equation}
When considering only one mutation, $r_0$ can be set as $1$ without loss of generality. Why are the four parameters, $r_{1/n}$, $r_{1/2}$, $p_{1/n}$, and $p_{1/2}$, sufficient to quantify the condition for invasion of the mutant allele, $a$? Since we consider invasion of $a$, the frequency of the mutant allele is low. Therefore, almost all colonies are of type $AA,0$, which means a wild-type queen, $AA$, has mated with $n$ wild-type males, $A$, and 0 mutant males, $a$. In addition, the colonies $Aa,0$ and $AA,1$ are relevant. These are all colony types that include exactly one mutant allele. Colony types that include more than one mutant allele (such as $Aa,1$ or $AA,2$) are too rare to contribute to the invasion dynamics. For an $Aa,0$ colony, half of all workers are policing, and therefore the parameters $r_{1/2}$ and $p_{1/2}$ occur in Equation~\eqref{eqn:dom_r}. For an $AA,1$ colony, $1/n$ of all workers are policing, which explains the occurrence of $r_{1/n}$ and $p_{1/n}$ in Equation~\eqref{eqn:dom_r}.

Next, we ask the converse question: What happens if a population in which all workers are policing is perturbed by the introduction of a rare mutant allele that prevents workers from policing?  If the $a$ allele for worker policing is fully dominant, and if colony efficiency is affected by policing, then a resident policing population is stable against invasion by non-police workers if the following ``evolutionary stability condition'' holds:
\begin{equation}
\frac{r_1}{r_{(2n-1)/(2n)}} > \frac{(2+n)(2+p_1)+p_{(2n-1)/(2n)}(n-2)}{2(2+n+np_1)}
\label{eqn:dom_r_stable}
\end{equation}
What is the intuition behind the occurrence of the four parameters, $r_{1}$, $r_{(2n-1)/(2n)}$, $p_{1}$, and $p_{(2n-1)/(2n)}$? The condition applies to a population in which all workers are initially policing. Note that, because the allele, $a$, for policing is fully dominant in our treatment, non-policing behavior arises if at least two mutant $A$ alleles for non-policing are present in the genome of the colony, which is the combination of the queen's genome and the sperm she has stored. To study the invasion of a non-policing mutant allele, we must consider all colony types that have $0$, $1$, or $2$ mutant $A$ alleles; these are $aa,n$; $aa,n-1$; $Aa,n$; $aa,n-2$; $Aa,n-1$; and $AA,n$.  The colonies $aa,n$; $aa,n-1$; $Aa,n$; $aa,n-2$; and $AA,n$ do not contain non-police workers; the efficiency of those colonies is $r_1$, and the fraction of male eggs that originate from the queen in those colonies is $p_1$.  Both of these parameters occur in Equation~\eqref{eqn:dom_r_stable}. Colonies of type $Aa,n-1$ produce a fraction of $1/(2n)$ non-police workers, which explains the occurrence of $r_{(2n-1)/(2n)}$ and $p_{(2n-1)/(2n)}$ in Equation~\eqref{eqn:dom_r_stable}.

\begin{figure}
\centering
\begin{subfigure}{0.45\textwidth}
\centering
\caption{}
\includegraphics*[width=1\textwidth]{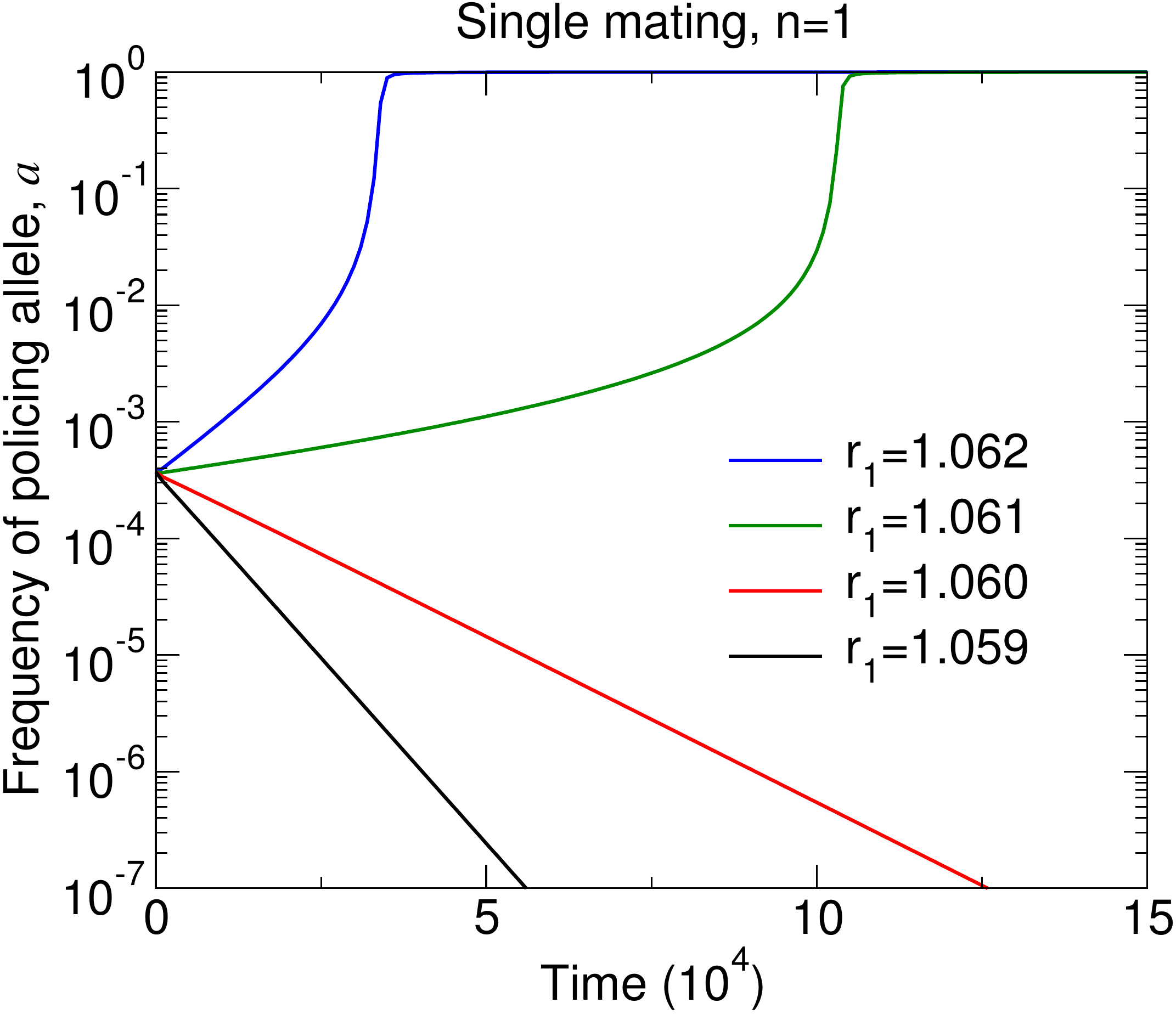}
\end{subfigure}
\qquad
\begin{subfigure}{0.45\textwidth}
\centering
\caption{}
\includegraphics*[width=1\textwidth]{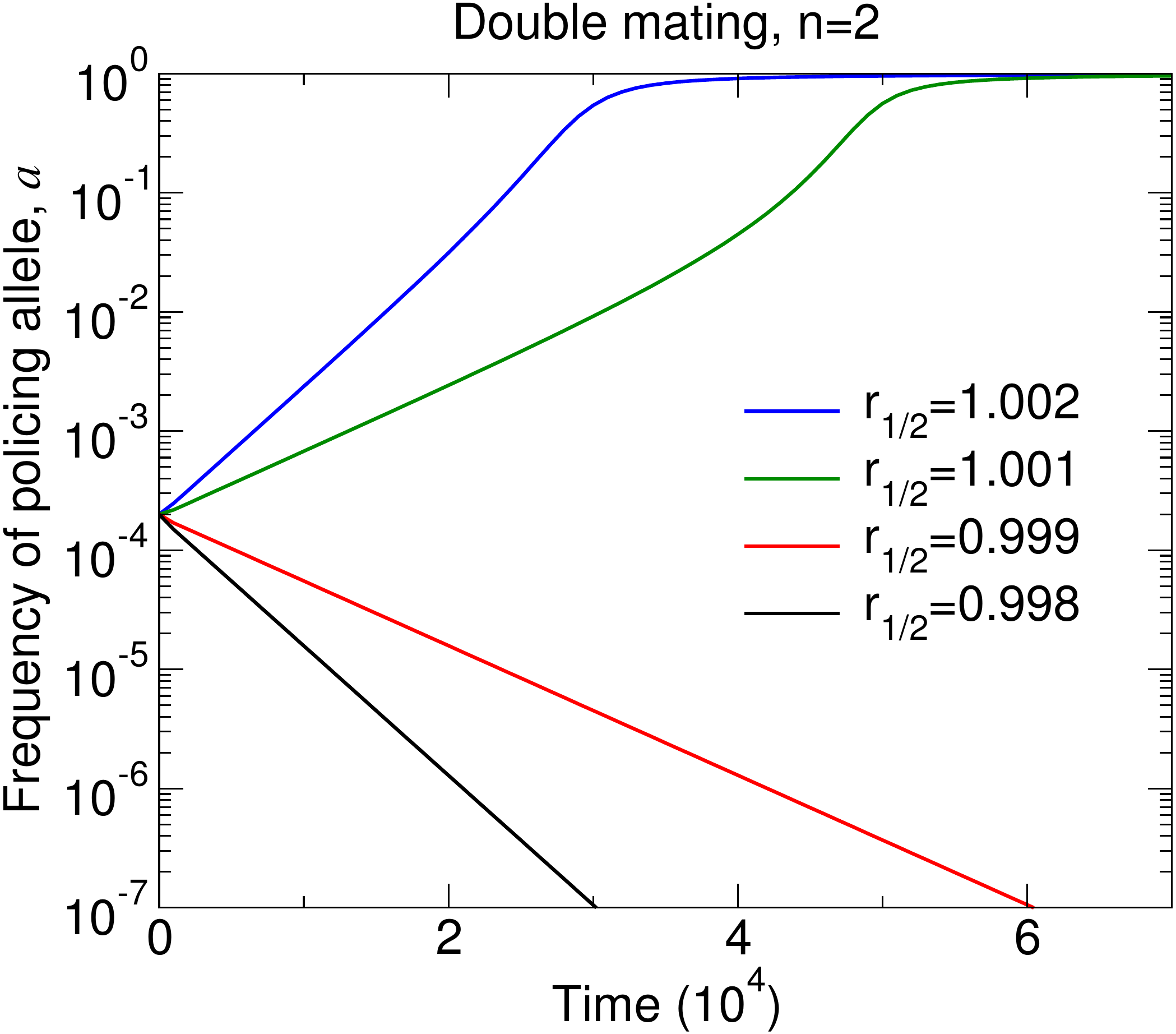}
\end{subfigure}
\caption{Numerical simulations of the evolutionary dynamics of worker policing confirm the condition given by Equation~\eqref{eqn:dom_r}.  The policing allele is dominant.  For numerically probing invasion, we use the initial condition $X_{AA,0}=1-10^{-3}$ and $X_{AA,1}=10^{-3}$.  We set $r_0=1$ without loss of generality.  Other parameters are: (a) $p_{1/2}=0.75$, $p_1=0.9$, and $r_{1/2}=1.01$; (b) $p_{1/2}=0.6$, $p_1=0.8$, $r_{3/4}=1.005$, and $r_1=1.01$.}
\label{fig:dom_sim}
\end{figure}

Numerical simulations of the evolutionary dynamics with a dominant police allele are shown in Figure~\ref{fig:dom_sim}.

\begin{figure}
\begin{center}
\includegraphics[width=0.9\textwidth]{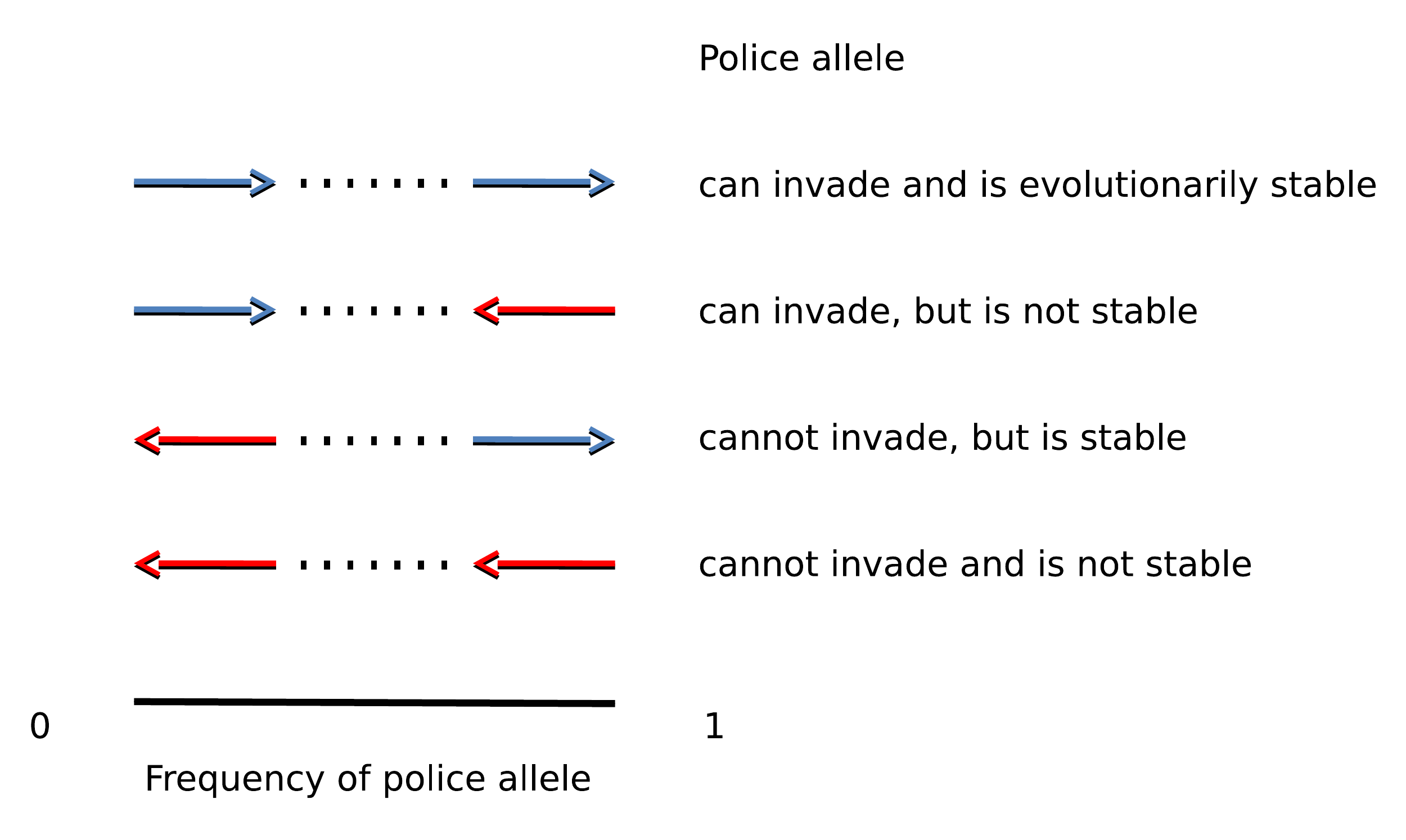}
\caption{There are four possibilities for the dynamical behavior in the proximity of two pure equilibria.}
\label{fig:arrows}
\end{center}
\end{figure}

Generally, four scenarios regarding the two pure equilibria are possible:  Policing may not be able to invade and be unstable, policing may not be able to invade but be stable, policing may be able to invade but be unstable, or policing may be able to invade and be stable.  The possibilities are shown in Figure~\ref{fig:arrows}.  In the cases where policing cannot invade but is stable, or where policing can invade but is unstable, Brouwer's fixed-point theorem guarantees the existence of at least one mixed equilibrium.  In the case where policing can invade but is unstable, police and non-police workers will coexist indefinitely.

We will now discuss the implications of our results for particular numbers of matings.

\vspace*{10mm}

\section{Single mating}
\label{sec:single}

For single mating, $n=1$, the invasion condition for a dominant police allele is
\begin{equation}
r_1 > \frac{2(2-r_{1/2})}{2(1-p_1)+(p_1+p_{1/2})r_{1/2}}
\label{eqn:dom_1_r}
\end{equation}
(Recall that $r_0=1$.)

The stability condition for a dominant police allele is 
\begin{equation}
r_1 > \frac{6-p_{1/2}+3p_1}{6+2p_1} r_{1/2}
\label{eqn:dom_1_r_stable}
\end{equation}

Evolution of policing is highly sensitive to changes in colony efficiency. For example, let us consider $p_{1/2}=0.99$ and $p_1=1$. This means that if half of all workers police then 99\% of all males come from the queen. If all workers police then all males come from the queen. In this case, efficiency values such as $r_{1/2}=1.001$ and $r_1=1.0031$ lead to the evolution of policing. In principle, arbitrarily small increases in colony efficiency can lead to the evolution of policing for single mating.

\begin{figure}
\centering
\begin{subfigure}{0.45\textwidth}
\centering
\caption{}
\includegraphics*[width=1\textwidth]{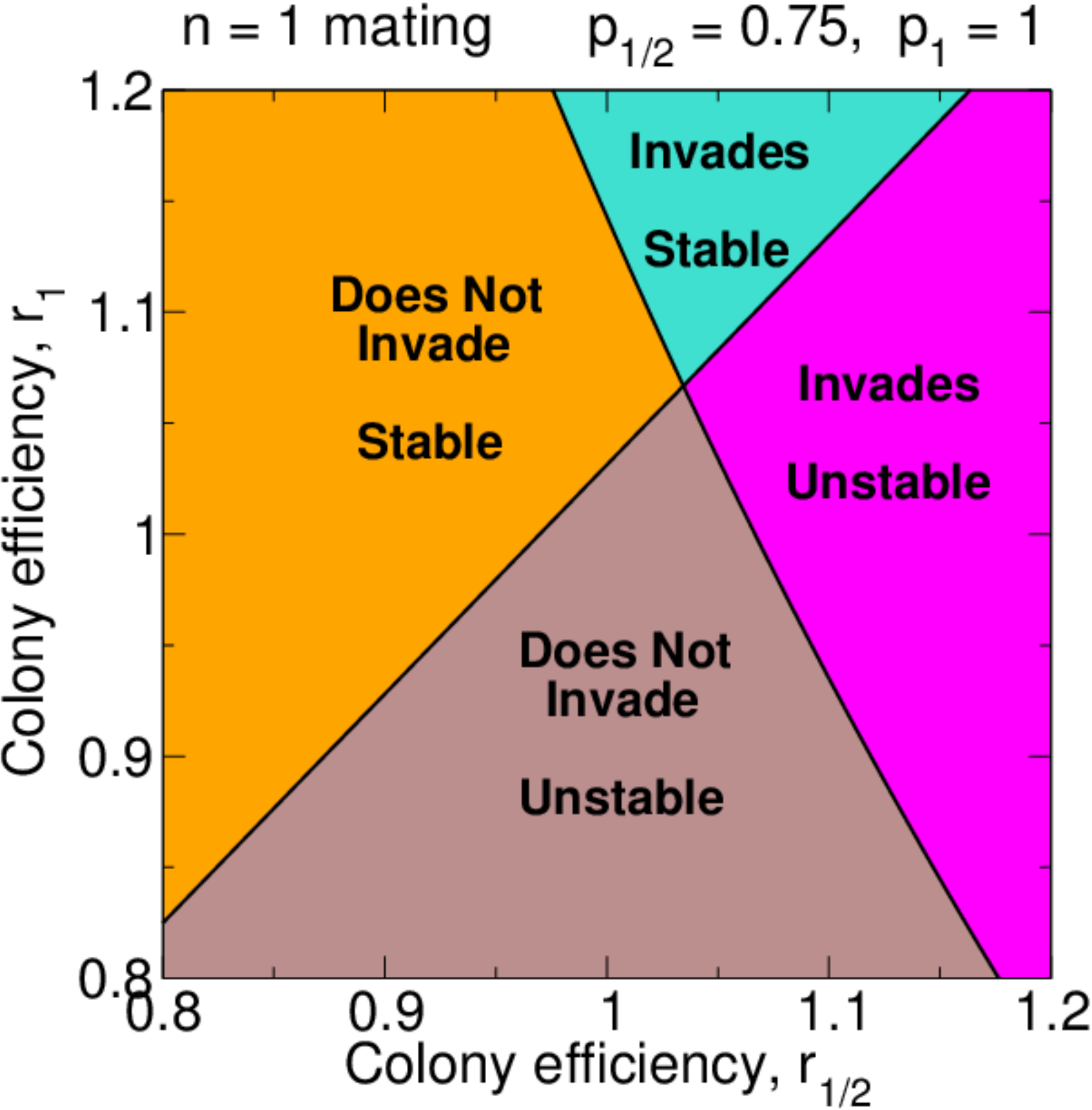}
\end{subfigure}
\qquad
\begin{subfigure}{0.45\textwidth}
\centering
\caption{}
\includegraphics*[width=1\textwidth]{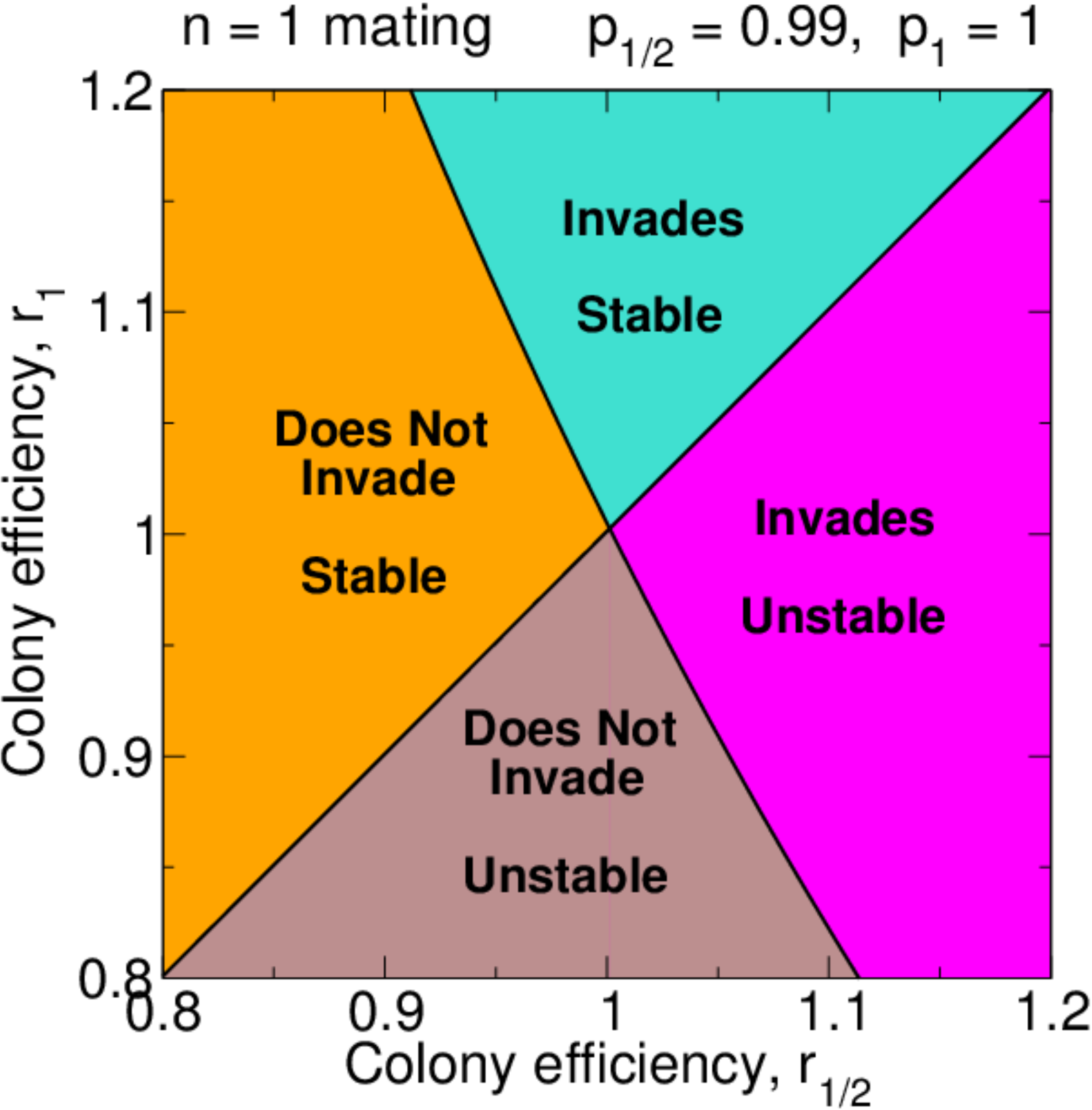}
\end{subfigure}
\caption{If queens are singly mated ($n=1$), then a plot of $r_1$ versus $r_{1/2}$ clearly shows all four possibilities for the behavior around the two pure equilibria.  For (a), we set $p_{1/2}=0.75$ and $p_1=1$.  For (b), we set $p_{1/2}=0.99$ and $p_1=1$.}
\label{fig:n1regions}
\end{figure}

\begin{figure}
\centering
\begin{subfigure}{0.45\textwidth}
\centering
\caption{}
\includegraphics*[width=1\textwidth]{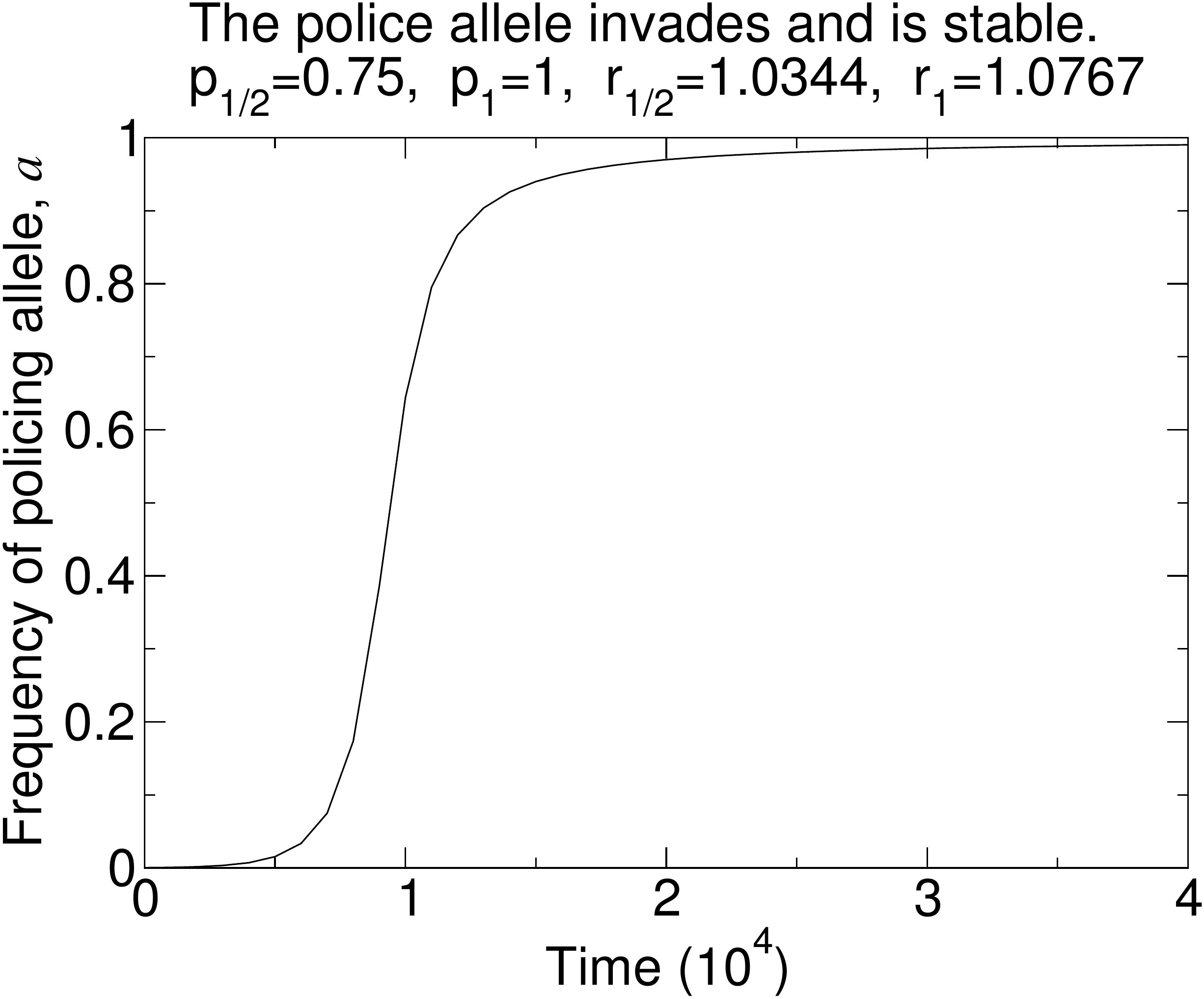}
\end{subfigure}
\qquad
\begin{subfigure}{0.45\textwidth}
\centering
\caption{}
\includegraphics*[width=1\textwidth]{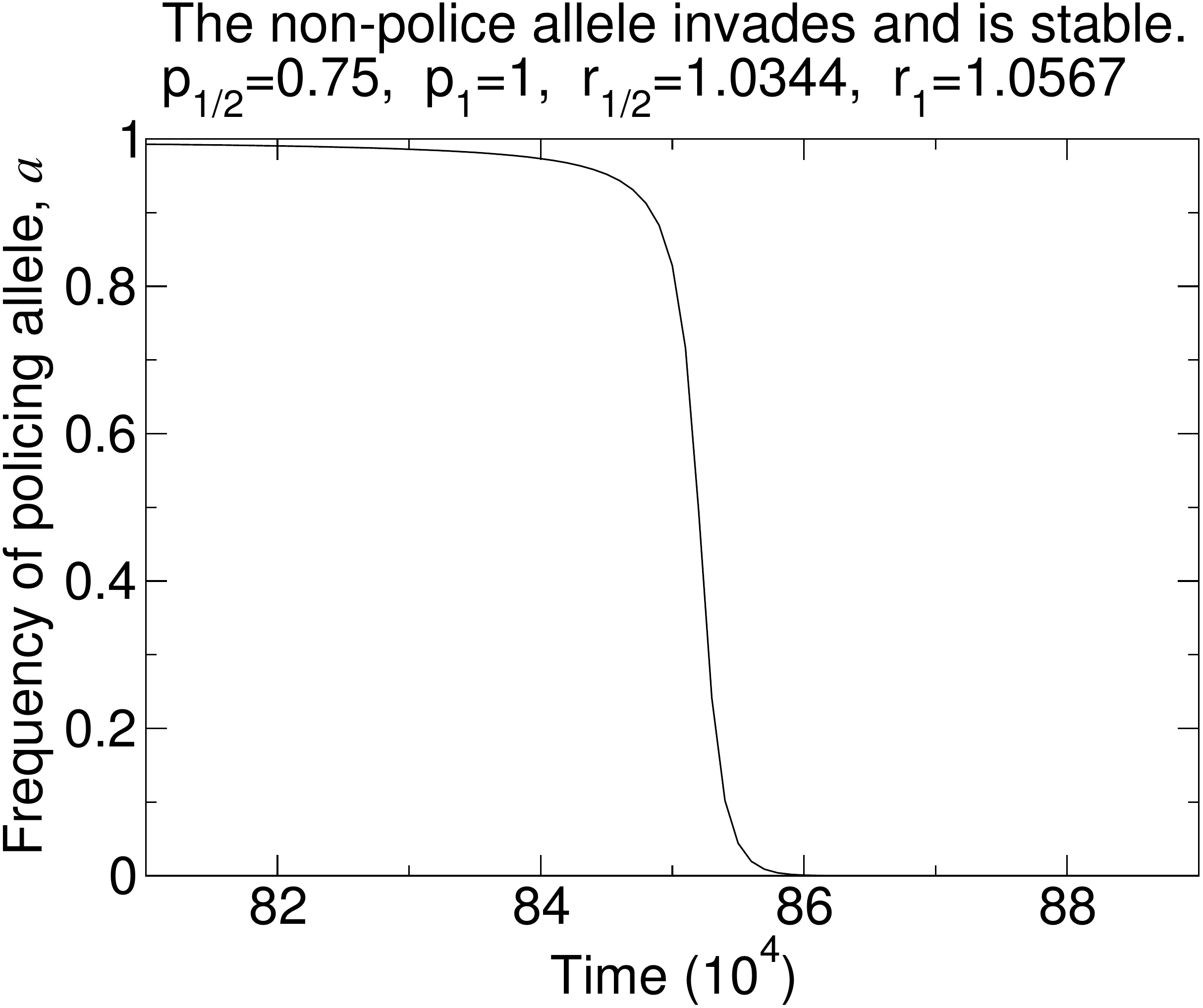}
\end{subfigure}
\par\medskip
\begin{subfigure}{0.45\textwidth}
\centering
\caption{}
\includegraphics*[width=1\textwidth]{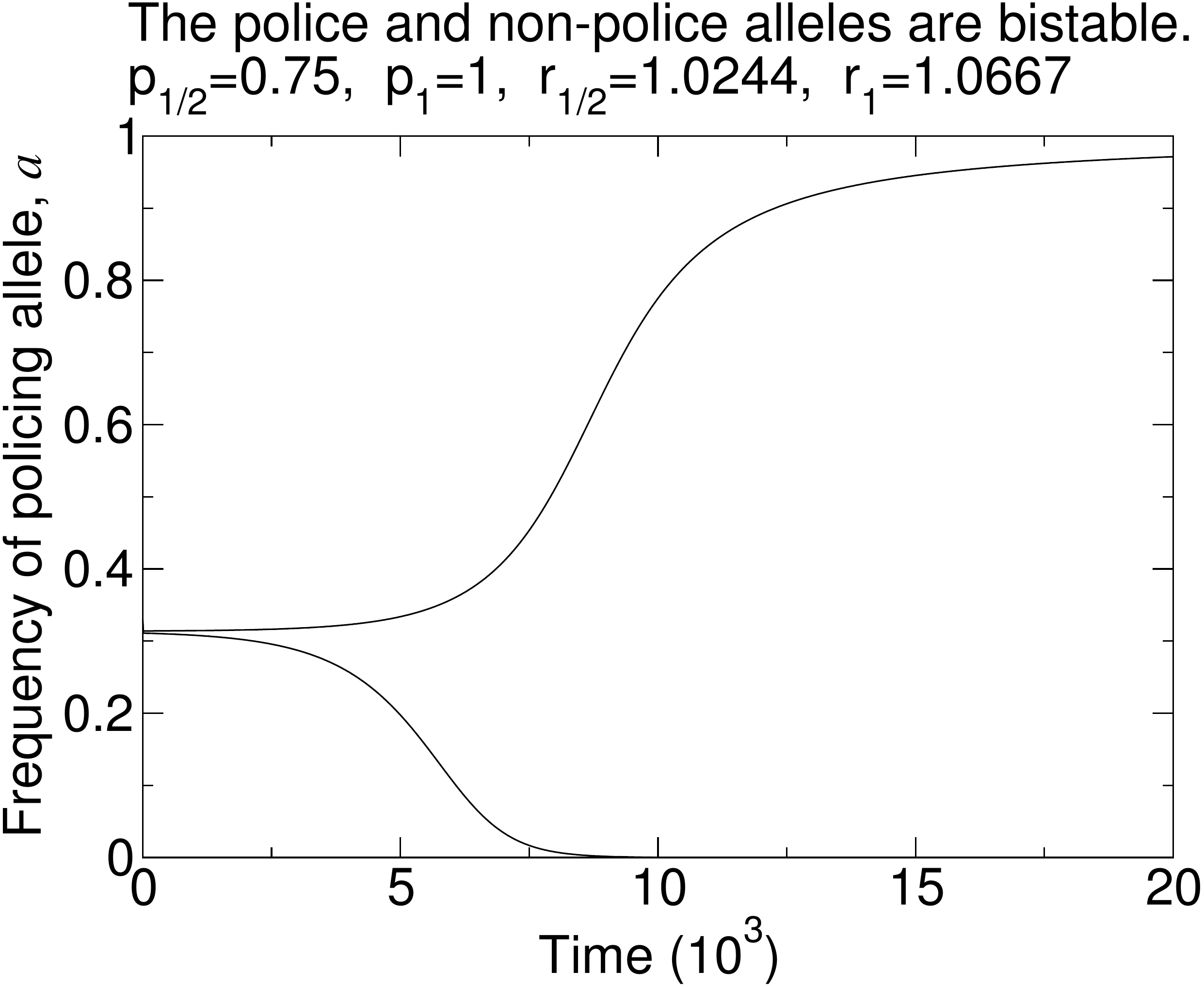}
\end{subfigure}
\qquad
\begin{subfigure}{0.45\textwidth}
\centering
\caption{}
\includegraphics*[width=1\textwidth]{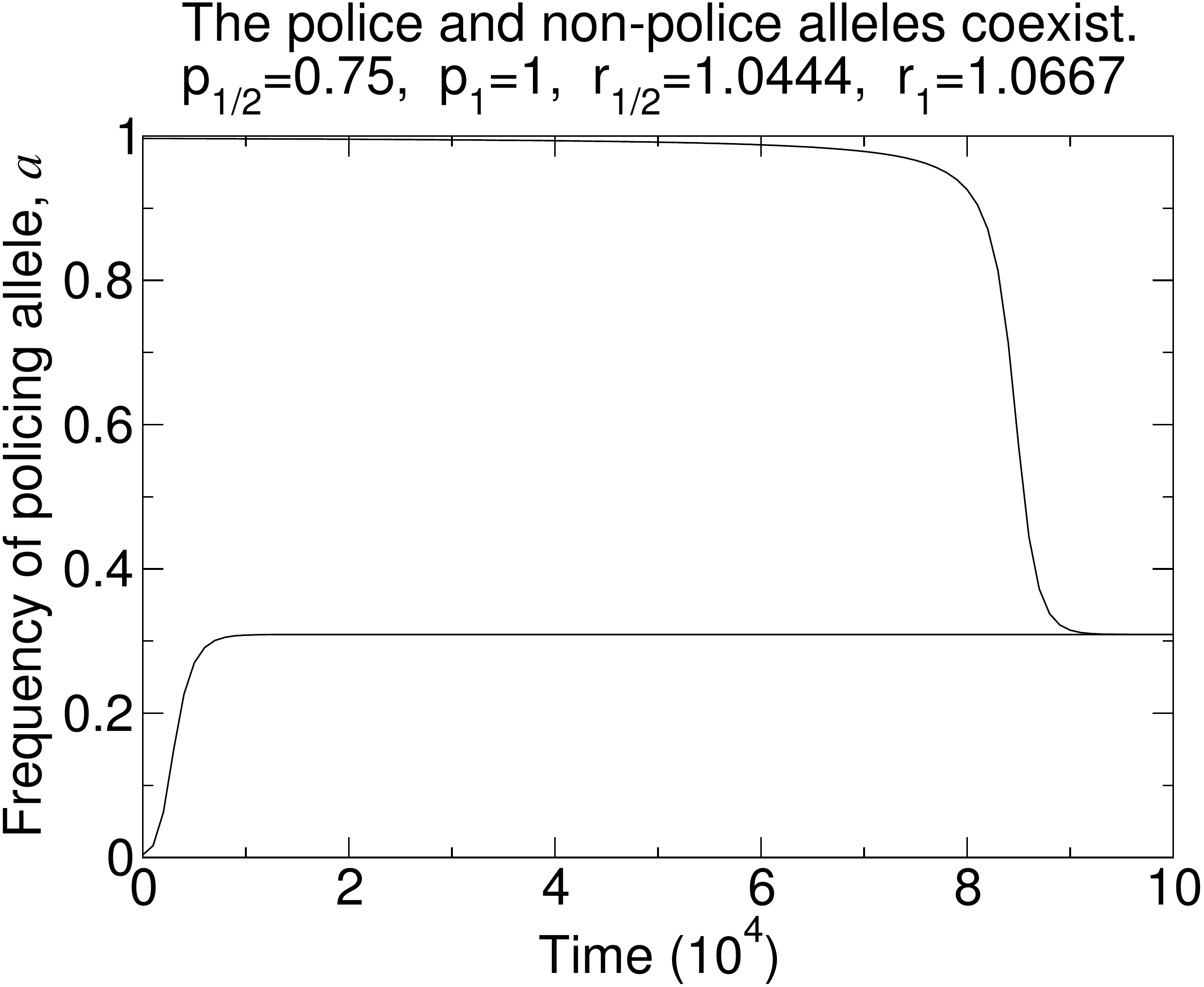}
\end{subfigure}
\caption{Numerical simulations of the evolutionary dynamics of worker policing that show the four behaviors in Figure~\ref{fig:n1regions}(a).  The policing allele is dominant.  For each of the four panels, we use the initial conditions: (a) $X_{AA,0}=1-10^{-3}$ and $X_{AA,1}=10^{-3}$; (b) $X_{aa,1}=1-10^{-3}$ and $X_{aa,0}=10^{-3}$; (c) $X_{AA,0}=0.02$ and $X_{AA,1}=0.98$ (lower curve), and $X_{AA,0}=0.01$ and $X_{AA,1}=0.99$ (upper curve); (d) $X_{AA,0}=1-10^{-2}$ and $X_{AA,1}=10^{-2}$ (lower curve), and $X_{aa,1}=1-10^{-2}$ and $X_{aa,0}=10^{-2}$ (upper curve).  We set $r_0=1$ without loss of generality.}
\label{fig:n1regions_sim}
\end{figure}

A plot of $r_1$ versus $r_{1/2}$ for singly mated queens (Figure~\ref{fig:n1regions}) illustrates the rich behavior highlighted in Figure~\ref{fig:arrows}.  Numerical simulations of the evolutionary dynamics are shown in Figure~\ref{fig:n1regions_sim}.

\begin{figure}
\begin{center}
\includegraphics[width=0.5\textwidth]{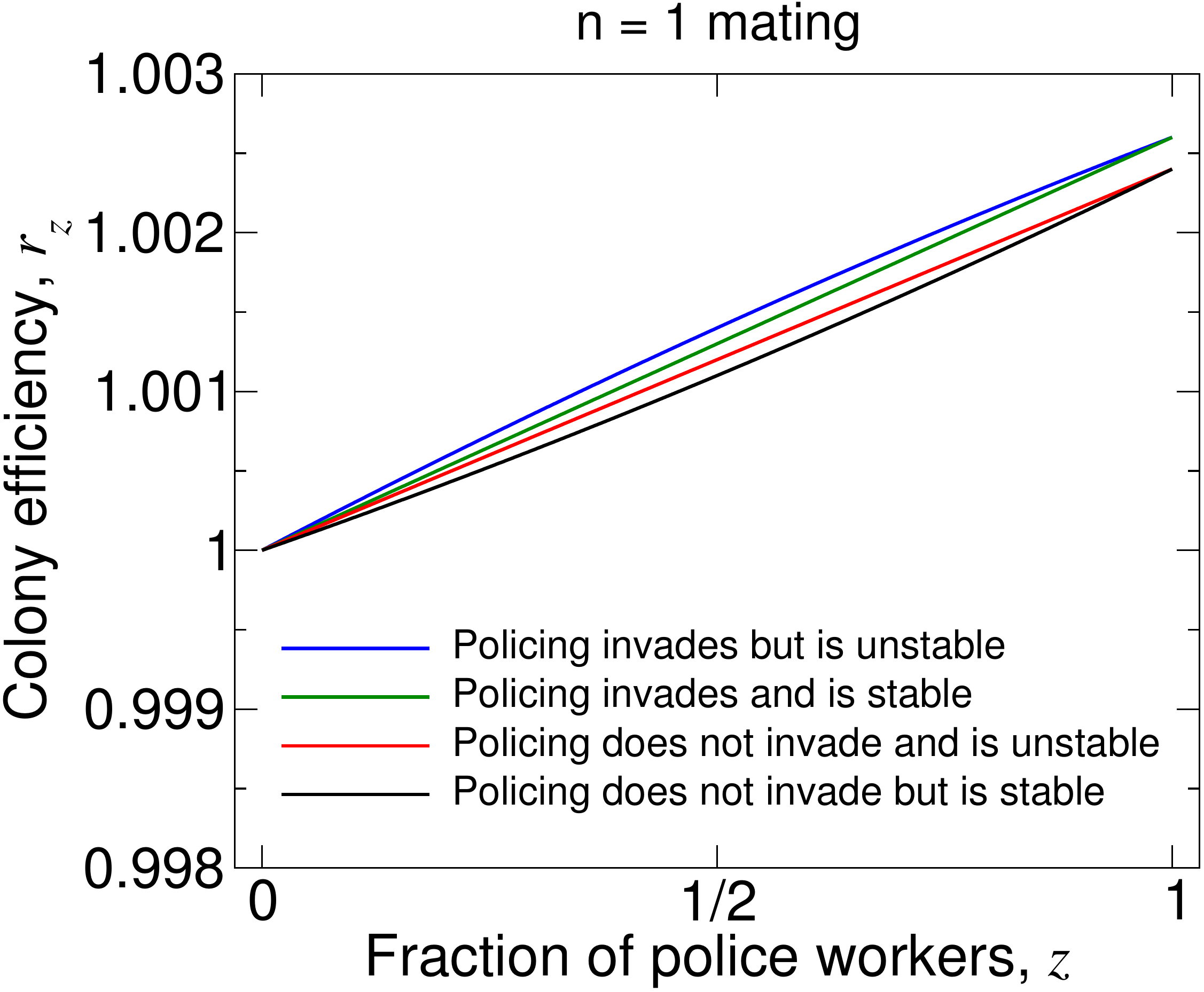}
\caption{Possible $r_z$ efficiency curves for $n=1$ mating which demonstrate different behaviors.  For this plot, we set $p_{1/2}=0.99$ and $p_1=1$.  Here, each curve has the functional form $r_z=1+\alpha z+\beta z^2$.  For example, we can have: (blue) policing invades but is unstable, $\alpha=0.003$, $\beta=-0.0004$; (green) policing invades and is stable, $\alpha=0.0026$, $\beta=0$; (red) policing does not invade and is unstable, $\alpha=0.0024$, $\beta=0$; (black) policing does not invade but is stable, $\alpha=0.002$, $\beta=0.0004$.}
\label{fig:curves_1}
\end{center}
\end{figure}

Another intriguing feature is that increases in colony efficiency due to policing do not necessarily result in a higher frequency of police workers at equilibrium.  Figure~\ref{fig:curves_1} illustrates this phenomenon.  Four possibilities for the efficiency function $r_z$ are shown.  Notice that the $r_z$ curve which results in coexistence of police workers and non-police workers (blue, top) is strictly greater than the $r_z$ curve which results in all workers policing (green, second from top).   How can increased efficiency due to policing possibly result in policing being less abundant at equilibrium?  If a mutation for non-policing behavior is introduced into a resident policing population, then the evolutionary success of the non-policing mutation depends on the success of $Aa,0$ colonies relative to $aa,1$, $aa,0$, $Aa,1$, and $AA,1$ colonies.  $Aa,0$ colonies have an efficiency parameter $r_{1/2}$, while the other four relevant colonies each have an efficiency parameter $r_1$.  Thus, if $r_{1/2}$ is too large relative to $r_1$, then the non-police allele can invade a resident policing population, and there is coexistence.  

Also notice that the $r_z$ curve which results in bistability of police workers and non-police workers (black, bottom) is strictly less than the $r_z$ curve which results in policing being dominated by non-policing (red, second from bottom).   This phenomenon arises in a similar way: if $r_{1/2}$ is too small relative to $r_1$, then the non-police allele cannot invade a resident policing population, and there is bistability.

\vspace*{10mm}

\section{Double mating}
\label{sec:double}

For double mating, $n=2$, the invasion condition for a dominant police allele is given by
\begin{equation}
r_{1/2} > 1
\label{eqn:dom_2_r}
\end{equation}
Thus, policing can invade if there is an infinitesimal increase in colony efficiency when half of all workers police.  Policing cannot invade if there is an infinitesimal decrease in colony efficiency when half of all workers police.

The stability condition for policing is given by
\begin{equation}
r_1 > r_{3/4}
\label{eqn:dom_2_r_stable}
\end{equation}
Therefore, the policing allele is stable if the colony efficiency is greater for $z=1$ (when all workers police) than for $z=3/4$ (when only three quarters of the workers police). 

\begin{figure}
\begin{center}
\includegraphics[width=0.5\textwidth]{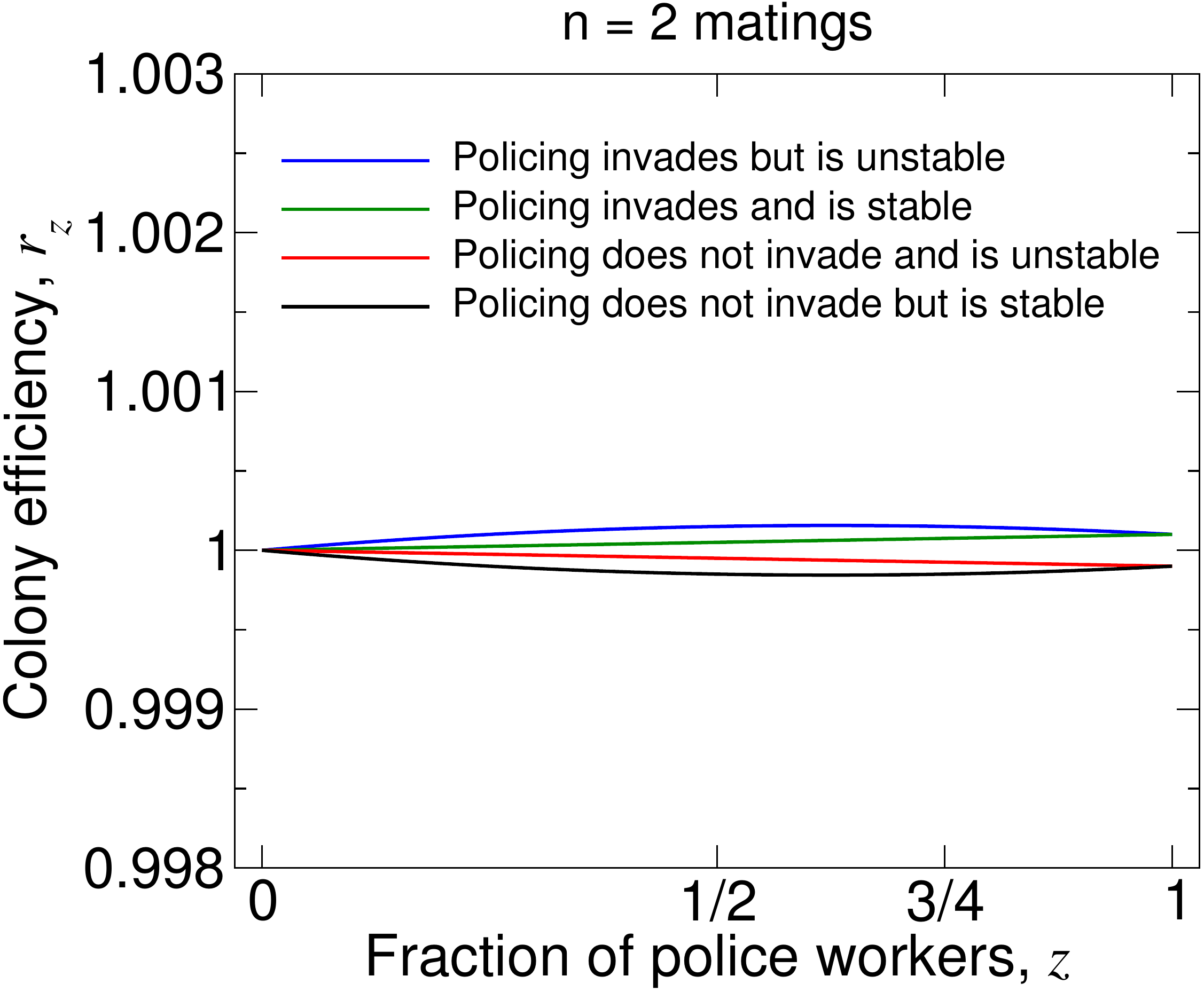}
\caption{Possible $r_z$ efficiency curves for $n=2$ matings which demonstrate different behaviors.  Here, each curve has the functional form $r_z=1+\alpha z+\beta z^2$.  For example, we can have: (blue) policing invades but is unstable, $\alpha=0.0005$, $\beta=-0.0004$; (green) policing invades and is stable, $\alpha=0.0001$, $\beta=0$; (red) policing does not invade and is unstable, $\alpha=-0.0001$, $\beta=0$; (black) policing does not invade but is stable, $\alpha=-0.0005$, $\beta=0.0004$.}
\label{fig:curves_2}
\end{center}
\end{figure}

Four possible efficiency curves $r_z$ and the corresponding behavior of the police allele are shown in Figure~\ref{fig:curves_2}.

\vspace*{10mm}

\section{Triple mating}
\label{sec:triple}

For triple mating, $n=3$, the invasion condition for a dominant police allele is given by
\begin{equation}
r_{1/2} > \frac{4-2(1-p_{1/3})r_{1/3}}{2+(p_{1/3}+p_{1/2})r_{1/3}}
\label{eqn:dom_3_r}
\end{equation}

The stability condition for policing is given by
\begin{equation}
r_1 > \frac{10+p_{5/6}+5p_1}{10+6p_1} r_{5/6}
\label{eqn:dom_3_r_stable}
\end{equation}

As a numerical example, let us consider $p_{1/3}=0.98$ and $p_{1/2}=0.99$. If $z=1/3$ of workers police, then $98\%$ of males come from the queen. If $z=1/2$ of workers police, then $99\%$ of males come from the queen. In this case, policing cannot invade if $r_{1/3}=0.9990$ and $r_{1/2}=0.9979$. In principle, arbitrarily small reductions in colony efficiency can prevent evolution of policing for triple mating. 

\begin{figure}
\begin{center}
\includegraphics[width=0.5\textwidth]{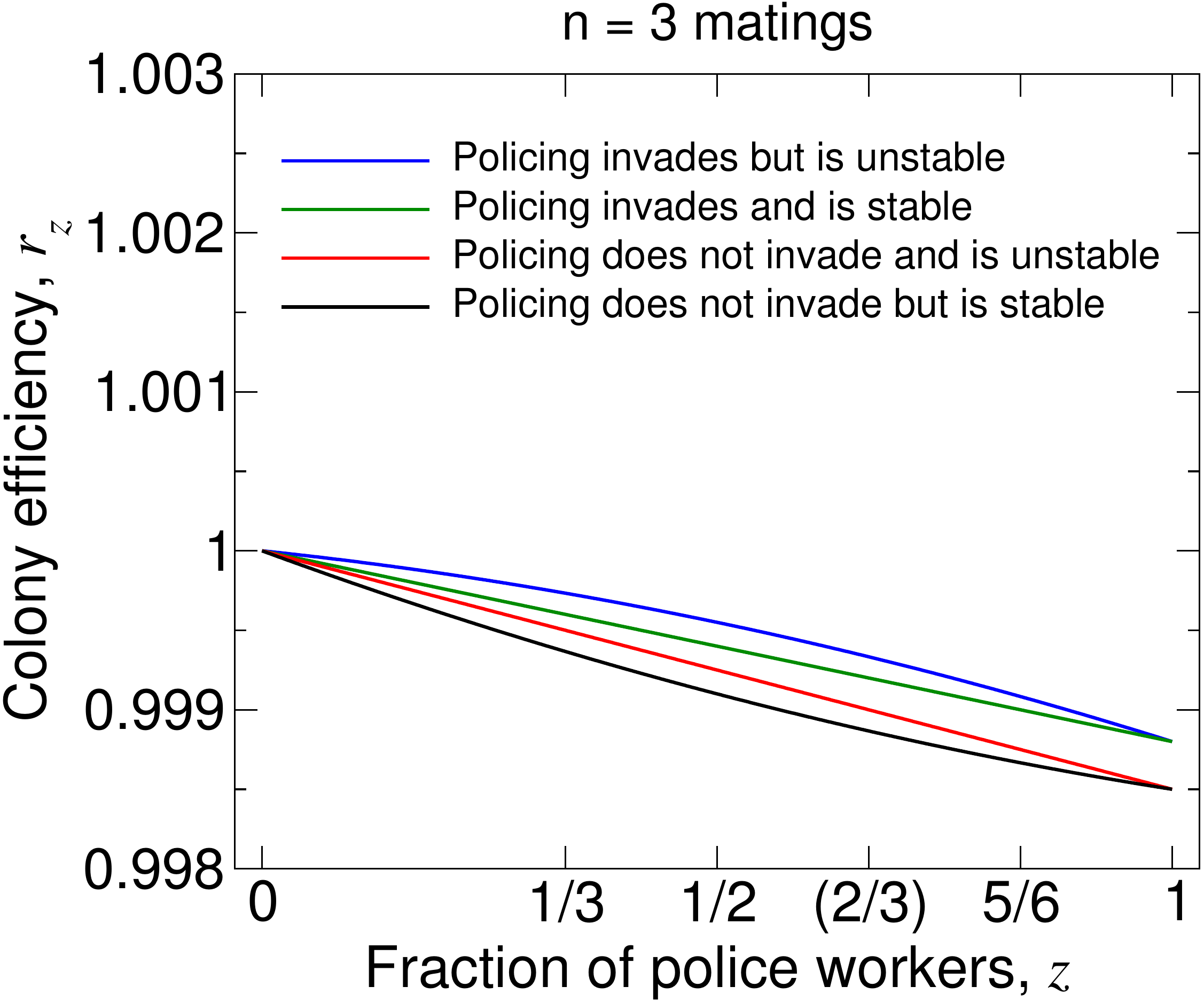}
\caption{Possible $r_z$ efficiency curves for $n=3$ matings which demonstrate different behaviors.  For this plot, we set $p_{1/3}=0.98\overline{6}$, $p_{1/2}=0.99$, $p_{5/6}=0.99\overline{6}$, and $p_1=1$.  Here, each curve has the functional form $r_z=1+\alpha z+\beta z^2$.  For example, we can have: (blue) policing invades but is unstable, $\alpha=-0.0006$, $\beta=-0.0006$; (green) policing invades and is stable, $\alpha=-0.0012$, $\beta=0$; (red) policing does not invade and is unstable, $\alpha=-0.0015$, $\beta=0$; (black) policing does not invade but is stable, $\alpha=-0.0021$, $\beta=0.0006$.  Note that the value $r_{2/3}$ affects the population dynamics but does not appear in the conditions for invasion and stability of the police allele, hence the parentheses on the horizontal axis.}
\label{fig:curves_3}
\end{center}
\end{figure}

Just as for single mating, we observe the intriguing feature that increases in colony efficiency due to policing do not necessarily result in a higher frequency of police workers at equilibrium.  Figure~\ref{fig:curves_3} illustrates this phenomenon.  Four possibilities for the efficiency function $r_z$ are shown.  Notice that the $r_z$ curve which results in coexistence of police workers and non-police workers (blue, top) is strictly greater than the $r_z$ curve which results in all workers policing (green, second from top).  Also notice that the $r_z$ curve which results in bistability of police workers and non-police workers (black, bottom) is strictly less than the $r_z$ curve which results in policing being dominated by non-policing (red, second from bottom).

\vspace*{10mm}

\section{Recessive police allele}
\label{sec:recessive}

We have also derived the conditions for the emergence and evolutionary stability of worker policing if the police allele is fully recessive.  In this case, $AA$ and $Aa$ workers are phenotypically identical and do not police, while $aa$ workers do police.  (Alternatively, $AA$ and $Aa$ workers police with intensity $Z_{AA}=Z_{Aa}$, while $aa$ workers police with intensity $Z_{aa}=Z_{AA}+w=Z_{Aa}+w$. We consider this case in Section~\ref{sec:gradual}.)

\vspace*{10mm}

\subsection*{Emergence of worker policing}

The invasion condition for a recessive police allele, $a$, is given by
\begin{equation}
\frac{r_{1/(2n)}}{r_0} > \frac{2(2+n+np_0)}{(2+n)(2+p_0)+p_{1/(2n)}(n-2)}
\label{eqn:rec_r}
\end{equation}
Note that Equation~\eqref{eqn:rec_r} for invasion of a recessive police allele has the same mathematical form as Equation~\eqref{eqn:dom_r_stable} for evolutionary stability of a dominant police allele.  Starting from Equation~\eqref{eqn:dom_r_stable}, making the substitution $z \rightarrow 1-z$, and reversing the inequality, we recover Equation~\eqref{eqn:rec_r}.  The intuition behind this correspondence is described in \ref{appendix:stability}.

\vspace*{10mm}

\subsection*{Stability of worker policing}

A recessive police allele, $a$, is evolutionarily stable if
\begin{equation}
\left(\frac{r_1}{r_{(n-1)/n}}\right)\left[2\left(\frac{r_1}{r_{1/2}}\right)-1\right]-\left(1-p_{(n-1)/n}\right)\left(\frac{r_1}{r_{1/2}}\right) > \frac{p_{(n-1)/n}+p_{1/2}}{2}
\label{eqn:rec_r_stable}
\end{equation}
Note that Equation~\eqref{eqn:rec_r_stable} for evolutionary stability of a recessive police allele has the same mathematical form as Equation~\eqref{eqn:dom_r} for invasion of a dominant police allele.  Starting from Equation~\eqref{eqn:dom_r}, making the substitution $z \rightarrow 1-z$, and reversing the inequality, we recover Equation~\eqref{eqn:rec_r_stable}.  Again, the intuition behind this correspondence is described in \ref{appendix:stability}.

\begin{figure}
\centering
\begin{subfigure}{0.45\textwidth}
\centering
\caption{}
\includegraphics*[width=1\textwidth]{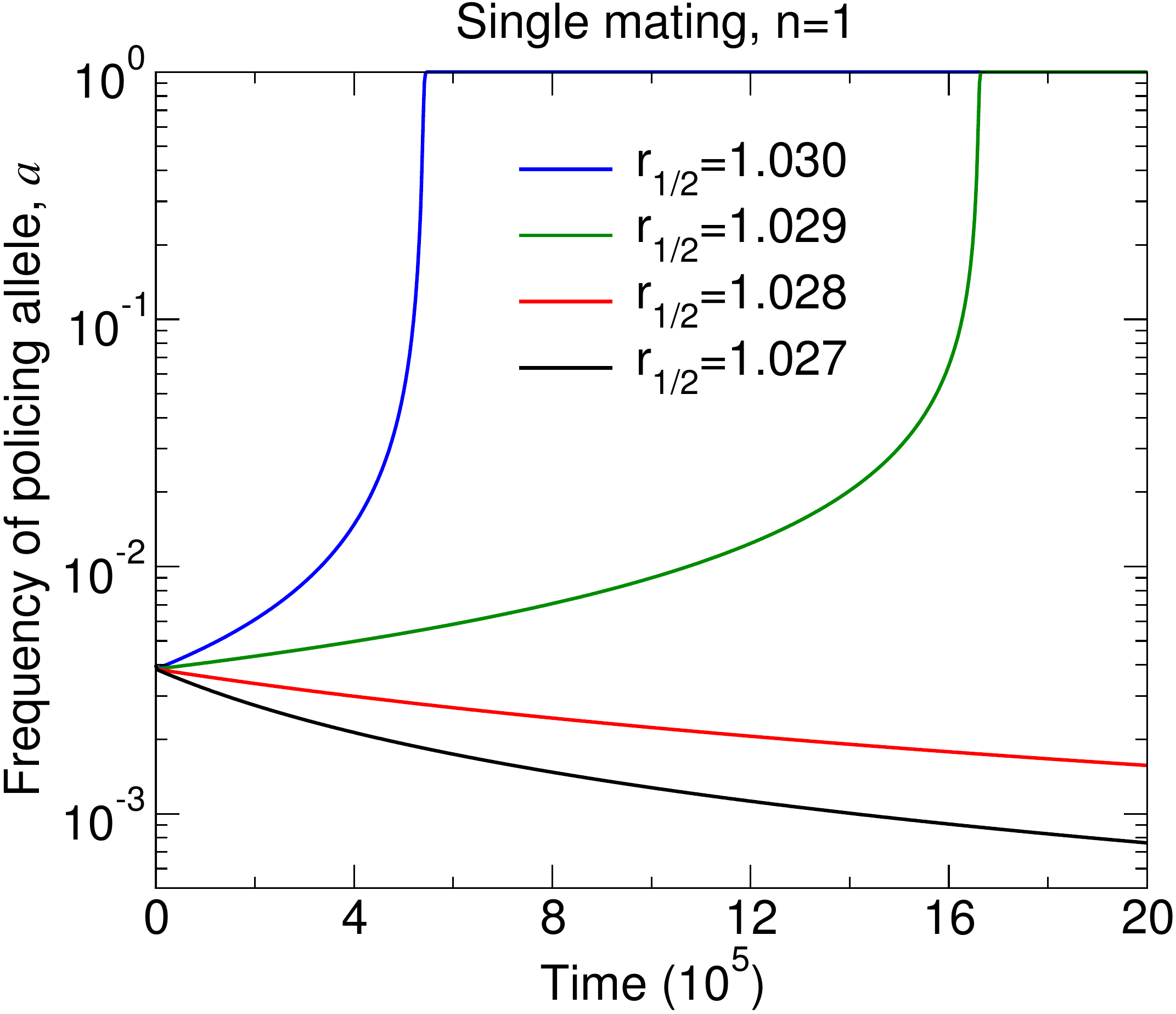}
\end{subfigure}
\qquad
\begin{subfigure}{0.45\textwidth}
\centering
\caption{}
\includegraphics*[width=1\textwidth]{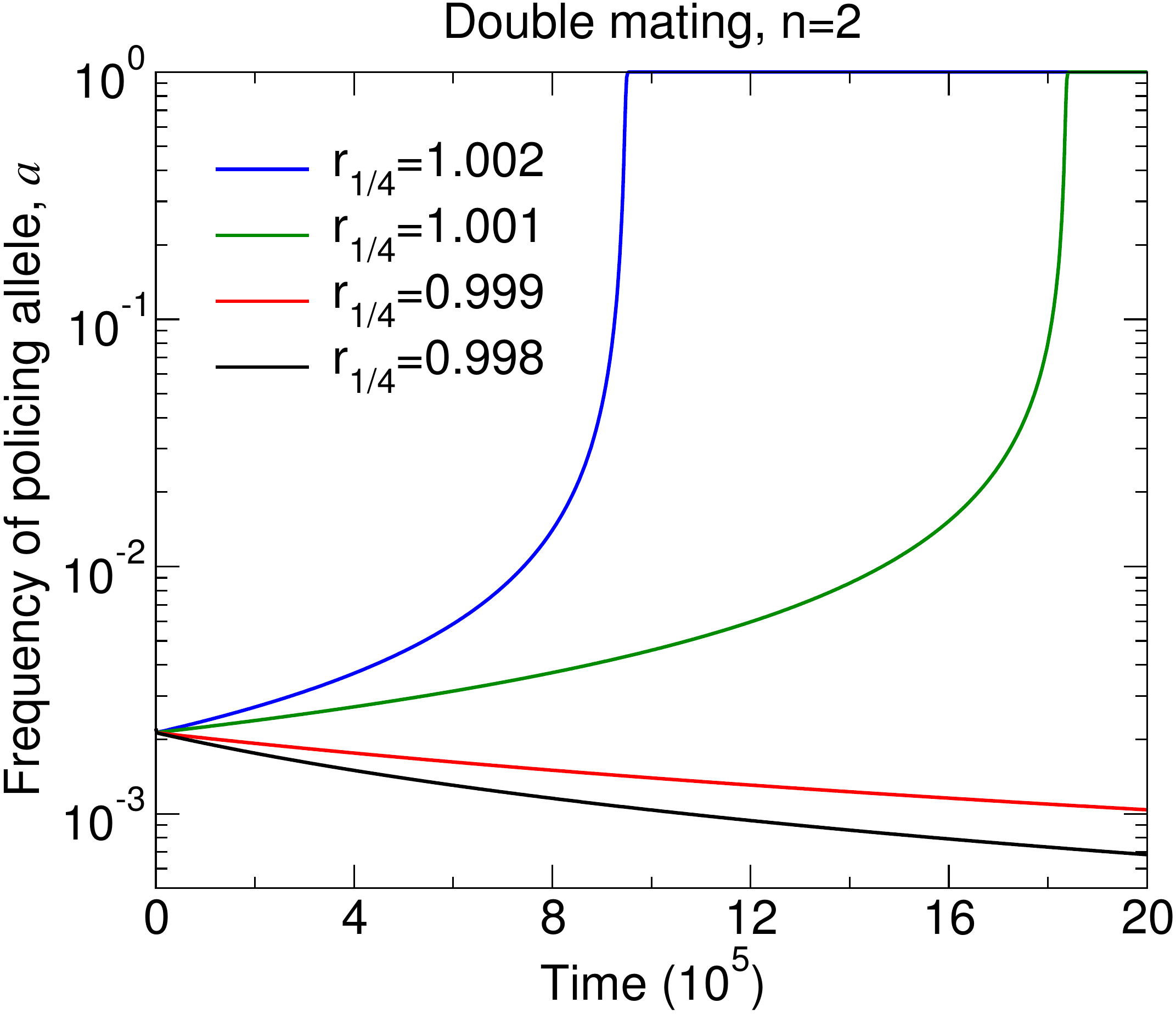}
\end{subfigure}
\caption{Numerical simulations of the evolutionary dynamics of worker policing confirm the condition given by Equation~\eqref{eqn:rec_r}.  The policing allele is recessive.  For numerically probing invasion, we use the initial condition $X_{AA,0}=1-10^{-2}$ and $X_{AA,1}=10^{-2}$.  We set $r_0=1$ without loss of generality.  Other parameters are: (a) $p_0=0.6$, $p_{1/2}=0.8$, and $r_1=1.06$; (b) $p_0=0.35$, $p_{1/2}=0.9$, $r_{1/2}=1.004$, and $r_1=1.012$.}
\label{fig:rec_sim}
\end{figure}

Numerical simulations of the evolutionary dynamics with a recessive police allele are shown in Figure~\ref{fig:rec_sim}.

\vspace*{10mm}

\section{Shape of the efficiency function, $r_z$}
\label{sec:shape}

The shape of the efficiency function, $r_z$, determines whether policing is more likely to evolve for single mating or multiple matings. Recall that $r_z$ is the colony efficiency (defined as the rate of generation of reproductives) if a fraction, $z$, of all workers perform policing. The variable $z$ can assume values between 0 and 1. If no workers police, $z=0$, then the colony efficiency is at baseline, which we set to one; therefore, we have $r_0=1$. Policing can in principle increase or decrease colony efficiency.  

We have the following results regarding the invasion and stability of police workers. We discuss single ($n=1$), double ($n=2$), and triple ($n=3$) mating. All results apply to both dominant and recessive police alleles. They can be instantiated with arbitrarily small changes in colony efficiency. 

\vspace*{10mm}

\subsection*{Evolutionary invasion of policing}

(i) If $r_z$ is strictly constant (which is ungeneric), then policing does not invade for single mating, is neutral for double mating, and does invade for triple mating. 

(ii) If $r_z$ is monotonically decreasing, then policing either invades not at all or only for triple mating.  

(iii) If $r_z$ is monotonically increasing, then policing either invades for single, double, and triple mating or only for double and triple mating.  

(iv) If $r_z$ reaches an intermediate maximum (which means colony efficiency is highest for an intermediate fraction of police workers), then policing can invade for $n=1,2,3$ or $n=2,3$ or $n=3$ or not at all.

(v) If $r_z$ reaches an intermediate minimum (which means colony efficiency is lowest for an intermediate fraction of police workers), then policing can invade with any pattern of matings. For example, it is possible that policing invades only for single mating but neither for double nor triple mating. Or it invades for single and double mating but not for triple mating.

\vspace*{10mm}

\subsection*{Evolutionary stability of policing}

(i) If $r_z$ is constant, then policing is unstable for single mating, is neutral for double mating, and is stable for triple mating.

(ii) If $r_z$ is monotonically decreasing, then policing is unstable for single and double mating. For triple mating it can be stable or unstable.

(iii) If $r_z$ is monotonically increasing, then policing either is always stable or is stable only for double and triple mating.

(iv) If $r_z$ reaches an intermediate maximum, then policing can be stable for any pattern of matings. For example, policing can be stable for single mating but neither for double nor triple mating. 

(v) If $r_z$ reaches an intermediate minimum, then policing can be stable for $n=1,2,3$ or $n=2,3$ or $n=3$ or not at all.

\vspace*{10mm}

\begin{figure}
\begin{center}
\includegraphics[width=0.95\textwidth]{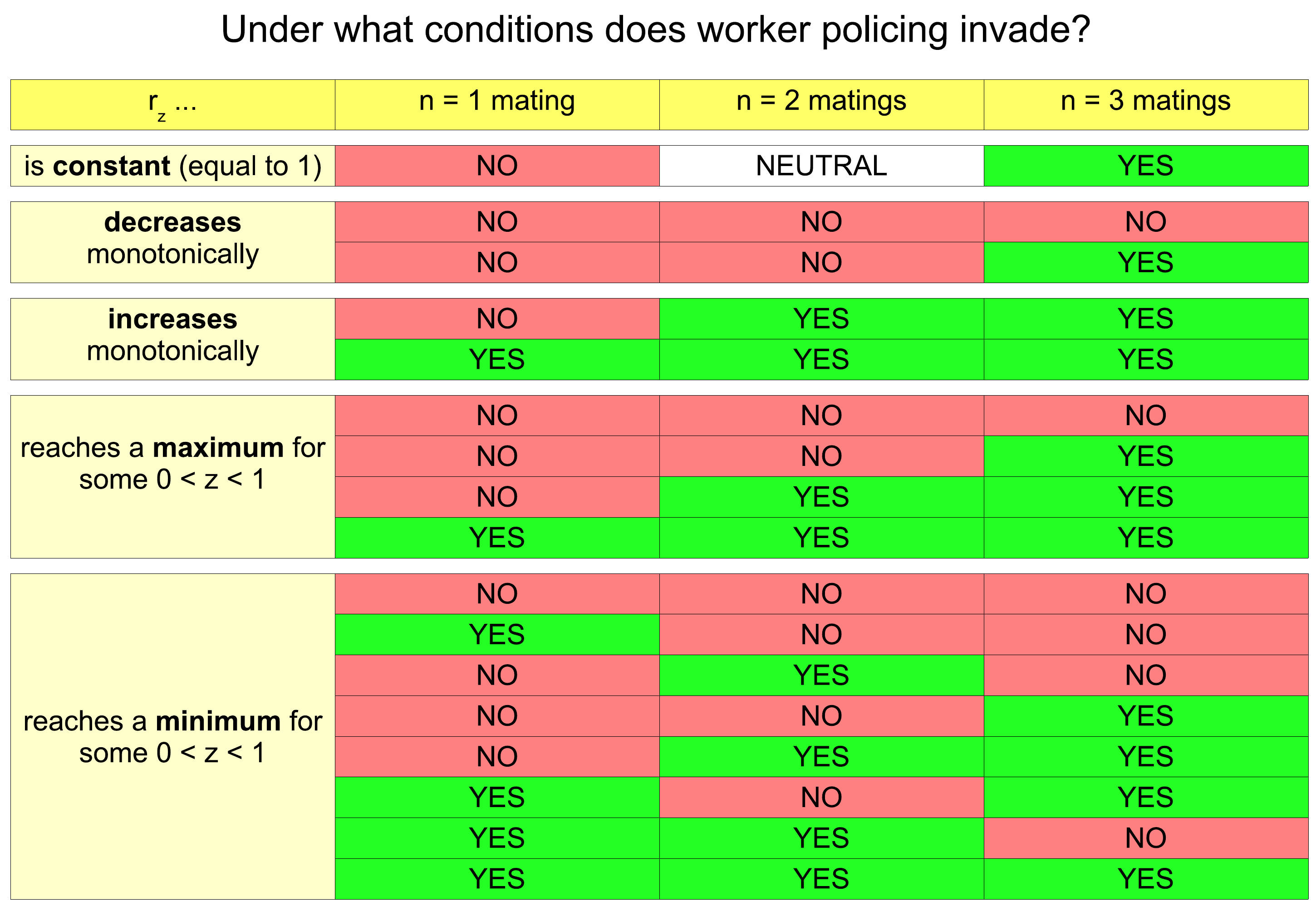}
\caption{Depending on the functional form of colony efficiency, $r_z$, on the fraction of police workers, $z$,  policing alleles may or may not invade for single, double, or triple mating. Various possibilities of $r_z$ are shown. The outcomes hold both for dominant and recessive police alleles.  If $r_z$ is constant, then policing does not invade for single mating, is neutral for double mating, and invades for triple mating. If $r_z$ decreases monotonically, then policing does not invade or invades only for triple mating. If $r_z$ increases monotonically, then policing either invades only for double and triple mating or for single, double, and triple mating. If $r_z$ reaches a maximum at an intermediate value $0<z<1$, then policing does not invade or may invade for triple mating only, for double and triple mating, or for single, double, and triple mating. If $r_z$ reaches a minimum at an intermediate value $0<z<1$, then any pattern is possible.}
\label{fig:cases}
\end{center}
\end{figure}

\vspace*{10mm}

\subsection*{Examples for single and double mating}

\begin{figure}
\centering
\begin{subfigure}{0.45\textwidth}
\centering
\caption{}
\includegraphics*[width=1\textwidth]{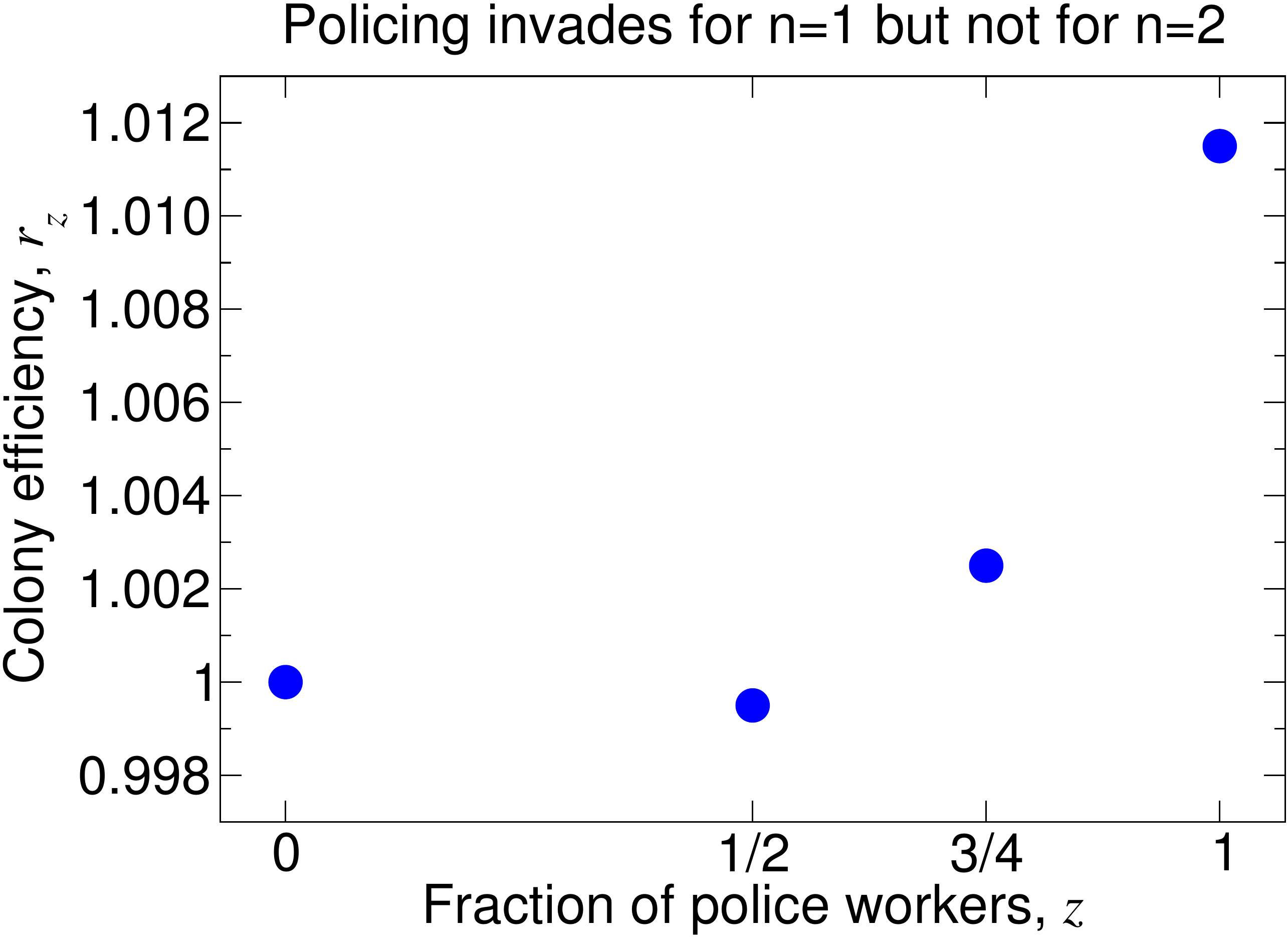}
\end{subfigure}
\qquad
\begin{subfigure}{0.45\textwidth}
\centering
\caption{}
\includegraphics*[width=1\textwidth]{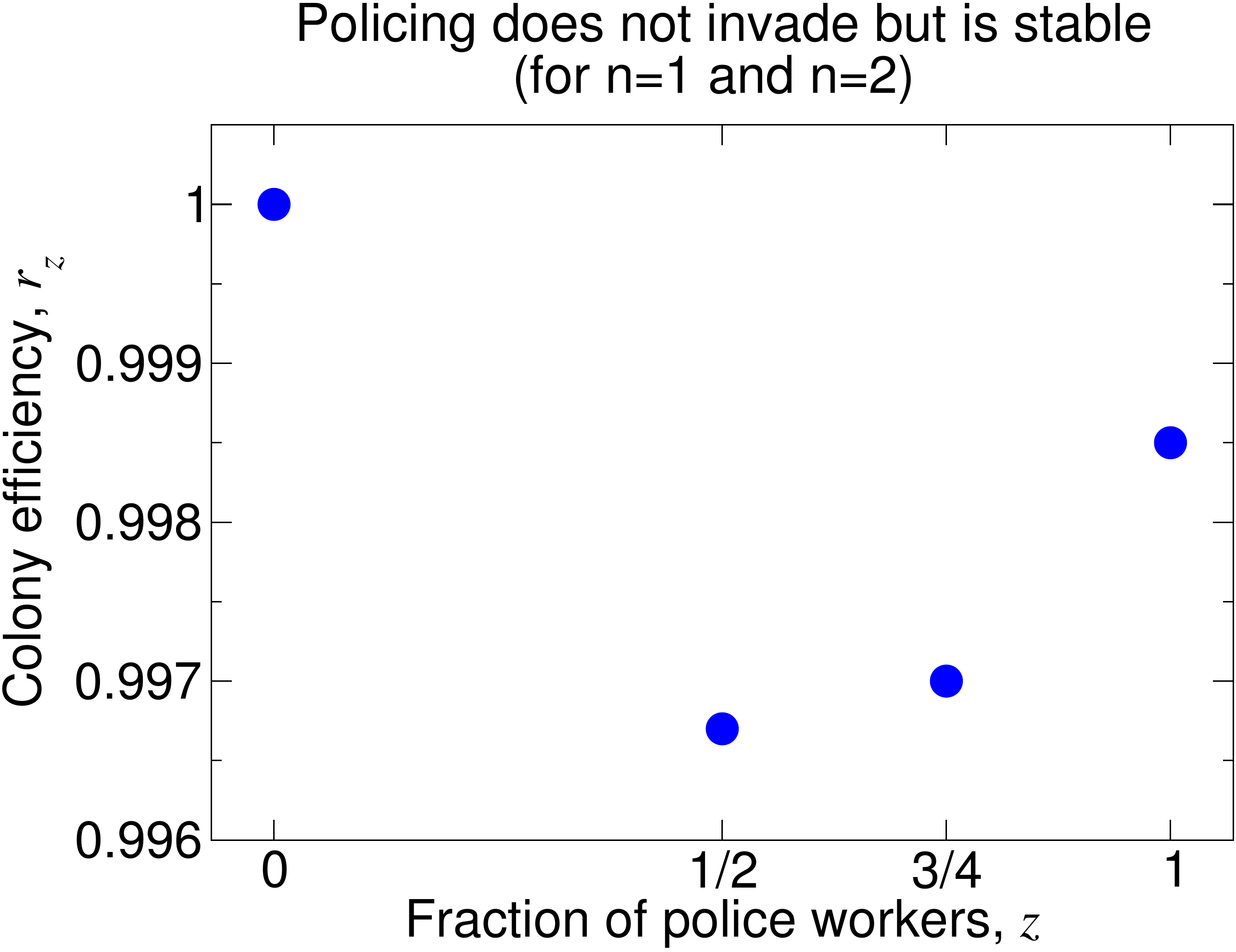}
\end{subfigure}
\par\medskip
\begin{subfigure}{0.45\textwidth}
\centering
\caption{}
\includegraphics*[width=1\textwidth]{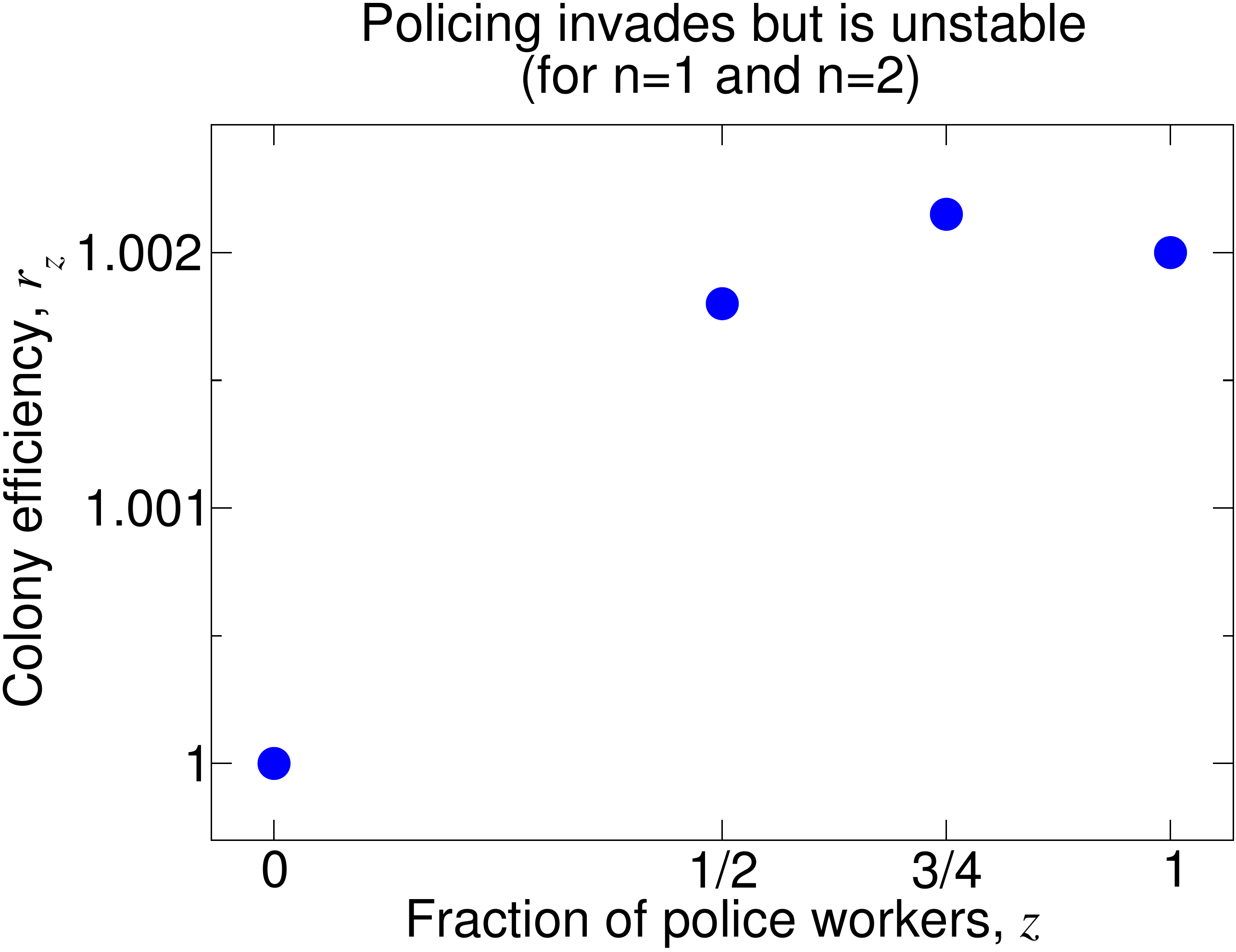}
\end{subfigure}
\qquad
\begin{subfigure}{0.45\textwidth}
\centering
\caption{}
\includegraphics*[width=1\textwidth]{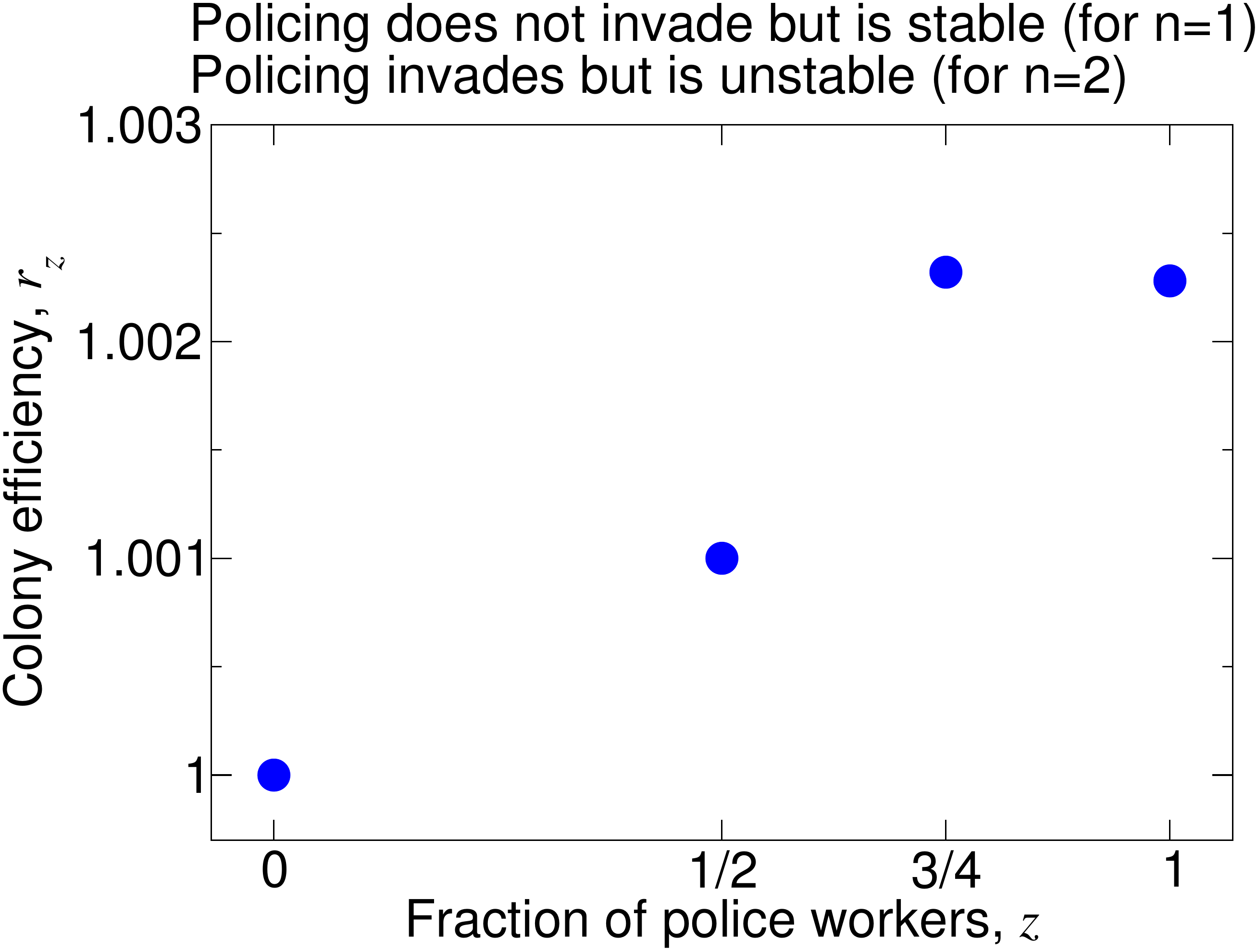}
\end{subfigure}
\caption{Non-monotonic efficiency functions can lead to rich and counterintuitive behavior. We consider invasion and stability of a dominant police allele for single ($n=1$) and double ($n=2$) mating. The baseline colony efficiency without policing is $r_0=1$. Three other values must be specified:  $r_{1/2}$, $r_{3/4}$, and $r_1$.  Moreover, we need to specify two values for how the presence of police workers affects the fraction of male offspring coming from the queen; we choose $p_{1/2}=0.99$ and $p_1=1$.  A variety of behaviors can be realized with very small variation in colony efficiency.  (a) Policing invades for single mating but not for double mating.  (b) Policing does not invade but is stable for single and double mating.  (c) Policing invades but is unstable for single and double mating.  (d) Policing does not invade but is stable for single mating, while policing invades but is unstable for double mating.}
\label{fig:examples}
\end{figure}

Figure~\ref{fig:examples} gives some interesting examples for how non-monotonic efficiency functions can influence the evolution of policing for single ($n=1$) and double ($n=2$) mating. In order to discuss the invasion and stability of a dominant police allele for single and double mating, we need to specify efficiency at three discrete values for the fraction of police workers present in a colony:  $r_{1/2}$, $r_{3/4}$, and $r_1$. Note that $r_0=1$ is the baseline. Moreover, we need to specify the fraction of male offspring coming from the queen at two values: $p_{1/2}$ and $p_1$. For all examples in Figure~\ref{fig:examples}, we assume $p_{1/2}=0.99$ and $p_1=1$. We show four cases: (a) policing invades for single mating but not for double mating; (b) for both single and double mating, policing does not invade but is stable; (c) for both single and double mating, policing invades but is unstable (leading to coexistence of policing and non-policing alleles); (d) policing does not invade but is stable for single mating; policing invades but is unstable for double mating. These cases demonstrate the rich behavior of the system, which goes beyond the simple view that multiple matings are always favorable for the evolution of policing.  

\vspace*{10mm}

\section{Gradual evolution of worker policing}
\label{sec:gradual}

Our main calculation applies to mutations of any effect size.  In this section, we calculate the limit of incremental mutation (small mutational effect size).  Our calculations in this section are reminiscent of adaptive dynamics \citep{Nowak_1990,HofbauerAD,Dieckmann,Metz,Geritz}, which is usually formulated for asexual and haploid models.  The analysis in this section applies both to the case of small phenotypic effect and to the case of weak penetrance.

Mathematically, we consider the evolutionary dynamics of policing if the phenotypic mutations induced by the $a$ allele are small.  If an allele affecting intensity of policing is dominant, then it is intuitive to think of wild-type workers as policing with intensity $Z_{AA}$, while mutant workers police with intensity $Z_{Aa}=Z_{aa}=Z_{AA}+w$.  If an allele affecting intensity of policing is recessive, then it is intuitive to think of wild-type workers as policing with intensity $Z_{AA}=Z_{Aa}$, while mutant workers police with intensity $Z_{aa}=Z_{AA}+w=Z_{Aa}+w$.  In the limit of incremental mutation, the fraction, $p$, of queen-derived males and the colony efficiency, $r$, become functions of the average intensity of policing in the colony, which is $Z+wz$, where $z$ is the fraction of mutant workers in the colony.  We have
\begin{equation}
\begin{aligned}
p_z \rightarrow P(Z+wz) &= P(Z) + P'(Z)wz + \frac{1}{2}P''(Z)w^2z^2 + \mathcal{O}(w^3) \\
r_z \rightarrow R(Z+wz) &= R(Z) + R'(Z)wz + \frac{1}{2}R''(Z)w^2z^2 + \mathcal{O}(w^3) \\
\end{aligned}
\label{eqn:weak}
\end{equation}
We have made the substitutions $p_z \rightarrow P(Z+wz)$ and $r_z \rightarrow R(Z+wz)$, and \eqref{eqn:weak} gives the Taylor expansions of these quantities in terms of their first and second derivatives at intensity $Z$.  (For conciseness, we will often omit the argument $Z$ from the functions $P$ and $R$ and their derivatives.)  Here, $|w|\ll1$, so that workers with the phenotype corresponding to the mutant allele only have an incremental effect on colony dynamics.  Thus, the expansions \eqref{eqn:weak} are accurate approximations.  We assume that $P'>0$.  The sign of $w$ can be positive or negative.  If $w$ is positive, then the mutant allele's effect is to increase the intensity of policing.  If $w$ is negative, then the mutant allele's effect is to decrease the intensity of policing.  Note that this formalism could also be interpreted as describing the case of weak penetrance, in which only a small fraction of all workers that have the mutant genotype express the mutant phenotype.

For considering the dynamics of a dominant police allele with weak phenotypic mutation, we introduce the quantity
\begin{equation}
C_\mathrm{dom} = \frac{p_{1/n}+p_{1/2}}{2}\left(\frac{r_{1/n}}{r_0}\right)\left(\frac{r_{1/2}}{r_0}\right) - \left[2-\left(\frac{r_{1/2}}{r_0}\right)-\left(1-p_{1/n}\right)\left(\frac{r_{1/n}}{r_0}\right)\right]
\label{eqn:weak_dom}
\end{equation}
If $C_\mathrm{dom}>0$, then increased intensity of policing is selected, and if $C_\mathrm{dom}<0$, then increased intensity of policing is not selected.  This is just a different way of writing \eqref{eqn:dom_r}.

We substitute \eqref{eqn:weak} into \eqref{eqn:weak_dom} and collect powers of $w$.  To first order in $w$, we get
\begin{equation}
C_\mathrm{dom} = w \left[ \frac{(n-2)P'R+2(2+n+nP)R'}{4nR} \right] + \mathcal{O}(w^2)
\label{eqn:weak_dom_order_1}
\end{equation}

For considering the dynamics of a recessive police allele with weak phenotypic mutation, we introduce the quantity
\begin{equation}
C_\mathrm{rec} = \frac{r_{1/(2n)}}{r_0} - \frac{2(2+n+np_0)}{(2+n)(2+p_0)+p_{1/(2n)}(n-2)}
\label{eqn:weak_rec}
\end{equation}
If $C_\mathrm{rec}>0$, then increased intensity of policing is selected, and if $C_\mathrm{rec}<0$, then increased intensity of policing is not selected.  This is just a different way of writing \eqref{eqn:rec_r}.

We substitute \eqref{eqn:weak} into \eqref{eqn:weak_rec} and collect powers of $w$.  To first order in $w$, we get
\begin{equation}
C_\mathrm{rec} = w \left[ \frac{(n-2)P'R+2(2+n+nP)R'}{4nR(2+n+nP)} \right] + \mathcal{O}(w^2)
\label{eqn:weak_rec_order_1}
\end{equation}
Notice that \eqref{eqn:weak_dom_order_1} and \eqref{eqn:weak_rec_order_1} are, up to a multiplicative factor, the same to first order in $w$.

Using Equations~\eqref{eqn:weak_dom_order_1}~and~\eqref{eqn:weak_rec_order_1}, the condition for policing to increase from a given level $Z$ is
\begin{equation}
\frac{R'(Z)}{P'(Z)} > -(n-2)\frac{R(Z)}{2(2+n+nP(Z))}
\label{eqn:RPincrease}
\end{equation}
Policing decreases from a given level $Z$ if the opposite inequality holds.  We have explicitly written the $Z$ dependencies in Equation~\eqref{eqn:RPincrease} to emphasize that the quantities $P$, $P'$, $R$, and $R'$ are all functions of the intensity of policing, $Z$.

The left-hand side of Equation~\eqref{eqn:RPincrease} can be understood as a ratio of marginal effects.  To be specific, the left-hand side gives the ratio of the marginal change in efficiency over the marginal increase in the proportion of queen-derived males, if policing were to increase by a small amount.  For selection to favor increased policing, this ratio of marginals must exceed a quantity depending on the current values of $R$ and $P$.

Notice that the sign of the right-hand side is determined by $n-2$.  So we get different behavior for different numbers of matings:
\begin{itemize}
\item
For $n=2$ (double mating), policing increases from $Z$ if and only if $R'(Z)>0$.  This means that evolution maximizes the value of $R$, regardless of the behavior of $P$.  In other words, for double mating, evolution maximizes colony efficiency regardless of the effect on the number of queen-derived males.
\item
For $n=1$ (single mating), the right-hand side of Equation~\eqref{eqn:RPincrease} is positive.  So the condition for $Z$ to increase is more stringent than in the $n=2$ case.  Increases in policing may be disfavored even if they increase colony efficiency.  
\item
For $n \geq 3$ (triple mating or more than three matings), the right-hand side of Equation~\eqref{eqn:RPincrease} is negative. So the condition for $Z$ to increase is less stringent than in the $n=2$ case.  Any increase in policing that improves colony efficiency will be favored, and even increases in policing that reduce colony efficiency may be favored.
\end{itemize}

Equations~\eqref{eqn:weak_dom_order_1}~and~\eqref{eqn:weak_rec_order_1} also allow us to determine the location(s) of evolutionarily singular strategies \citep{Geritz}.  Intuitively, a singular strategy is a particular intensity of policing, denoted by $Z^*$, at which rare workers with slightly different policing behavior are, to first order in $w$, neither favored nor disfavored by natural selection.  The parameter measuring intensity of policing, $Z$, can take values between $0$ (corresponding to no policing) and $1$ (corresponding to full policing).  There are several possibilities:  There may not exist a singular strategy for intermediate intensity of policing; in this case, there is either no policing ($Z^*=0$) or full policing ($Z^*=1$).  If there exists a singular strategy for $0<Z^*<1$, then there are additional considerations:  There may be convergent evolution toward intensity $Z^*$ or divergent evolution away from intensity $Z^*$.  In a small neighborhood for which $Z \approx Z^*$, further analysis is needed to determine if the singular strategy corresponding to $Z^*$ is an ESS.

To determine the location(s) of evolutionarily singular strategies, we set the quantity in square brackets that multiplies $w$ in \eqref{eqn:weak_dom_order_1} and \eqref{eqn:weak_rec_order_1} to zero, yielding
\begin{equation}
\frac{R'(Z^*)}{P'(Z^*)} + (n-2)\frac{R(Z^*)}{2(2+n+nP(Z^*))} = 0
\label{eqn:weak_singular}
\end{equation}
Equation~\eqref{eqn:weak_singular} gives the location(s) of singular strategies for both dominant and recessive mutations that affect policing.

For a given singular strategy $Z^*$, there is convergent evolution toward $Z^*$ if
\begin{equation*}
\frac{d}{dZ} \left[ \frac{R'(Z)}{P'(Z)} + (n-2)\frac{R(Z)}{2(2+n+nP(Z))} \right] \bigg |_{Z=Z^*} < 0
\end{equation*}
There is divergent evolution away from $Z^*$ if the opposite inequality holds.

\begin{figure}
\centering
\begin{subfigure}{0.45\textwidth}
\centering
\caption{}
\includegraphics*[width=1\textwidth]{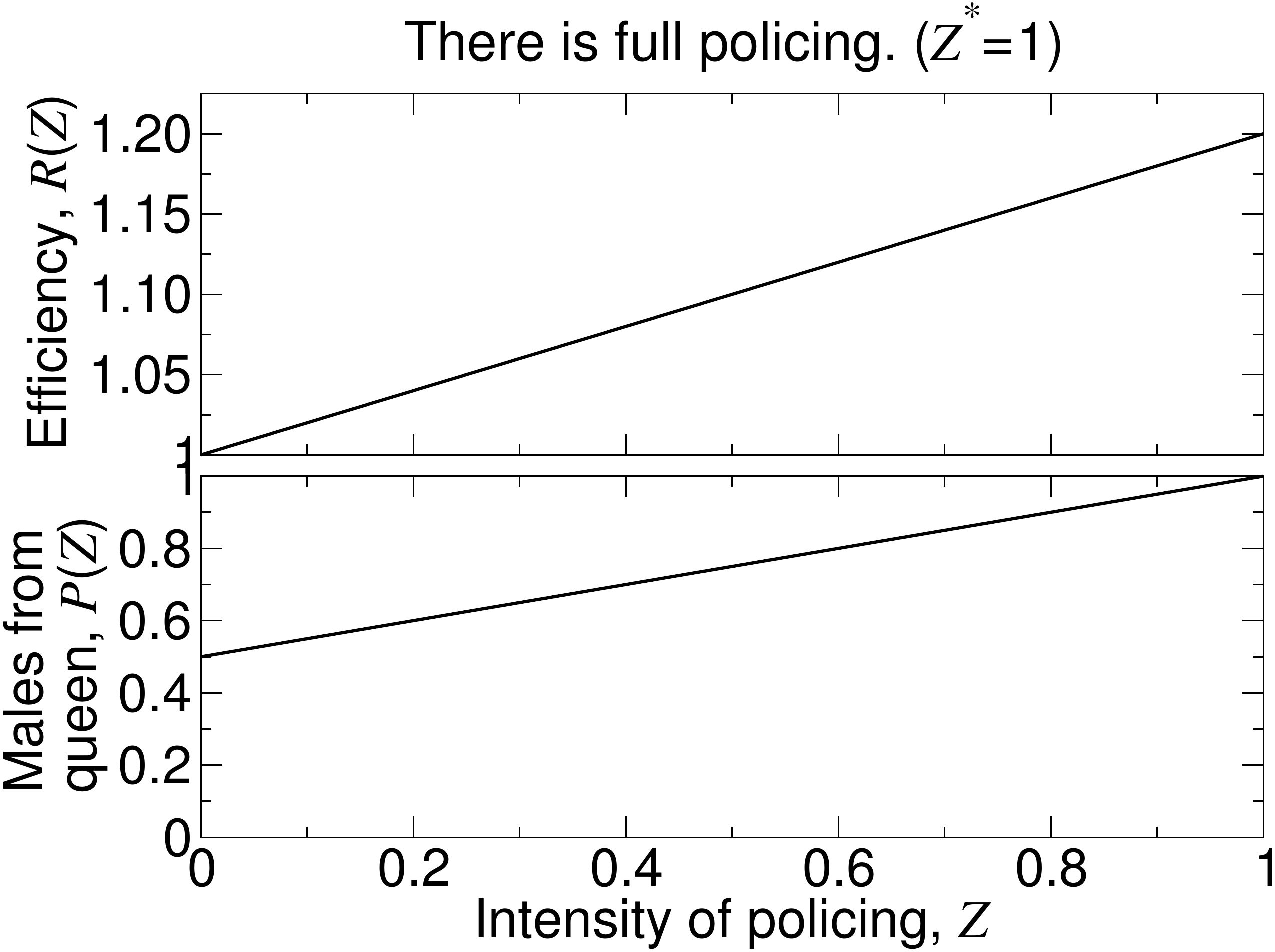}
\end{subfigure}
\qquad
\begin{subfigure}{0.45\textwidth}
\centering
\caption{}
\includegraphics*[width=1\textwidth]{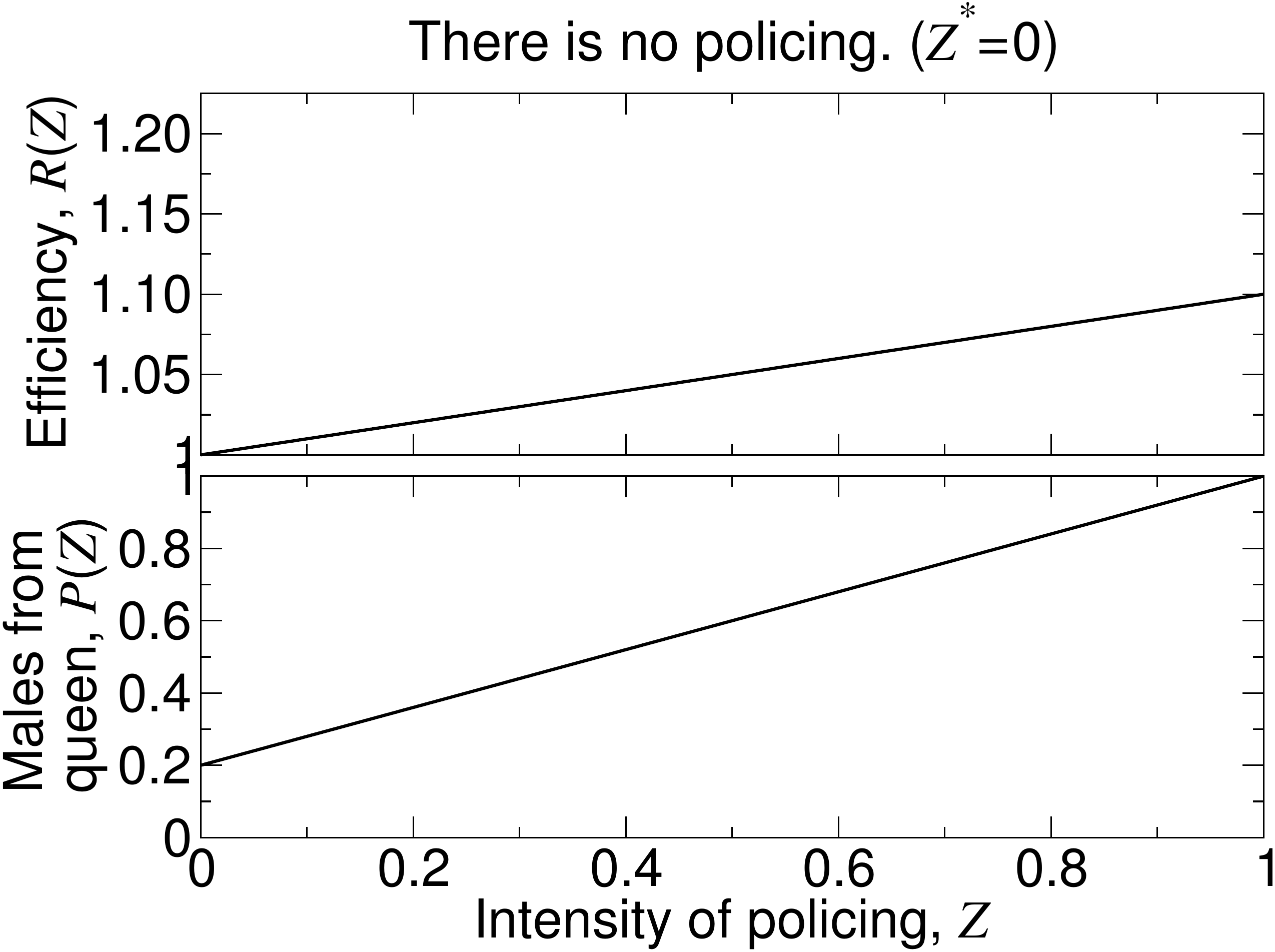}
\end{subfigure}
\par\medskip
\begin{subfigure}{0.45\textwidth}
\centering
\caption{}
\includegraphics*[width=1\textwidth]{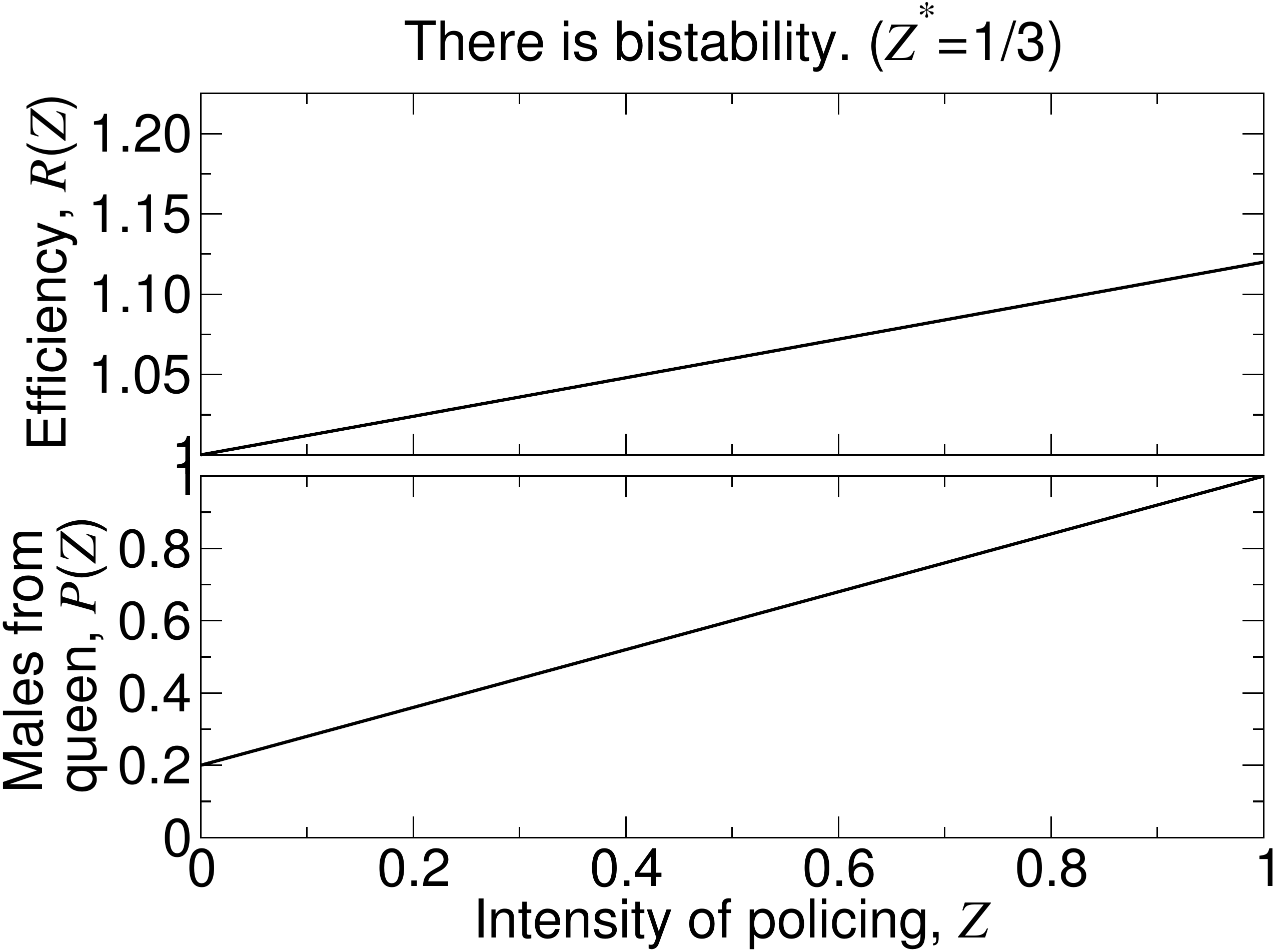}
\end{subfigure}
\qquad
\begin{subfigure}{0.45\textwidth}
\centering
\caption{}
\includegraphics*[width=1\textwidth]{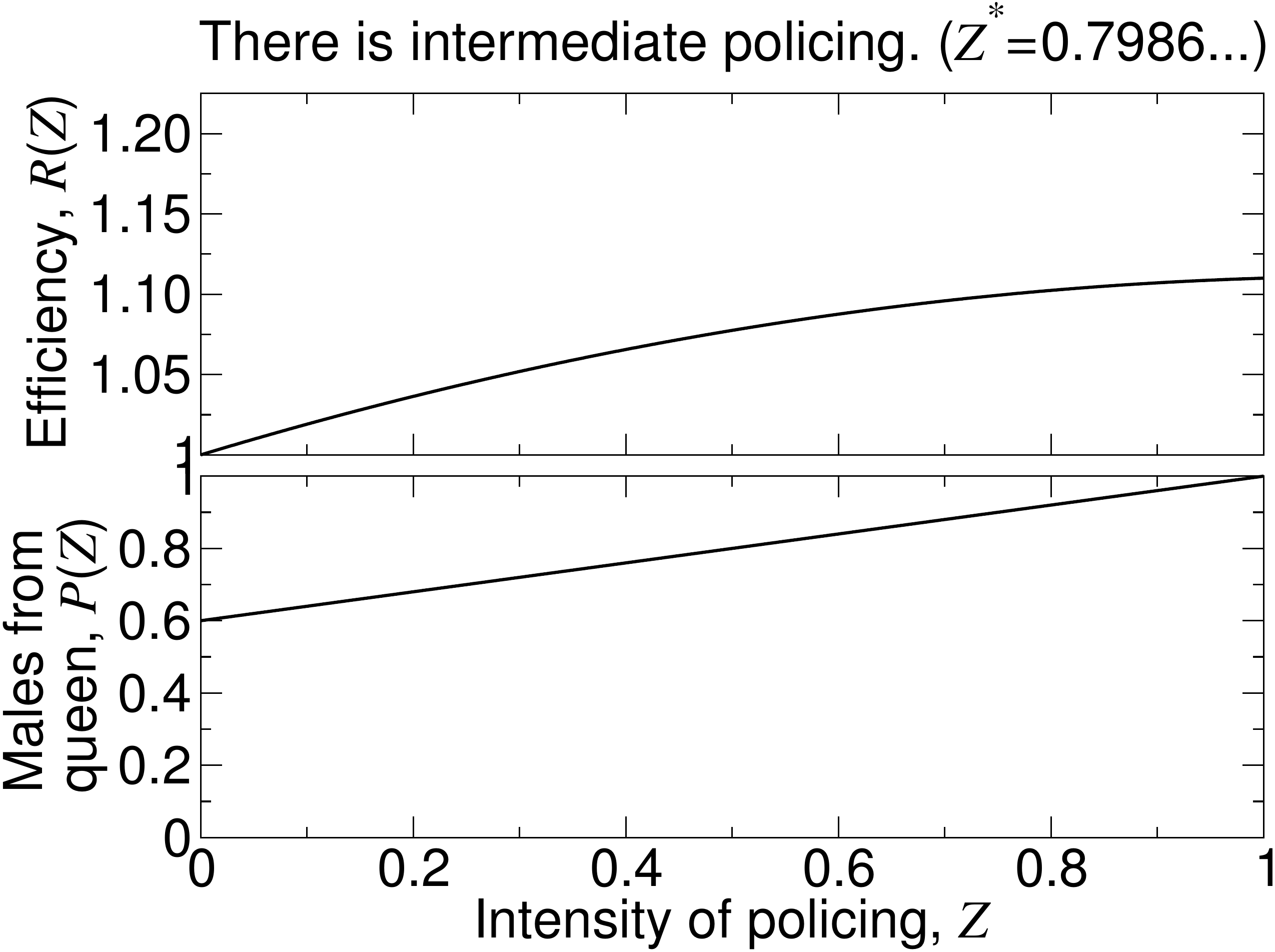}
\end{subfigure}
\caption{Several simple examples of functions $P(Z)$ and $R(Z)$ are shown.  For single mating, the corresponding dynamics of policing intensity with gradual evolution are also shown.  We use the forms $P(Z)=1-P'+P'Z$ and $R(Z)=1+C_1Z+(1/2)C_2Z^2$.  For each of the four panels, we set: (a) $P'=0.5$, $C_1=0.2$, $C_2=0$, corresponding to $Z^*=1$; (b) $P'=0.8$, $C_1=0.1$, $C_2=0$, corresponding to $Z^*=0$; (c) $P'=0.8$, $C_1=0.12$, $C_2=0$, corresponding to bistability around $Z^*=1/3$; (d) $P'=0.4$, $C_1=0.2$, $C_2=-0.18$, corresponding to an intermediate level of policing around $Z^*\approx0.7986\ldots$.}
\label{fig:intensity}
\end{figure}

It is helpful to consider some examples.  If the functions $P(Z)$ and $R(Z)$ are known for a given species, then the behavior of worker policing with gradual evolution can be studied.  It is possible that policing is at maximal intensity, $Z^*=1$ (Figure~\ref{fig:intensity}(a)), is nonexistent, $Z^*=0$ (Figure~\ref{fig:intensity}(b)), is bistable around a critical value of intensity, $0<Z^*<1$ (Figure~\ref{fig:intensity}(c)), or exists at an intermediate value of intensity, $0<Z^*<1$ (Figure~\ref{fig:intensity}(d)).

Note that a singular strategy may or may not be an evolutionarily stable strategy (ESS).  (For example, it is possible that there is convergent evolution toward a particular singular strategy $Z^*$ which is not an ESS.  In this case, once $Z \approx Z^*$, evolutionary branching may occur; \citealp{Geritz})  To determine if \eqref{eqn:weak_singular} is an ESS, we must look at second-order terms in \eqref{eqn:weak_dom} and \eqref{eqn:weak_rec}.

For a dominant police allele, we return to \eqref{eqn:weak_dom} with the substitutions \eqref{eqn:weak}.  We focus on a singular strategy given by \eqref{eqn:weak_singular}.  For a singular strategy, $C_\mathrm{dom}$ is zero to first order in $w$.  To second order in $w$, we get
\begin{equation}
\begin{aligned}
C_\mathrm{dom} = w^2 & \left[ \frac{(n^2-4)P''R^2+2(n^2+4n-4)P'R'R}{16n^2R^2} \right. \\
          & \left.+\frac{8nPR'^2+2(n^2+n^2P+4)R''R}{16n^2R^2} \right] + \mathcal{O}(w^3)
\label{eqn:weak_dom_order_2}
\end{aligned}
\end{equation}
We may alternatively write \eqref{eqn:weak_dom_order_2} by substituting for $R'$ using \eqref{eqn:weak_singular}:
\begin{equation}
\begin{aligned}
C_\mathrm{dom} = w^2 & \left[ \frac{(2+n+nP)^2[(n^2-4)P''R+2(n^2+n^2P+4)R'']}{16n^2R(2+n+nP)^2} \right. \\
          & \left.-\frac{(n^2-4)(n^2+n^2P+4n-4)P'^2R}{16n^2R(2+n+nP)^2} \right] + \mathcal{O}(w^3)
\label{eqn:weak_dom_order_2_substitution}
\end{aligned}
\end{equation}

For a recessive police allele, we return to \eqref{eqn:weak_rec} with the substitutions \eqref{eqn:weak}.  We focus on a singular strategy given by \eqref{eqn:weak_singular}.  For a singular strategy, $C_\mathrm{rec}$ is zero to first order in $w$.  To second order in $w$, we get
\begin{equation}
\begin{aligned}
C_\mathrm{rec} = w^2 & \left[ \frac{(n-2)(2+n+nP)P''R-(n-2)^2P'^2R}{16n^2R(2+n+nP)^2} \right. \\
          & \left.+\frac{2(2+n+nP)^2R''}{16n^2R(2+n+nP)^2} \right] + \mathcal{O}(w^3)
\label{eqn:weak_rec_order_2}
\end{aligned}
\end{equation}

Inspection of \eqref{eqn:weak_dom_order_2} and \eqref{eqn:weak_rec_order_2} allows us to determine if a singular strategy is an ESS.  If the bracketed quantity multiplying $w^2$ is negative, then mutations that change policing in either direction are disfavored.  If the bracketed quantity multiplying $w^2$ is positive, then mutations that change policing in either direction are favored.  Thus, for a dominant allele that affects intensity of policing, the singular strategy \eqref{eqn:weak_singular} represents a local ESS if
\begin{equation}
(n^2-4)P''R^2+2(n^2+4n-4)P'R'R+8nPR'^2+2(n^2+n^2P+4)R''R < 0
\label{eqn:weak_dom_ESS}
\end{equation}
We may alternatively write \eqref{eqn:weak_dom_ESS} by substituting for $R'$ using \eqref{eqn:weak_singular}:
\begin{equation}
(2+n+nP)^2[(n^2-4)P''R+2(n^2+n^2P+4)R'']-(n^2-4)(n^2+n^2P+4n-4)P'^2R < 0
\label{eqn:weak_dom_ESS_substitution}
\end{equation}
Similarly, for a recessive allele that affects intensity of policing, the singular strategy 
\eqref{eqn:weak_singular} represents a local ESS if
\begin{equation}
(n-2)(2+n+nP)P''R-(n-2)^2P'^2R+2(2+n+nP)^2R'' < 0
\label{eqn:weak_rec_ESS}
\end{equation}

Here, $P$, $P'$, $P''$, $R$, $R'$, and $R''$ are all functions of the intensity of policing, $Z$.  The local ESS conditions \eqref{eqn:weak_dom_ESS_substitution} and \eqref{eqn:weak_rec_ESS} are quite opaque and do not allow for simple analysis.  Notice that, although the locations of evolutionarily singular strategies are the same for dominant and recessive mutations that influence policing, the conditions for a singular strategy to be a local ESS are different.

\vspace*{10mm}

\section{Policing and inclusive fitness theory}
\label{sec:IF}

It has been claimed that policing is a test case of inclusive fitness theory \citep{Abbot_2011}. But the first two papers to theoretically establish the phenomenon \citep{Woyciechowski_1987,Ratnieks_1988} use standard population genetics; they do not mention the term ``inclusive fitness'', and they do not calculate inclusive fitness. Therefore, the claims that theoretical investigations of worker policing emerge from inclusive fitness theory or that empirical studies of policing test predictions of inclusive fitness theory are incorrect.  

In light of known and mathematically proven limitations of inclusive fitness theory \citep{Nowak_2010,Allen_2013}, it is unlikely that inclusive fitness theory can be used to study general questions of worker policing.  Inclusive fitness theory assumes that each individual contributes a separate, well-defined portion of fitness to itself and to every other individual.  It has been shown repeatedly \citep{Cavalli-Sforza_1978,Uyenoyama_1982,Matessi_1984,Nowak_2010,vanVeelen_2014}, that this assumption does not hold for general evolutionary processes.  Therefore, inclusive fitness is a limited concept that does not exist in most biological situations.

Our work shows that the evolution of worker policing depends on the effectiveness of egg removal ($p_z$) and the consequences of colony efficiency ($r_z$).  Each of these effects can be nonlinear (not the sum of contributions from separate individuals), with important consequences for the fate of a policing allele.  Moreover, the invasion and stability conditions involve the product of $p$- and $r$-values, indicating a nontrivial interaction between these two effects which does not reduce to a simple sum of costs and benefits.  We also found that there are separate conditions for invasion and stability, with neither implying the other.  Inclusive fitness theory, which posits a single, linear condition for the success of a trait, is not equipped to deal with these considerations.

Attempts to extend inclusive fitness theory to more general evolutionary processes \citep{Queller_1992,Frank_1998,Gardner_2011} rely on the incorrect interpretation of linear regression coefficients (\citealp{Allen_2013}; see also \citealp{Birch_2014}).  This misuse of statistical inference tools is unique to inclusive fitness theory, and differs from legitimate uses of linear regression in quantitative genetics and other areas of science.  It was also recently discovered that even in situations where inclusive fitness does exist, it can give the wrong result as to the direction of natural selection \citep{Tarnita_2014}. 

Relatedness-based arguments are often seen in conjunction with inclusive fitness, but there is a crucial difference. Consider the following statement: if the queen is singly mated, then workers share more genetic material with sons of other workers than with sons of the queen.  This statement is not wrong and could be useful in formulating evolutionary hypotheses. Such hypotheses can then be checked using exact mathematical methods.

The problem arises when one attempts to formulate the quantity of inclusive fitness by partitioning fitness into contributions from different individuals and reassigning these contributions from recipient to actor.  A worker does not make separate contributions to fitnesses of others, and therefore does not have ``inclusive fitness''.  Arguments such as ``the worker maximizes her inclusive fitness by not policing'' are meaningless, since they are based on maximizing a nonexistent quantity.  Moreover, even when evolution leads individuals to maximize some quantity, that quantity is not necessarily inclusive fitness \citep{Okasha_2015,Lehmann_2015}.

It is true that genes (alleles) can be favored by natural selection if they enhance the reproduction of copies of themselves in other individuals. But that argument works out on the level of genes and can be fully analyzed using population genetics. Inclusive fitness only arises when the individual is chosen as the level of analysis, which is a problematic choice for many cases of complex family or population structure \citep{akccay2016there}.

\cite{Bourke_2011} has proposed that inclusive fitness remains valid as a concept even when it is nonexistent as a quantity.  But why is such an uninstantiable concept useful?  The mathematical theory of evolution is clear and powerful. Exact calculations of evolutionary dynamics \citep{Antal_2009,Allen_2014,Fu_2014,Hauert_2004,Szabo_2007,Antal_2006,Traulsen_2008,vanVeelen_2014,Simon_2013} demonstrate that inclusive fitness is not needed for understanding any phenomenon in evolutionary biology. This realization is good news for all whose primary goal is to understand evolution rather than to insist on a particular method of analysis.  By releasing ourselves from the confines of a mathematically limited theory, we expand the possibilities of scientific discovery.

\vspace*{10mm}

\section{Discussion}
\label{sec:discussion}

We have derived analytical conditions for the invasion and stability of policing in situations where queens mate once or several times and where colony efficiency can be affected by policing. In the special case where policing has no effect on colony efficiency, our results confirm the traditional view that policing does not evolve for single mating, is neutral for double mating, and does evolve for triple mating or more than three matings. If colony efficiency depends linearly or monotonically on the fraction of workers that are policing, then our results support the view that multiple mating is favorable to evolution of policing \citep{Ratnieks_1988}. Our results also show that non-monotonic relations in colony dynamics and small changes in colony efficiency necessitate a more careful analysis.

We find that policing can evolve in species with singly mated queens if it causes minute increases in colony efficiency. We find that policing does not evolve in species with multiply mated queens if it causes minute decreases in colony efficiency. For non-monotonic efficiency functions, it is possible that single mating allows evolution of policing, while multiple mating opposes evolution of policing.

Our analysis is the first to give precise conditions for both the invasion and stability of policing for both dominant and recessive mutations that effect policing. We study the evolutionary invasion and evolutionary stability of policing both analytically and numerically. For any number of matings, there are four possible outcomes (see Figure~\ref{fig:arrows}): (i) policing can invade and is stable; (ii) policing can invade but is unstable, leading to coexistence; (iii) policing cannot invade but is stable, leading to bistability; (iv) policing cannot invade and is unstable. We give precise conditions for all outcomes for both dominant and recessive police alleles. All outcomes can be achieved with arbitrarily small changes in colony efficiency.

Our calculations are not based on any assumption about the strength of phenotypic mutation induced by an allele. The conditions \eqref{eqn:dom_r}, \eqref{eqn:dom_r_stable}, \eqref{eqn:rec_r}, and \eqref{eqn:rec_r_stable} also describe the dynamics of mutations that have an arbitrarily small phenotypic effect on colony dynamics. This facilitates investigation of the evolution of complex social behaviors that result from gradual accumulation of many mutations \citep{Kapheim_2015}. We derive a simple relation, Equation \eqref{eqn:weak_singular}, for the location(s) of evolutionarily singular strategies. We also derive precise conditions for a singular strategy to be an ESS. These results are applicable for understanding both the case of weak phenotypic effect and the case of weak penetrance.

Our analysis does not use inclusive fitness theory. Given the known limitations of inclusive fitness \citep{Nowak_2010,Allen_2013}, it is unlikely that inclusive fitness theory could provide a general framework for analyzing the evolution of worker policing.

In summary, the main conclusions of our paper are: (i) The prevalent relatedness-based argument that policing evolves under multiple mating but not under single mating is not robust with respect to arbitrarily small variations in colony efficiency; (ii) For non-monotonic efficiency functions, it is possible that policing evolves for single mating, but not for double or triple mating; (iii) Careful measurements of colony efficiency and the fraction of queen-derived males are needed to understand how natural selection acts on policing; (iv) Contrary to what has been claimed \citep{Abbot_2011}, the phenomenon of worker policing is no empirical confirmation of inclusive fitness theory; the first two mathematical papers on worker policing \citep{Woyciechowski_1987,Ratnieks_1988} do not use inclusive fitness theory. The present paper, which also does not use inclusive fitness theory, is the first detailed analysis of policing for any number of matings and taking into account effects on colony efficiency.

\vspace*{10mm}

\section*{Acknowledgements}

We are grateful to the referees and editor for helpful comments that have significantly benefited this manuscript.  This publication was made possible through the support of a grant from the John Templeton Foundation.  The opinions expressed in this publication are those of the authors and do not necessarily reflect the views of the John Templeton Foundation.

\vspace*{10mm}

\appendix

\section{Stability Analysis}
\label{appendix:stability}

In this Supplementary Information, we introduce a population genetics-based model for 
insect dynamics in the Hymenoptera, and we calculate precise conditions that must 
be satisfied if worker policing is to evolve.  Our model is consistent with haplodiploid genetics.  
We postulate that a specific locus is responsible for worker policing.  The $A$ allele is wild-type, 
while the mutant $a$ allele leads to unmated females killing the male eggs of 
other unmated females.  If the $a$ allele is dominant, then 
workers that have at least one $a$ allele kill other workers' male eggs.  
If the $a$ allele is recessive, then workers that are homozygous in the $a$ allele 
kill other workers' male eggs.  
What conditions on the colony parameters are sufficient for worker policing to arise and be 
established in a population?  The mathematical analysis is similar to that in Olejarz et al. (2015).  
The calculations for both dominant and recessive alleles affecting 
policing are presented below.

\vspace*{10mm}

\section*{Description of the Model}

We study a population of insects that follows haplodiploid genetics, in which females have 
homologous pairs of maternal and paternal chromosomes, and males have a single set of 
chromosomes.  A queen produces female workers and gynes (future queens) from her own genotype 
and using sperm from the male drones that she has mated with.  A queen also produces drones.  
Female workers, though unfertilized, lay male eggs as well.  Each colony produces many offspring, 
and the population is made of many colonies.

We investigate the dynamics of two alleles, $A$ and $a$, that enable us to study the evolutionary 
emergence and stability of worker policing---i.e., the eating of worker-laid male eggs by workers.  
The $A$ allele generates a phenotype such that a worker does not police (or that 
policing occurs with intensity $Z$).  The $a$ allele generates a phenotype such that a worker 
performs policing (or that policing occurs with intensity $Z+w$).

Our analysis of the invasion of police workers is general and open to multiple 
interpretations.  For example, we have thus far considered that wild-type workers 
perform no policing, while workers carrying the mutant allele police.  In this 
view, $p_0$ represents the fraction of male eggs that originate from the queen 
when no workers are policing, while $p_z>p_0$ represents the fraction of male 
eggs that originate from the queen when a fraction $z>0$ of workers are policing.  
We may instead consider the case where all workers are policing and view our 
parameter $z$ as representing the intensity of policing.  It is possible that 
wild-type workers perform policing with intensity $Z$, while mutant 
workers perform policing with increased intensity 
$Z+w>Z$.  In this alternative view, $p_0$ represents 
the fraction of male eggs that originate from the queen when all workers police 
with intensity $Z$, while $p_z>p_0$ represents the fraction of male 
eggs that originate from the queen when a fraction $z>0$ of workers police with 
intensity $Z+w$, while the remaining fraction $1-z$ of workers 
police with intensity $Z$.  Regardless of interpretation, only 
changes in the resulting function $p_z$, which measures the fraction of male eggs 
that originate from the queen as a function of the fraction of mutant workers in 
the colony, affect the condition for invasion of worker policing in our model.  
Of course, there may be biological differences between adding more police workers 
or simply increasing the intensity at which existing workers police, but such 
biological differences would be studied by adding parameters to our model.  
To keep the presentation here as simple as possible, we will reserve such 
biological subtleties for study in a future publication.

The parameter $n$ is the number of males with which the colony's queen has mated.  
An illustration of the mating events is shown in Figure~1(a).  There are several possibilities:  
A type $AA$ gyne mates with $n-m$ type $A$ males and $m$ type $a$ males.  
A type $Aa$ gyne mates with $n-m$ type $A$ males and $m$ type $a$ males.  
A type $aa$ gyne mates with $n-m$ type $A$ males and $m$ type $a$ males.  

A reproductive female (queen) mates with $n$ randomly chosen males in the population.  For mating, the gynes and drones 
are well-mixed:  If a gyne from one colony mates with $n$ drones, then each of the $n$ drones is chosen randomly 
from among the colonies in the population.

The selection dynamics in continuous time are described by a simple system of differential equations:
\begin{equation}
\begin{aligned}
\dot{X}_{AA,m} = \frac{dX_{AA,m}}{dt} &= {n \choose m} x_{AA}y_A^{n-m}y_a^m - \phi X_{AA,m} \\
\dot{X}_{Aa,m} = \frac{dX_{Aa,m}}{dt} &= {n \choose m} x_{Aa}y_A^{n-m}y_a^m - \phi X_{Aa,m} \\
\dot{X}_{aa,m} = \frac{dX_{aa,m}}{dt} &= {n \choose m} x_{aa}y_A^{n-m}y_a^m - \phi X_{aa,m}
\end{aligned}
\label{eqn:SI_dynamics_unscaled}
\end{equation}
(In what follows, we use the overdot notation to denote the time derivative, $d/dt$.)

We make sense of Eqs. \eqref{eqn:SI_dynamics_unscaled} as follows.  
Each queen has one of three possible combinations of the $A$ and $a$ alleles in her own genotype; 
she can have type $AA$, type $Aa$, or type $aa$.  Each queen also has sperm stored from her matings, 
and $m$ represents the number of mutant males (type $a$ males) that a queen has mated with.  
Since each colony is headed by a single queen, 
the variables $X_{AA,m}$, $X_{Aa,m}$, and $X_{aa,m}$ can be used to refer to either a colony 
or to the queen that heads the colony; either intuition is acceptable.  
The number of colonies with a type $AA$ queen who has mated with $n-m$ type $A$ males 
and $m$ type $a$ males is represented by $X_{AA,m}$.
The number of colonies with a type $Aa$ queen who has mated with $n-m$ type $A$ males 
and $m$ type $a$ males is represented by $X_{Aa,m}$.
The number of colonies with a type $aa$ queen who has mated with $n-m$ type $A$ males 
and $m$ type $a$ males is represented by $X_{aa,m}$.
$x_{AA}$, $x_{Aa}$, and $x_{aa}$ represent the numbers of gynes (reproductive females) in the 
population with the three 
possible genotypes.  $y_A$ and $y_a$ represent the numbers of drones in the population 
with the two possible genotypes.  A gyne mates randomly with $n-m$ type $A$ males 
and $m$ type $a$ males in the population (i.e.; the population is well-mixed in terms of mating).  
The binomial coefficient measures all possible ways in which a gyne can mate with 
$m$ males of type $a$ out of $n$ total matings.

Due to environmental factors, the total number of colonies is constant (with value $c$) at all times:
\begin{equation}
\sum_{m=0}^n(X_{AA,m}+X_{Aa,m}+X_{aa,m})=c
\label{eqn:SI_density_unscaled}
\end{equation}
Since environmental resources that are necessary for colonies to thrive are limited, the requirement 
that the total number of colonies is constant in time is biologically justifiable.  
Consequently, $\phi$ in Eqs. \eqref{eqn:SI_dynamics_unscaled} represents a 
density-dependent colony death rate, and we use 
$\phi$ to model the limiting effects of environmental constraints on the total number of 
colonies.  To maintain the density constraint, Eq. \eqref{eqn:SI_density_unscaled}, on the colony 
variables, we choose $\phi$ to equal
\begin{equation}
\phi=c^{-1}(x_{AA}+x_{Aa}+x_{aa})(y_A+y_a)^n
\label{eqn:SI_phi_unscaled}
\end{equation}

We acknowledge that there may be different ways to perform the calculations that follow.  We feel 
that working in continuous time is easiest and simplifies the analysis.  When calculations must be 
performed to second order in a perturbation (for example, when considering the invasion of a 
recessive police allele or the stability of a dominant police allele), working in continuous time 
is quite convenient.

We now introduce the biological parameters of our model.  The appearance of police 
workers affects the fraction of male eggs in a colony that come from the 
queen.  If a fraction $z$ of workers in a colony are policing, then the 
fraction of male offspring that come from the queen is denoted by $p_z$.  
The form of the quantity $p_z$ for $0 \leq z \leq 1$ is likely species-dependent.  
However, it is reasonable to expect that $p_z$ is an increasing function of $z$, 
because an increase in the proportion of workers that are policing results in a larger 
proportion of queen-produced males.

Equally important is the efficiency, $r_z$, of a colony in 
which a fraction $z$ of workers are policing.  Intuitively, the parameter $r_z$ is the total 
number of offspring produced by a 
colony when a fraction $z$ of workers in the colony are policing.  The most important point 
conceptually is that the \emph{ratios} of colony efficiency values, $r_z$, for colonies with different 
genotypes---i.e., the \emph{relative} reproductive efficiencies of colonies with different 
genotypes---are the important quantities for understanding the evolutionary emergence of worker policing.

It is helpful to rewrite everything in Eqs. \eqref{eqn:SI_dynamics_unscaled} in terms of the colony 
variables, $X_{AA,m}$, $X_{Aa,m}$, and $X_{aa,m}$.  Specifically, we must rewrite the first terms on the 
right-hand sides of Eqs. \eqref{eqn:SI_dynamics_unscaled} in terms of the colony variables.  
We express each of the gyne and drone numbers, 
$x_{AA}$, $x_{Aa}$, $x_{aa}$, $y_A$, and $y_a$, as a simple linear sum of the colony variables, 
$X_{AA,m}$, $X_{Aa,m}$, and $X_{aa,m}$.  In these linear relationships, the coefficients depend on whether 
the allele, $a$, acting in a worker to induce that worker's policing behavior is dominant or recessive.

\vspace*{10mm}

\subsection*{Reproductives of Each Type of Colony}

For each colony following haplodiploid genetics, and with single mating of each queen, 
we have the reproduction events shown in Figure~1(b).  
For each colony following haplodiploid genetics, and with $n$ matings of each queen (with $n \geq 1$), 
we have the reproduction events shown in Figure~1(c).

Figure~1(c) can be interpreted as follows.

Consider the individuals produced by type $AA,m$ colonies.  The queen makes $n-m$ type $AA$ females for 
every $m$ type $Aa$ females that she makes.  Because the queen only has the $A$ allele, she 
can only make type $A$ drones.  A fraction $(n-m)/n$ of all workers make only type $A$ males, and 
a fraction $m/n$ of all workers make type $A$ males and type $a$ males in equal numbers.  In all, workers 
make $2n-m$ type $A$ males for every $m$ type $a$ males.

Consider the individuals produced by type $Aa,m$ colonies.  The queen makes $n-m$ type $AA$ females, 
$n$ type $Aa$ females, and $m$ type $aa$ females out of every $2n$ females that she makes.  
Since the queen carries the $A$ and $a$ alleles, she makes type $A$ drones and type $a$ drones in 
equal numbers.  A fraction $(n-m)/(2n)$ of all workers make only type $A$ males, 
a fraction $1/2$ of all workers make type $A$ males and type $a$ males in equal numbers, and a 
fraction $m/(2n)$ of all workers make only type $a$ males.  In all, workers make $3n-2m$ type $A$ 
males for every $n+2m$ type $a$ males.

Consider the individuals produced by type $aa,m$ colonies.  The queen makes $n-m$ type $Aa$ females for 
every $m$ type $aa$ females that she makes.  Because the queen only has the $a$ allele, she 
can only make type $a$ drones.  A fraction $(n-m)/n$ of all workers make type $A$ males and type $a$ 
males in equal numbers, and a fraction $m/n$ of all workers make only type $a$ males.  In all, workers 
make $n-m$ type $A$ males for every $n+m$ type $a$ males.

We do not consider stochastic effects here because the number of individuals produced by 
a colony is assumed to be very large.  Thus, the fractions of colony offspring with each possible 
genotype is always exactly the same for that type of colony.

\vspace*{10mm}

\subsection*{Reproductives with a Dominant Policing Allele}

We focus on the evolution of the $3(n+1)$ colony variables.  An important step is 
to write all quantities in terms of the colony variables.  Each 
type of reproductive of a colony 
($x_{AA}$, $x_{Aa}$, $x_{aa}$, $y_A$, and $y_a$) can be written as a 
simple weighted sum of colony variables.  
From looking at Figure~1(c), 
the numbers of unfertilized females ($x_{AA}$, $x_{Aa}$, and $x_{aa}$) 
and males ($y_A$ and $y_a$) in the population which are reproductive (i.e., capable of mating) can 
be expressed as:
\begin{equation}
\begin{aligned}
x_{AA} =& \sum_{m=0}^{n} \left[ \frac{n-m}{n}gr_\frac{m}{n}X_{AA,m} + \frac{n-m}{2n}gr_\frac{m+n}{2n}X_{Aa,m} \right] \\
x_{Aa} =& \sum_{m=0}^{n} \left[ \frac{m}{n}gr_\frac{m}{n}X_{AA,m} + \frac{1}{2}gr_\frac{m+n}{2n}X_{Aa,m} + \frac{n-m}{n}gr_1X_{aa,m} \right] \\
x_{aa} =& \sum_{m=0}^{n} \left[ \frac{m}{2n}gr_\frac{m+n}{2n}X_{Aa,m} + \frac{m}{n}gr_1X_{aa,m} \right] \\
y_{A}  =& \sum_{m=0}^{n} \left[ \frac{2n-m+mp_\frac{m}{n}}{2n}kr_\frac{m}{n}X_{AA,m} + \frac{3n-2m+(2m-n)p_\frac{m+n}{2n}}{4n}kr_\frac{m+n}{2n}X_{Aa,m} \right. \\
        & \left. + \frac{(n-m)(1-p_1)}{2n}kr_1X_{aa,m} \right] \\
y_{a}  =& \sum_{m=0}^{n} \left[ \frac{m\left(1-p_\frac{m}{n}\right)}{2n}kr_\frac{m}{n}X_{AA,m} + \frac{2m+n+(n-2m)p_\frac{m+n}{2n}}{4n}kr_\frac{m+n}{2n}X_{Aa,m} \right. \\
        & \left. + \frac{n+m+(n-m)p_1}{2n}kr_1X_{aa,m} \right] \\
\end{aligned}
\label{eqn:SI_dom_r_steady_state_unscaled}
\end{equation}
$0<g\leq1$ is the fraction of all females that are gynes.  Likewise, $0<k\leq1$ is 
the fraction of all males that are able to mate.  For instance, we might expect that 
$g\ll1$ and $k\ll1$, which means that only a small amount of the total number of males and 
females produced by each colony are capable of dispersing, mating, and starting new colonies.  
We explicitly write the parameters $g$ and $k$ here for conceptual clarity; they end up being 
irrelevant in the conditions for invasion and stability of police workers.

\vspace*{10mm}

\subsection*{Reproductives with a Recessive Policing Allele}

We focus on the evolution of the $3(n+1)$ colony variables.  An important step is 
to write all quantities in terms of the colony variables.  Each 
type of reproductive of a colony 
($x_{AA}$, $x_{Aa}$, $x_{aa}$, $y_A$, and $y_a$) can be written as a 
simple weighted sum of colony variables.  
From looking at Figure~1(c), 
the numbers of unfertilized females ($x_{AA}$, $x_{Aa}$, and $x_{aa}$) 
and males ($y_A$ and $y_a$) in the population which are reproductive (i.e., capable of mating) can 
be expressed as:
\begin{equation}
\begin{aligned}
x_{AA} =& \sum_{m=0}^{n} \left[ \frac{n-m}{n}gr_0X_{AA,m} + \frac{n-m}{2n}gr_\frac{m}{2n}X_{Aa,m} \right] \\
x_{Aa} =& \sum_{m=0}^{n} \left[ \frac{m}{n}gr_0X_{AA,m} + \frac{1}{2}gr_\frac{m}{2n}X_{Aa,m} + \frac{n-m}{n}gr_\frac{m}{n}X_{aa,m} \right] \\
x_{aa} =& \sum_{m=0}^{n} \left[ \frac{m}{2n}gr_\frac{m}{2n}X_{Aa,m} + \frac{m}{n}gr_\frac{m}{n}X_{aa,m} \right] \\
y_{A}  =& \sum_{m=0}^{n} \left[ \frac{2n-m+mp_0}{2n}kr_0X_{AA,m} + \frac{3n-2m+(2m-n)p_\frac{m}{2n}}{4n}kr_\frac{m}{2n}X_{Aa,m} \right. \\
        & \left. + \frac{(n-m)(1-p_\frac{m}{n})}{2n}kr_\frac{m}{n}X_{aa,m} \right] \\
y_{a}  =& \sum_{m=0}^{n} \left[ \frac{m\left(1-p_0\right)}{2n}kr_0X_{AA,m} + \frac{2m+n+(n-2m)p_\frac{m}{2n}}{4n}kr_\frac{m}{2n}X_{Aa,m} \right. \\
        & \left. + \frac{n+m+(n-m)p_\frac{m}{n}}{2n}kr_\frac{m}{n}X_{aa,m} \right] \\
\end{aligned}
\label{eqn:SI_rec_r_steady_state_unscaled}
\end{equation}
$0<g\leq1$ is the fraction of all females that are gynes.  Likewise, $0<k\leq1$ is 
the fraction of all males that are able to mate.  For instance, we might expect that 
$g\ll1$ and $k\ll1$, which means that only a small amount of the total number of males and 
females produced by each colony are capable of dispersing, mating, and starting new colonies.  
We explicitly write the parameters $g$ and $k$ here for conceptual clarity; they end up being 
irrelevant in the conditions for invasion and stability of police workers.

\vspace*{10mm}

\section*{Rescaling of the Model Variables}

We have presented the biological intuition for our mathematical model that describes the 
population genetics of Hymenopteran insect colonies.  
For the calculations of evolutionary dynamics of worker policing that follow, it is 
mathematically convenient to make the following substitutions:
\begin{equation}
\begin{aligned}
X_{AA,m} &\rightarrow cX_{AA,m}           \\
X_{Aa,m} &\rightarrow cX_{Aa,m}           \\
X_{aa,m} &\rightarrow cX_{aa,m}           \\
x_{AA}   &\rightarrow gcx_{AA}            \\
x_{Aa}   &\rightarrow gcx_{Aa}            \\
x_{aa}   &\rightarrow gcx_{aa}            \\
y_A      &\rightarrow kcy_A               \\
y_a      &\rightarrow kcy_a               \\
\phi     &\rightarrow gk^nc^n\phi         \\
t        &\rightarrow g^{-1}k^{-n}c^{-n}t \\
\end{aligned}
\label{eqn:SI_rescaled}
\end{equation}
Rescaling the model variables and parameters 
according to Eqs. \eqref{eqn:SI_rescaled} leads to simplifications in the mathematics.  
We substitute Eqs. \eqref{eqn:SI_rescaled} into Eqs. \eqref{eqn:SI_dynamics_unscaled} and 
get
\begin{equation}
\begin{aligned}
\dot{X}_{AA,m} = \frac{dX_{AA,m}}{dt} &= {n \choose m} x_{AA}y_A^{n-m}y_a^m - \phi X_{AA,m} \\
\dot{X}_{Aa,m} = \frac{dX_{Aa,m}}{dt} &= {n \choose m} x_{Aa}y_A^{n-m}y_a^m - \phi X_{Aa,m} \\
\dot{X}_{aa,m} = \frac{dX_{aa,m}}{dt} &= {n \choose m} x_{aa}y_A^{n-m}y_a^m - \phi X_{aa,m}
\end{aligned}
\label{eqn:SI_dynamics}
\end{equation}
We substitute Eqs. \eqref{eqn:SI_rescaled} into 
Eq. \eqref{eqn:SI_density_unscaled} and get
\begin{equation}
\sum_{m=0}^n(X_{AA,m}+X_{Aa,m}+X_{aa,m})=1
\label{eqn:SI_density}
\end{equation}
We substitute Eqs. \eqref{eqn:SI_rescaled} into 
Eq. \eqref{eqn:SI_phi_unscaled} and get
\begin{equation}
\phi=(x_{AA}+x_{Aa}+x_{aa})(y_A+y_a)^n
\label{eqn:SI_phi}
\end{equation}

\vspace*{10mm}

\subsection*{Reproductives (Rescaled) with a Dominant Policing Allele}

We substitute Eqs. \eqref{eqn:SI_rescaled} into 
Eqs. \eqref{eqn:SI_dom_r_steady_state_unscaled} and get
\begin{equation}
\begin{aligned}
x_{AA} =& \sum_{m=0}^{n} \left[ \frac{n-m}{n}r_\frac{m}{n}X_{AA,m} + \frac{n-m}{2n}r_\frac{m+n}{2n}X_{Aa,m} \right] \\
x_{Aa} =& \sum_{m=0}^{n} \left[ \frac{m}{n}r_\frac{m}{n}X_{AA,m} + \frac{1}{2}r_\frac{m+n}{2n}X_{Aa,m} + \frac{n-m}{n}r_1X_{aa,m} \right] \\
x_{aa} =& \sum_{m=0}^{n} \left[ \frac{m}{2n}r_\frac{m+n}{2n}X_{Aa,m} + \frac{m}{n}r_1X_{aa,m} \right] \\
y_{A}  =& \sum_{m=0}^{n} \left[ \frac{2n-m+mp_\frac{m}{n}}{2n}r_\frac{m}{n}X_{AA,m} + \frac{3n-2m+(2m-n)p_\frac{m+n}{2n}}{4n}r_\frac{m+n}{2n}X_{Aa,m} \right. \\
        & \left. + \frac{(n-m)(1-p_1)}{2n}r_1X_{aa,m} \right] \\
y_{a}  =& \sum_{m=0}^{n} \left[ \frac{m\left(1-p_\frac{m}{n}\right)}{2n}r_\frac{m}{n}X_{AA,m} + \frac{2m+n+(n-2m)p_\frac{m+n}{2n}}{4n}r_\frac{m+n}{2n}X_{Aa,m} \right. \\
        & \left. + \frac{n+m+(n-m)p_1}{2n}r_1X_{aa,m} \right] \\
\end{aligned}
\label{eqn:SI_dom_r_steady_state}
\end{equation}

\vspace*{10mm}

\subsection*{Reproductives (Rescaled) with a Recessive Policing Allele}

We substitute Eqs. \eqref{eqn:SI_rescaled} into 
Eqs. \eqref{eqn:SI_rec_r_steady_state_unscaled} and get
\begin{equation}
\begin{aligned}
x_{AA} =& \sum_{m=0}^{n} \left[ \frac{n-m}{n}r_0X_{AA,m} + \frac{n-m}{2n}r_\frac{m}{2n}X_{Aa,m} \right] \\
x_{Aa} =& \sum_{m=0}^{n} \left[ \frac{m}{n}r_0X_{AA,m} + \frac{1}{2}r_\frac{m}{2n}X_{Aa,m} + \frac{n-m}{n}r_\frac{m}{n}X_{aa,m} \right] \\
x_{aa} =& \sum_{m=0}^{n} \left[ \frac{m}{2n}r_\frac{m}{2n}X_{Aa,m} + \frac{m}{n}r_\frac{m}{n}X_{aa,m} \right] \\
y_{A}  =& \sum_{m=0}^{n} \left[ \frac{2n-m+mp_0}{2n}r_0X_{AA,m} + \frac{3n-2m+(2m-n)p_\frac{m}{2n}}{4n}r_\frac{m}{2n}X_{Aa,m} \right. \\
        & \left. + \frac{(n-m)(1-p_\frac{m}{n})}{2n}r_\frac{m}{n}X_{aa,m} \right] \\
y_{a}  =& \sum_{m=0}^{n} \left[ \frac{m\left(1-p_0\right)}{2n}r_0X_{AA,m} + \frac{2m+n+(n-2m)p_\frac{m}{2n}}{4n}r_\frac{m}{2n}X_{Aa,m} \right. \\
        & \left. + \frac{n+m+(n-m)p_\frac{m}{n}}{2n}r_\frac{m}{n}X_{aa,m} \right] \\
\end{aligned}
\label{eqn:SI_rec_r_steady_state}
\end{equation}

\vspace*{10mm}

\section*{Differences Between Our Model of Policing and the Model in Olejarz et al. (2015)}

A key point distinguishes our model of policing from the model of non-reproductive workers 
presented in Olejarz et al. (2015).  In our model of policing, mutant workers lay male eggs.  In 
the model of non-reproductive workers in Olejarz et al. (2015), mutant workers do not lay male eggs.  
Therefore, the ``Workers' Sons'' column in Figure~1(c) differs from 
the ``Workers' Sons'' column in Figure~1(b,c) of Olejarz et al. (2015).  Consequently, 
$y_A$ and $y_a$ of Eqs. \eqref{eqn:SI_dom_r_steady_state} differ from $y_A$ and $y_a$ of 
Eqs. (14) in Olejarz et al. (2015).  Also, $y_A$ and $y_a$ of Eqs. \eqref{eqn:SI_rec_r_steady_state} 
differ from $y_A$ and $y_a$ of Eqs. (15) in Olejarz et al. (2015).

\vspace*{10mm}

\section*{Conditions for Evolutionary Invasion and Evolutionary Stability of Worker Policing: Perturbative Analysis}

After rescaling the model variables and parameters according to 
Eqs. \eqref{eqn:SI_rescaled}, the evolutionary dynamics are mathematically unchanged:  
Eqs. \eqref{eqn:SI_dynamics_unscaled} are identical in form to Eqs. \eqref{eqn:SI_dynamics}, 
Eq. \eqref{eqn:SI_density_unscaled} is identical in form to Eq. \eqref{eqn:SI_density}, 
Eq. \eqref{eqn:SI_phi_unscaled} is identical in form to Eq. \eqref{eqn:SI_phi}, 
Eqs. \eqref{eqn:SI_dom_r_steady_state_unscaled} are identical in form to Eqs. \eqref{eqn:SI_dom_r_steady_state}, and 
Eqs. \eqref{eqn:SI_rec_r_steady_state_unscaled} are identical in form to Eqs. \eqref{eqn:SI_rec_r_steady_state}.  
But the rescalings \eqref{eqn:SI_rescaled} are helpful in doing calculations.  
Notice that when the right-hand sides of \eqref{eqn:SI_dynamics} are expressed in terms of the colony frequency 
variables $X_{AA,m}$, $X_{Aa,m}$, and $X_{aa,m}$, the parameters $g$, $k$, and $c$, which are not 
necessary for understanding the evolutionary invasion or evolutionary stability of police 
workers, disappear from the calculations.  This simplifies presentation and clarity in the 
calculations that follow.

To begin, notice that our model admits only two pure equilibria:
\begin{itemize}
\item $X_{AA,0}=1$ with all other colony variables equal to zero.  In this case, the 
$a$ allele is absent from every individual in the population.
\item $X_{aa,n}=1$ with all other colony variables equal to zero.  In this case, the 
$A$ allele is absent from every individual in the population.
\end{itemize}
As seen from Eqs. \eqref{eqn:SI_dynamics}, if any mixed equilibria exist, 
then they will correspond to all $3(n+1)$ colony frequency variables being nonzero.  

\vspace*{10mm}

\subsection{Invasion of a Dominant Worker Policing Allele}
\label{sec:SI_dom_invasion}

We start with an infinitesimal quantity of the mutant allele, $a$, and we 
perturb the $X_{AA,0}=1$ pure equilibrium: 
$X_{AA,0}\rightarrow1-\epsilon\delta^{(1)}_{AA,0}$?  
Does a dominant worker policing allele spread in the population, or is it eliminated?  

There are a total of $3n+3$ types of colonies, and with the density constraint, there are 
$3n+2$ independent colony variables.  However, the calculation simplifies.  
If the perturbation is small 
(i.e. if $\epsilon \ll 1$), then 
only three colony types, $AA,0$, $AA,1$, and $Aa,0$, determine whether 
or not the dominant worker policing allele invades.  Any other colony type has 
a queen that contains at least two mutant $a$ alleles (from her own genotype 
combined with the sperm she has stored), but such queens are so rare that they are negligible.  
The relevant equations among \eqref{eqn:SI_dynamics} for understanding invasion of a dominant 
police allele are
\begin{equation}
\begin{aligned}
\dot{X}_{AA,0} &= x_{AA}y_A^n - \phi X_{AA,0} \\
\dot{X}_{AA,1} &= n x_{AA}y_A^{n-1}y_a - \phi X_{AA,1} \\
\dot{X}_{Aa,0} &= x_{Aa}y_A^n - \phi X_{Aa,0}
\end{aligned}
\label{eqn:SI_dom_dynamics}
\end{equation}
Formally keeping track of powers of $\epsilon$, and 
neglecting higher-order terms, we have:
\begin{equation}
\begin{aligned}
X_{AA,0} &= 1 & -\epsilon\delta^{(1)}_{AA,0} & -\mathcal{O}(\epsilon^2) \\[0.1cm]
X_{AA,1} &=   & +\epsilon\delta^{(1)}_{AA,1} & +\mathcal{O}(\epsilon^2) \\[0.1cm]
X_{Aa,0} &=   & +\epsilon\delta^{(1)}_{Aa,0} & +\mathcal{O}(\epsilon^2)
\end{aligned}
\label{eqn:SI_dom_epsilon}
\end{equation}
We must simplify the density constraint \eqref{eqn:SI_density}.  
We substitute \eqref{eqn:SI_dom_epsilon} into \eqref{eqn:SI_density} 
and collect powers of $\epsilon$.  We get
\begin{equation}
\delta^{(1)}_{AA,0}=\delta^{(1)}_{AA,1}+\delta^{(1)}_{Aa,0}
\label{eqn:SI_density_1}
\end{equation}
Next, we substitute \eqref{eqn:SI_dom_epsilon} 
into \eqref{eqn:SI_dom_r_steady_state}, substituting the density constraint \eqref{eqn:SI_density_1} 
and keeping track of terms only up to order $\epsilon$:
\begin{equation}
\begin{aligned}
x_{AA} =& \; r_0 + \epsilon \left[ \frac{(n-1)r_\frac{1}{n}-nr_0}{n}\delta^{(1)}_{AA,1} + \frac{-2r_0+r_\frac{1}{2}}{2}\delta^{(1)}_{Aa,0} \right] +\mathcal{O}(\epsilon^2) \\
x_{Aa} =& \; \epsilon \left[ \frac{r_\frac{1}{n}}{n}\delta^{(1)}_{AA,1} + \frac{r_\frac{1}{2}}{2}\delta^{(1)}_{Aa,0} \right] +\mathcal{O}(\epsilon^2) \\
y_{A}  =& \; r_0 + \epsilon \left[ \frac{-2nr_0-\left(1-2n-p_\frac{1}{n}\right)r_\frac{1}{n}}{2n}\delta^{(1)}_{AA,1} + \frac{-4r_0+\left(3-p_\frac{1}{2}\right)r_\frac{1}{2}}{4}\delta^{(1)}_{Aa,0} \right] +\mathcal{O}(\epsilon^2) \\
y_{a}  =& \; \epsilon \left[ \frac{1-p_\frac{1}{n}}{2n}r_\frac{1}{n}\delta^{(1)}_{AA,1} + \frac{1+p_\frac{1}{2}}{4}r_\frac{1}{2}\delta^{(1)}_{Aa,0} \right] +\mathcal{O}(\epsilon^2) \\
\end{aligned}
\label{eqn:SI_dom_r_steady_state_epsilon}
\end{equation}
By substituting \eqref{eqn:SI_dom_r_steady_state_epsilon} and \eqref{eqn:SI_dom_epsilon} 
into \eqref{eqn:SI_dom_dynamics}, using the density constraint \eqref{eqn:SI_density_1}, 
and collecting powers of $\epsilon$, we find 
\begin{equation*}
\begin{aligned}
\dot{\delta}^{(1)}_{AA,1} =& \; \frac{-2r_0^{n+1}+\left(1-p_\frac{1}{n}\right)r_\frac{1}{n}r_0^n}{2}\delta^{(1)}_{AA,1}+\frac{n\left(1+p_\frac{1}{2}\right)r_\frac{1}{2}r_0^n}{4}\delta^{(1)}_{Aa,0} \\
\dot{\delta}^{(1)}_{Aa,0} =& \; \frac{r_\frac{1}{n}r_0^n}{n}\delta^{(1)}_{AA,1}+\frac{-2r_0^{n+1}+r_\frac{1}{2}r_0^n}{2}\delta^{(1)}_{Aa,0}
\end{aligned}
\end{equation*}
The equations for $\dot{\delta}^{(1)}_{AA,1}$ and $\dot{\delta}^{(1)}_{Aa,0}$ can be 
written in matrix form as 
\begin{equation*}
\begin{pmatrix} \dot{\delta}^{(1)}_{AA,1} \\ \dot{\delta}^{(1)}_{Aa,0} \end{pmatrix} = \begin{pmatrix} \frac{-2r_0^{n+1}+\left(1-p_\frac{1}{n}\right)r_\frac{1}{n}r_0^n}{2} & \frac{n\left(1+p_\frac{1}{2}\right)r_\frac{1}{2}r_0^n}{4} \\ \frac{r_\frac{1}{n}r_0^n}{n} & \frac{-2r_0^{n+1}+r_\frac{1}{2}r_0^n}{2} \end{pmatrix} \begin{pmatrix} \delta^{(1)}_{AA,1} \\ \delta^{(1)}_{Aa,0} \end{pmatrix}
\end{equation*}
Setting the dominant eigenvalue to be greater than zero and simplifying, 
we find that the dominant allele for worker policing increases in frequency if
\begin{equation}
\frac{p_\frac{1}{n}+p_\frac{1}{2}}{2}\left(\frac{r_\frac{1}{n}}{r_0}\right)\left(\frac{r_\frac{1}{2}}{r_0}\right) > 2-\left(\frac{r_\frac{1}{2}}{r_0}\right)-\left(1-p_\frac{1}{n}\right)\left(\frac{r_\frac{1}{n}}{r_0}\right)
\label{eqn:appendix_dom_r}
\end{equation}

\vspace*{10mm}

\subsection{Invasion of a Recessive Worker Policing Allele}
\label{sec:SI_rec_invasion}

We start with an infinitesimal quantity of the mutant allele, $a$, and we 
perturb the $X_{AA,0}=1$ pure equilibrium: 
$X_{AA,0}\rightarrow1-\epsilon\delta^{(1)}_{AA,0}$?  
Does a recessive worker policing allele spread in the population, or is it eliminated?  

There are a total of $3n+3$ types of colonies, and with the density constraint, there are 
$3n+2$ independent colony variables.  However, the calculation again simplifies.  
If the perturbation is small (i.e. if $\epsilon \ll 1$), then 
only six colony types, $AA,0$, $AA,1$, $Aa,0$, $AA,2$, 
$Aa,1$, and $aa,0$, determine whether 
or not the recessive worker policing allele invades.  Any other colony type has 
a queen that contains at least three mutant $a$ alleles (from her own genotype 
combined with the sperm she has stored), but such queens are so rare that they are negligible.  
The relevant equations among \eqref{eqn:SI_dynamics} for understanding invasion of a recessive 
police allele are
\begin{equation}
\begin{aligned}
\dot{X}_{AA,0} &= x_{AA}y_A^n - \phi X_{AA,0} \\
\dot{X}_{AA,1} &= n x_{AA}y_A^{n-1}y_a - \phi X_{AA,1} \\
\dot{X}_{Aa,0} &= x_{Aa}y_A^n - \phi X_{Aa,0} \\
\dot{X}_{AA,2} &= \frac{n(n-1)}{2} x_{AA}y_A^{n-2}y_a^2 - \phi X_{AA,2} \\
\dot{X}_{Aa,1} &= n x_{Aa}y_A^{n-1}y_a - \phi X_{Aa,1} \\
\dot{X}_{aa,0} &= x_{aa}y_A^n - \phi X_{aa,0}
\end{aligned}
\label{eqn:SI_rec_dynamics}
\end{equation}
Recall that for investigation of the dominant allele, it was only necessary to consider 
terms of order $\epsilon$ to obtain conditions for invasion of the mutant allele.  
For analysis of the recessive allele, terms of order $\epsilon$ 
do not provide all information needed for determining if the allele invades, 
making the calculation more involved.  
Formally keeping track of powers of $\epsilon$ and $\epsilon^2$, and neglecting 
higher-order terms, we have:
\begin{equation}
\begin{aligned}
X_{AA,0} &= 1 & -\epsilon\delta^{(1)}_{AA,0} & -\epsilon^2\delta^{(2)}_{AA,0} & -\mathcal{O}(\epsilon^3) \\[0.1cm]
X_{AA,1} &=   & +\epsilon\delta^{(1)}_{AA,1} & +\epsilon^2\delta^{(2)}_{AA,1} & +\mathcal{O}(\epsilon^3) \\[0.1cm]
X_{Aa,0} &=   & +\epsilon\delta^{(1)}_{Aa,0} & +\epsilon^2\delta^{(2)}_{Aa,0} & +\mathcal{O}(\epsilon^3) \\[0.1cm]
X_{AA,2} &=   &                              & +\epsilon^2\delta^{(2)}_{AA,2} & +\mathcal{O}(\epsilon^3) \\[0.1cm]
X_{Aa,1} &=   &                              & +\epsilon^2\delta^{(2)}_{Aa,1} & +\mathcal{O}(\epsilon^3) \\[0.1cm]
X_{aa,0} &=   &                              & +\epsilon^2\delta^{(2)}_{aa,0} & +\mathcal{O}(\epsilon^3)
\end{aligned}
\label{eqn:SI_rec_epsilon}
\end{equation}
The simplified density constraint, Eq. \eqref{eqn:SI_density_1}, holds 
for the cases of dominant and recessive police alleles.  
We must further simplify the density constraint \eqref{eqn:SI_density} for the case of 
a recessive police allele.  We substitute \eqref{eqn:SI_rec_epsilon} into 
\eqref{eqn:SI_density} and collect powers of $\epsilon^2$.  We get
\begin{equation}
\delta^{(2)}_{AA,0}=\delta^{(2)}_{AA,1}+\delta^{(2)}_{Aa,0}+\delta^{(2)}_{AA,2}+\delta^{(2)}_{Aa,1}+\delta^{(2)}_{aa,0}
\label{eqn:SI_density_2}
\end{equation}
Next, we substitute \eqref{eqn:SI_rec_epsilon} into \eqref{eqn:SI_rec_r_steady_state}, 
substituting the density constraints \eqref{eqn:SI_density_1} and \eqref{eqn:SI_density_2} 
and keeping track of terms up to order $\epsilon^2$:
\begin{equation}
\begin{aligned}
x_{AA}r_0^{-1} =& \; 1 + \epsilon \left[ \frac{-1}{n}\delta^{(1)}_{AA,1} - \frac{1}{2}\delta^{(1)}_{Aa,0} \right] + \epsilon^2 \left[ \frac{-1}{n}\delta^{(2)}_{AA,1} - \frac{1}{2}\delta^{(2)}_{Aa,0} \right. \\
	        & \; \left. - \frac{2}{n}\delta^{(2)}_{AA,2} + \frac{-2n+(n-1)r_\frac{1}{2n}r_0^{-1}}{2n}\delta^{(2)}_{Aa,1} - \delta^{(2)}_{aa,0} \right] +\mathcal{O}(\epsilon^3) \\
x_{Aa}r_0^{-1} =& \; \epsilon \left[ \frac{1}{n}\delta^{(1)}_{AA,1} + \frac{1}{2}\delta^{(1)}_{Aa,0} \right] + \epsilon^2 \left[ \frac{1}{n}\delta^{(2)}_{AA,1} + \frac{1}{2}\delta^{(2)}_{Aa,0} \right. \\
        	& \; \left. + \frac{2}{n}\delta^{(2)}_{AA,2} + \frac{r_\frac{1}{2n}r_0^{-1}}{2}\delta^{(2)}_{Aa,1} + \delta^{(2)}_{aa,0} \right] +\mathcal{O}(\epsilon^3) \\
x_{aa}r_0^{-1} =& \; \epsilon^2 \left[ \frac{r_\frac{1}{2n}r_0^{-1}}{2n}\delta^{(2)}_{Aa,1} \right] +\mathcal{O}(\epsilon^3) \\
y_Ar_0^{-1}    =& \; 1 + \epsilon \left[ - \frac{1-p_0}{2n}\delta^{(1)}_{AA,1} - \frac{1+p_0}{4}\delta^{(1)}_{Aa,0} \right] + \epsilon^2 \left[ - \frac{1-p_0}{2n}\delta^{(2)}_{AA,1} - \frac{1+p_0}{4}\delta^{(2)}_{Aa,0} \right. \\
        	& \; \left. - \frac{1-p_0}{n}\delta^{(2)}_{AA,2} + \frac{-4n+\left(3n-2-(n-2)p_\frac{1}{2n}\right)r_\frac{1}{2n}r_0^{-1}}{4n}\delta^{(2)}_{Aa,1} \right. \\
        	& \; \left. - \frac{1+p_0}{2}\delta^{(2)}_{aa,0} \right] +\mathcal{O}(\epsilon^3) \\
y_ar_0^{-1}    =& \; \epsilon \left[ \frac{1-p_0}{2n}\delta^{(1)}_{AA,1} + \frac{1+p_0}{4}\delta^{(1)}_{Aa,0} \right] + \epsilon^2 \left[ \frac{1-p_0}{2n}\delta^{(2)}_{AA,1} + \frac{1+p_0}{4}\delta^{(2)}_{Aa,0} \right. \\
        	& \; \left. + \frac{1-p_0}{n}\delta^{(2)}_{AA,2} + \frac{\left(n+2+(n-2)p_\frac{1}{2n}\right)r_\frac{1}{2n}r_0^{-1}}{4n}\delta^{(2)}_{Aa,1} \right. \\
        	& \; \left. + \frac{1+p_0}{2}\delta^{(2)}_{aa,0} \right] +\mathcal{O}(\epsilon^3) \\
\end{aligned}
\label{eqn:SI_rec_r_steady_state_epsilon}
\end{equation}
By substituting \eqref{eqn:SI_rec_r_steady_state_epsilon} and \eqref{eqn:SI_rec_epsilon} 
into \eqref{eqn:SI_rec_dynamics}, using the density constraint \eqref{eqn:SI_density_1}, 
and collecting powers of $\epsilon$, we find 
\begin{equation*}
\begin{aligned}
 \dot{\delta}^{(1)}_{AA,1} =& \; \frac{-(1+p_0)}{2}r_0^{n+1}\delta^{(1)}_{AA,1}+\frac{n(1+p_0)}{4}r_0^{n+1}\delta^{(1)}_{Aa,0} \\
 \dot{\delta}^{(1)}_{Aa,0} =& \; \frac{1}{n}r_0^{n+1}\delta^{(1)}_{AA,1}-\frac{1}{2}r_0^{n+1}\delta^{(1)}_{Aa,0}
\end{aligned}
\end{equation*}
The equations for $\dot{\delta}^{(1)}_{AA,1}$ and $\dot{\delta}^{(1)}_{Aa,0}$ can be 
written in matrix form as 
\begin{equation*}
\begin{pmatrix} \dot{\delta}^{(1)}_{AA,1} \\ \dot{\delta}^{(1)}_{Aa,0} \end{pmatrix} = r_0^{n+1} \begin{pmatrix} \frac{-(1+p_0)}{2} & \frac{n(1+p_0)}{4} \\ \frac{1}{n} & \frac{-1}{2} \end{pmatrix} \begin{pmatrix} \delta^{(1)}_{AA,1} \\ \delta^{(1)}_{Aa,0} \end{pmatrix}
\end{equation*}
The two eigenvectors ($v_0$ and $v_-$) and their corresponding eigenvalues 
($\lambda_0$ and $\lambda_-$) are
\begin{equation*}
\begin{aligned}
v_0 =& \begin{pmatrix} n \\ 2 \end{pmatrix} & \qquad & \lambda_0 = 0 \\
v_- =& \begin{pmatrix} n(1+p_0) \\ -2 \end{pmatrix} & \qquad & \lambda_- = \frac{-(2+p_0)}{2} r_0^{n+1}
\end{aligned}
\end{equation*}
Because the dominant eigenvalue is equal to zero, a calculation to leading 
order in $\epsilon$ cannot provide a condition for the invasion of the recessive 
police allele.  

To see this more formally, an arbitrary initial perturbation to a 
resident $A$ population can be expressed as a linear superposition of the eigenvectors 
$v_0$ and $v_-$:
\begin{equation}
\begin{pmatrix} \delta^{(1)}_{AA,1} \\ \delta^{(1)}_{Aa,0} \end{pmatrix} = C_0 \begin{pmatrix} n \\ 2 \end{pmatrix} + C_- \begin{pmatrix} n(1+p_0) \\ -2 \end{pmatrix} \exp\left(\frac{-(2+p_0)}{2}r_0^{n+1}t\right) \\
\label{eqn:SI_rec_r_superposition}
\end{equation}
Here, $C_0$ and $C_-$ are constants.  
We can substitute \eqref{eqn:SI_rec_r_steady_state_epsilon} and \eqref{eqn:SI_rec_epsilon} 
into \eqref{eqn:SI_rec_dynamics}, substituting the density constraints \eqref{eqn:SI_density_1} and 
\eqref{eqn:SI_density_2}, keeping track of terms of order $\epsilon$ and $\epsilon^2$, and dividing 
each term by one factor of $\epsilon$.  We get
\begin{equation}
\begin{aligned}
&       \left[-\dot{\delta}^{(1)}_{AA,0}-\epsilon\dot{\delta}^{(2)}_{AA,0}\right]r_0^{-(n+1)} \\
&       = \frac{2-n-np_0}{4n}\left(-2\delta^{(1)}_{AA,1}+n\delta^{(1)}_{Aa,0}\right) \\
& \quad + \epsilon \left[ \frac{2-n-np_0}{4n}\left(-2\delta^{(2)}_{AA,1}+n\delta^{(2)}_{Aa,0}\right) \right. \\
& \quad \left. +\frac{-2+np_0}{n}\delta^{(2)}_{AA,2} \right. \\
& \quad +\frac{-2r_\frac{1}{2n}r_0^{-1}+2n\left[2-\left(2-p_\frac{1}{2n}\right)r_\frac{1}{2n}r_0^{-1}\right]-n^2\left(1+p_\frac{1}{2n}\right)r_\frac{1}{2n}r_0^{-1}}{4n}\delta^{(2)}_{Aa,1} \\
& \quad \left. -\frac{n(1+p_0)}{2}\delta^{(2)}_{aa,0} \right. \\
& \quad \left. +\frac{(1-p_0)[3+n(1-p_0)+p_0]}{8n}\left[\delta^{(1)}_{AA,1}\right]^2 \right. \\
& \quad \left. +\frac{n(1+p_0)[3+n+(n-1)p_0]}{32}\left[\delta^{(1)}_{Aa,0}\right]^2 \right. \\
& \quad \left. +\frac{3+n-(n-1)p_0^2}{8}\delta^{(1)}_{AA,1}\delta^{(1)}_{Aa,0} \right]
\label{eqn:SI_rec_r_time_scales}
\end{aligned}
\end{equation}
We can again substitute the density constraints \eqref{eqn:SI_density_1} and 
\eqref{eqn:SI_density_2} to rewrite the left-hand side of 
\eqref{eqn:SI_rec_r_time_scales}.  We can also substitute the general solution 
for the quantities $\delta^{(1)}_{AA,1}$ and $\delta^{(1)}_{Aa,0}$, 
Eq. \eqref{eqn:SI_rec_r_superposition}, into the right-hand side of 
\eqref{eqn:SI_rec_r_time_scales}:
\begin{equation}
\begin{aligned}
&       \left[-\dot{\delta}^{(1)}_{AA,1}-\dot{\delta}^{(1)}_{Aa,0}\right]r_0^{-(n+1)} + \epsilon \left[ -\dot{\delta}^{(2)}_{AA,1}-\dot{\delta}^{(2)}_{Aa,0}-\dot{\delta}^{(2)}_{AA,2}-\dot{\delta}^{(2)}_{Aa,1}-\dot{\delta}^{(2)}_{aa,0} \right]r_0^{-(n+1)} \\
&       = \frac{2-n-np_0}{4n}\left[-2\left(nC_0+n(1+p_0)C_-\exp\left(\frac{-(2+p_0)}{2}r_0^{n+1}t\right)\right) \right. \\
& \quad \left. +n\left(2C_0-2C_-\exp\left(\frac{-(2+p_0)}{2}r_0^{n+1}t\right)\right)\right] \\
& \quad + \epsilon \left[ \frac{2-n-np_0}{4n}\left(-2\delta^{(2)}_{AA,1}+n\delta^{(2)}_{Aa,0}\right) \right. \\
& \quad \left. +\frac{-2+np_0}{n}\delta^{(2)}_{AA,2} \right. \\
& \quad +\frac{-2r_\frac{1}{2n}r_0^{-1}+2n\left[2-\left(2-p_\frac{1}{2n}\right)r_\frac{1}{2n}r_0^{-1}\right]-n^2\left(1+p_\frac{1}{2n}\right)r_\frac{1}{2n}r_0^{-1}}{4n}\delta^{(2)}_{Aa,1} \\
& \quad \left. -\frac{n(1+p_0)}{2}\delta^{(2)}_{aa,0} \right. \\
& \quad \left. +\frac{(1-p_0)[3+n(1-p_0)+p_0]}{8n}\left[\delta^{(1)}_{AA,1}\right]^2 \right. \\
& \quad \left. +\frac{n(1+p_0)[3+n+(n-1)p_0]}{32}\left[\delta^{(1)}_{Aa,0}\right]^2 \right. \\
& \quad \left. +\frac{3+n-(n-1)p_0^2}{8}\delta^{(1)}_{AA,1}\delta^{(1)}_{Aa,0} \right]
\end{aligned}
\label{eqn:SI_rec_r_time_scales_substitution}
\end{equation}
Notice that each term in \eqref{eqn:SI_rec_r_time_scales_substitution} 
involving the quantities $\delta^{(2)}_{AA,1}$, 
$\delta^{(2)}_{Aa,0}$, $\delta^{(2)}_{AA,2}$, $\delta^{(2)}_{Aa,1}$, and 
$\delta^{(2)}_{aa,0}$ is multiplied by $\epsilon$.  In the limit 
$\epsilon\rightarrow0$, the quantities $\delta^{(2)}_{AA,1}$, 
$\delta^{(2)}_{Aa,0}$, $\delta^{(2)}_{AA,2}$, $\delta^{(2)}_{Aa,1}$, and 
$\delta^{(2)}_{aa,0}$ are irrelevant to the dynamics of the quantities 
$\delta^{(1)}_{AA,1}$ and $\delta^{(1)}_{Aa,0}$.  However, the quantities 
$\delta^{(1)}_{AA,1}$ and $\delta^{(1)}_{Aa,0}$ alone do not provide information about 
whether or not the recessive police allele invades a resident $A$ 
population.  Therefore, it is necessary to consider the terms of order $\epsilon^2$ in our 
dynamical equations \eqref{eqn:SI_rec_dynamics} to determine if a rare $a$ allele can 
invade a resident $A$ population.  In what follows, we use the 
eigenvector $v_0$ corresponding to the zero eigenvalue, i.e.
\begin{equation}
\begin{pmatrix} \delta^{(1)}_{AA,1} \\ \delta^{(1)}_{Aa,0} \end{pmatrix} = \frac{\delta^{(1)}_{AA,0}}{n+2} \begin{pmatrix} n \\ 2 \end{pmatrix} \\
\label{eqn:SI_rec_r_v_0}
\end{equation}

Substituting \eqref{eqn:SI_rec_r_steady_state_epsilon}, \eqref{eqn:SI_rec_epsilon}, 
and \eqref{eqn:SI_rec_r_v_0} into \eqref{eqn:SI_rec_dynamics}, substituting the density 
constraints \eqref{eqn:SI_density_1} and \eqref{eqn:SI_density_2}, and keeping track of 
terms of order $\epsilon^2$, we obtain
\begin{equation}
\begin{aligned}
-\dot{\delta}^{(2)}_{AA,0}r_0^{-(n+1)} =& \; \frac{2-n-np_0}{4n}\left(-2\delta^{(2)}_{AA,1}+n\delta^{(2)}_{Aa,0}\right) \\
                            & +\frac{-2+np_0}{n}\delta^{(2)}_{AA,2} \\
                            & +\frac{-2r_\frac{1}{2n}r_0^{-1}+2n\left[2-\left(2-p_\frac{1}{2n}\right)r_\frac{1}{2n}r_0^{-1}\right]-n^2\left(1+p_\frac{1}{2n}\right)r_\frac{1}{2n}r_0^{-1}}{4n}\delta^{(2)}_{Aa,1} \\
                            & -\frac{n(1+p_0)}{2}\delta^{(2)}_{aa,0} \\
                            & +\frac{n(n+3)}{2(n+2)^2}\left[\delta^{(1)}_{AA,0}\right]^2
\label{eqn:SI_rec_r_d_dt_AA0}
\end{aligned}
\end{equation}
We also obtain
\begin{equation*}
\begin{aligned}
 \dot{\delta}^{(2)}_{AA,1}r_0^{-(n+1)} =& \; \frac{1+p_0}{4}\left(-2\delta^{(2)}_{AA,1}+n\delta^{(2)}_{Aa,0}\right) \\
                                        & +(1-p_0)\delta^{(2)}_{AA,2} \\
                                        & +\frac{n+2+(n-2)p_\frac{1}{2n}}{4}r_\frac{1}{2n}r_0^{-1}\delta^{(2)}_{Aa,1} \\
                                        & +\frac{n(1+p_0)}{2}\delta^{(2)}_{aa,0} \\
                                        & -\frac{n(n+1)}{(n+2)^2}\left[\delta^{(1)}_{AA,0}\right]^2 \\
 \dot{\delta}^{(2)}_{Aa,0}r_0^{-(n+1)} =& \; \frac{-1}{2n}\left(-2\delta^{(2)}_{AA,1}+n\delta^{(2)}_{Aa,0}\right) \\
                                        & +\frac{2}{n}\delta^{(2)}_{AA,2} \\
                                        & +\frac{1}{2}r_\frac{1}{2n}r_0^{-1}\delta^{(2)}_{Aa,1} \\
                                        & +\delta^{(2)}_{aa,0} \\
                                        & -\frac{2n}{(n+2)^2}\left[\delta^{(1)}_{AA,0}\right]^2 \\
 \dot{\delta}^{(2)}_{AA,2}r_0^{-(n+1)} =& \; -\delta^{(2)}_{AA,2}+\frac{n(n-1)}{2(n+2)^2}\left[\delta^{(1)}_{AA,0}\right]^2 \\
 \dot{\delta}^{(2)}_{Aa,1}r_0^{-(n+1)} =& \; -\delta^{(2)}_{Aa,1}+\frac{2n}{(n+2)^2}\left[\delta^{(1)}_{AA,0}\right]^2 \\
 \dot{\delta}^{(2)}_{aa,0}r_0^{-(n+1)} =& \; -\delta^{(2)}_{aa,0}+\frac{1}{2n}r_\frac{1}{2n}r_0^{-1}\delta^{(2)}_{Aa,1}
\end{aligned}
\label{eqn:SI_rec_derivative_AAA_2_r}
\end{equation*}
We can directly integrate the equation for $\dot{\delta}^{(2)}_{AA,2}$.  We get
\begin{equation}
\delta^{(2)}_{AA,2} = \frac{n(n-1)}{2(n+2)^2}\left[\delta^{(1)}_{AA,0}\right]^2[1-\exp(-r_0^{n+1}t)]
\label{eqn:SI_rec_r_AA2}
\end{equation}
We can also directly integrate the equation for $\dot{\delta}^{(2)}_{Aa,1}$.  We get
\begin{equation}
\delta^{(2)}_{Aa,1} = \frac{2n}{(n+2)^2}\left[\delta^{(1)}_{AA,0}\right]^2[1-\exp(-r_0^{n+1}t)]
\label{eqn:SI_rec_r_Aa1}
\end{equation}
We can use the solution for $\delta^{(2)}_{Aa,1}$ to solve for $\delta^{(2)}_{aa,0}$.  We get
\begin{equation}
\delta^{(2)}_{aa,0} = \frac{r_\frac{1}{2n}}{r_0(n+2)^2}\left[\delta^{(1)}_{AA,0}\right]^2[1-(1+r_0^{n+1}t)\exp(-r_0^{n+1}t)]
\label{eqn:SI_rec_r_aa0}
\end{equation}
Manipulating the equations for $\dot{\delta}^{(2)}_{AA,1}$ and 
$\dot{\delta}^{(2)}_{Aa,0}$, we obtain
\begin{equation*}
\begin{aligned}
r_0^{-(n+1)}\frac{d}{dt}\left(-2\delta^{(2)}_{AA,1}+n\delta^{(2)}_{Aa,0}\right) =& \; \frac{-(2+p_0)}{2}\left(-2\delta^{(2)}_{AA,1}+n\delta^{(2)}_{Aa,0}\right) \\
                                                                                 & +2p_0\delta^{(2)}_{AA,2} \\
                                                                                 & -\frac{2+(n-2)p_\frac{1}{2n}}{2}r_\frac{1}{2n}r_0^{-1}\delta^{(2)}_{Aa,1} \\
                                                                                 & -np_0\delta^{(2)}_{aa,0} \\
                                                                                 & +\frac{2n}{(n+2)^2}\left[\delta^{(1)}_{AA,0}\right]^2 \\
\end{aligned}
\end{equation*}
We can integrate this equation to solve for the quantity 
$-2\delta^{(2)}_{AA,1}+n\delta^{(2)}_{Aa,0}$.  We obtain
\begin{equation}
\begin{aligned}
&       -2\delta^{(2)}_{AA,1}+n\delta^{(2)}_{Aa,0} \\
&       = \frac{2n\left[2+(n-1)p_0-\left(2+p_0+(n-2)p_\frac{1}{2n}\right)r_\frac{1}{2n}r_0^{-1}\right]}{(n+2)^2(2+p_0)}\left[\delta^{(1)}_{AA,0}\right]^2 \\
& \quad +\frac{2n\left[(n-2)p_\frac{1}{2n}r_\frac{1}{2n}r_0^{-1}-p_0\left(n-1-(1+r_0^{n+1}t)r_\frac{1}{2n}r_0^{-1}\right)\right]}{(n+2)^2p_0}\left[\delta^{(1)}_{AA,0}\right]^2\exp(-r_0^{n+1}t) \\
& \quad +\frac{4n(n-2)\left(p_0-p_\frac{1}{2n}r_\frac{1}{2n}r_0^{-1}\right)}{(n+2)^2p_0(2+p_0)}\left[\delta^{(1)}_{AA,0}\right]^2\exp\left(\frac{-(2+p_0)}{2}r_0^{n+1}t\right) \\
\end{aligned}
\label{eqn:SI_rec_r_AA1_Aa0}
\end{equation}

To see if the $a$ allele is able to invade the resident $A$ population, 
we must consider the regime $1 \ll t \ll \epsilon^{-1}$.  Notice that on a short time 
scale, each of the terms with time-dependence in Eqs. \eqref{eqn:SI_rec_r_AA2}, 
\eqref{eqn:SI_rec_r_Aa1}, \eqref{eqn:SI_rec_r_aa0}, and \eqref{eqn:SI_rec_r_AA1_Aa0} 
will approach zero.  
The important consideration is the sign of $\dot{\delta}^{(2)}_{AA,0}$ in the limit of large times 
$t \gg 1$ but before the terms in \eqref{eqn:SI_rec_epsilon} become comparable in 
magnitude.  Therefore, our condition for invasion of the police allele is 
\begin{equation}
\lim_{\substack{\epsilon t \rightarrow 0 \\ t \rightarrow \infty}}\dot{\delta}^{(2)}_{AA,0}>0
\label{eqn:SI_rec_r_limit}
\end{equation}
Substituting \eqref{eqn:SI_rec_r_d_dt_AA0}, \eqref{eqn:SI_rec_r_AA2}, 
\eqref{eqn:SI_rec_r_Aa1}, \eqref{eqn:SI_rec_r_aa0}, and \eqref{eqn:SI_rec_r_AA1_Aa0} 
into \eqref{eqn:SI_rec_r_limit}, we find that the recessive allele for worker policing 
increases in frequency if
\begin{equation}
\frac{r_\frac{1}{2n}}{r_0} > \frac{2(2+n+np_0)}{(2+n)(2+p_0)+p_\frac{1}{2n}(n-2)}
\label{eqn:appendix_rec_r}
\end{equation}

\vspace*{10mm}

\subsection{Stability of a Dominant Worker Policing Allele}

If we start with a population in which all workers are policing, and if we introduce a small amount of the non-policing allele, $A$, and if the policing allele, $a$, is dominant, then is policing evolutionarily stable?

There is a shortcut to obtaining the evolutionary stability condition for a dominant policing allele.  Recall what the calculations of \ref{sec:SI_rec_invasion} for invasion of a recessive policing allele are describing:  We start with a homogeneous population of colonies in which all individuals carry only the $A$ allele.  A fraction $p_0$ of males are queen-derived, and each colony has reproductive efficiency $r_0$.  The only effects of the mutant allele are to shift the value of $p$ such that $p_z>p_0$ for $z>0$ and to modify the colony efficiency, $r_z \neq r_0$ for $z>0$.  $z$ here is the fraction of workers in a colony that are homozygous in the $a$ allele.

Now consider the evolutionary stability of a dominant $a$ allele that effects policing.  Again, in the calculation, we start with a homogeneous population of colonies, but in this case, all individuals carry only the $a$ allele.  A fraction $p_1$ of males are queen-derived, and each colony has reproductive efficiency $r_1$.  The only effects of the mutant allele are to shift the value of $p$ such that $p_{1-z}<p_1$ for $z>0$ and to modify the colony efficiency, $r_{1-z} \neq r_1$ for $z>0$.  $z$ here is the fraction of workers in a colony that are homozygous in the $A$ allele.  (Since the $a$ allele for policing is dominant, $AA$ workers have a distinct phenotype, while $Aa$ and $aa$ workers have the same phenotype.)

Notice that, for worker policing, the mathematical analysis of invasion of a recessive policing allele necessarily also provides the condition for evolutionary stability of a dominant policing allele.  Start with Condition \eqref{eqn:appendix_rec_r}.  Because the genotype of the initial population that we consider is reversed (i.e., all individuals initially carry only the $a$ allele as opposed to the $A$ allele), we make the replacement $z \rightarrow 1-z$ in \eqref{eqn:appendix_rec_r}.  Also, evolutionary stability of the $a$ allele means that the mutant allele for non-policing, $A$, is unable to invade a resident $a$ population, so we reverse the sign of the inequality in \eqref{eqn:appendix_rec_r}.

Thus, if the police allele, $a$, is dominant, then worker policing is evolutionarily stable if
\begin{equation*}
\frac{r_1}{r_\frac{2n-1}{2n}} > \frac{(2+n)(2+p_1)+p_\frac{2n-1}{2n}(n-2)}{2(2+n+np_1)}
\end{equation*}

\vspace*{10mm}

\subsection{Stability of a Recessive Worker Policing Allele}

If we start with a population in which all workers are policing, and if we introduce a small amount of the non-policing allele, $A$, and if the policing allele, $a$, is recessive, then is policing evolutionarily stable?

There is a shortcut to obtaining the evolutionary stability condition for a recessive policing allele.  Recall what the calculations of \ref{sec:SI_dom_invasion} for invasion of a dominant policing allele are describing:  We start with a homogeneous population of colonies in which all individuals carry only the $A$ allele.  A fraction $p_0$ of males are queen-derived, and each colony has reproductive efficiency $r_0$.  The only effects of the mutant allele are to shift the value of $p$ such that $p_z>p_0$ for $z>0$ and to modify the colony efficiency, $r_z \neq r_0$ for $z>0$.  $z$ here is the fraction of workers in a colony that carry at least one copy of the $a$ allele.

Now consider the evolutionary stability of a recessive $a$ allele that effects policing.  Again, in the calculation, we start with a homogeneous population of colonies, but in this case, all individuals carry only the $a$ allele.  A fraction $p_1$ of males are queen-derived, and each colony has reproductive efficiency $r_1$.  The only effects of the mutant allele are to shift the value of $p$ such that $p_{1-z}<p_1$ for $z>0$ and to modify the colony efficiency, $r_{1-z} \neq r_1$ for $z>0$.  $z$ here is the fraction of workers in a colony that carry at least one copy of the $A$ allele.  (Since the $a$ allele for policing is recessive, $AA$ and $Aa$ workers have the same phenotype, while $aa$ workers have a distinct phenotype.)

Notice that, for worker policing, the mathematical analysis of invasion of a dominant policing allele necessarily also provides the condition for evolutionary stability of a recessive policing allele.  Start with Condition \eqref{eqn:appendix_dom_r}.  Because the genotype of the initial population that we consider is reversed (i.e., all individuals initially carry only the $a$ allele as opposed to the $A$ allele), we make the replacement $z \rightarrow 1-z$ in \eqref{eqn:appendix_dom_r}.  Also, evolutionary stability of the $a$ allele means that the mutant allele for non-policing, $A$, is unable to invade a resident $a$ population, so we reverse the sign of the inequality in \eqref{eqn:appendix_dom_r}.

Thus, if the police allele, $a$, is recessive, then worker policing is evolutionarily stable if
\begin{equation*}
\left(\frac{r_1}{r_\frac{n-1}{n}}\right)\left[2\left(\frac{r_1}{r_\frac{1}{2}}\right)-1\right]-\left(1-p_\frac{n-1}{n}\right)\left(\frac{r_1}{r_\frac{1}{2}}\right) > \frac{p_\frac{n-1}{n}+p_\frac{1}{2}}{2}
\end{equation*}

\end{document}